\documentclass[usenatbib,usegraphicx,letterpaper]{mn2e}
\usepackage[totalwidth=480pt,totalheight=680pt]{geometry}
\pdfoutput=1

\usepackage{amssymb}
\usepackage{epsfig}
\usepackage{amsmath}
\usepackage{color}
\usepackage{epsfig}  
\usepackage{graphicx}  
 \usepackage{rotating}
\newcommand{\unit}[1]{\mathrm{#1}}

\newcommand{\wprp}{w_{\mathrm{p}}(r_{\mathrm{p}})}

\newcommand{\kms}{\unit{km \ s^{-1}}}

\newcommand{\hmpc}{h^{-1}\mathrm{Mpc}}

\newcommand{\hkpc}{h^{-1}\mathrm{kpc}}
\newcommand{\hMsun}{h^{-1}M_{\odot}}

\newcommand{\Mmin}{M_\mathrm{min}}
\newcommand{\Mvir}{M_\mathrm{vir}} 
\newcommand{\Mhost}{M_\mathrm{host}} 
\newcommand{\Mstar}{M_{\ast}}

\newcommand{\Msun}{M_{\odot}}

\newcommand{\vmax}{\mathrm{V}_\mathrm{max}}

\newcommand{\vpeak}{\mathrm{V}_\mathrm{peak}}

\newcommand{\zform}{z_{\mathrm{form}}}

\newcommand{\zstarve}{z_{\mathrm{starve}}}

\newcommand{\Nsat}{N_\mathrm{sat}}
\newcommand{\Ncen}{N_\mathrm{cen}}

\newcommand{\Ngal}{\langle N_{\mathrm{gal}}\rangle}

\newcommand{\Psdss}{P_{\mathrm{SDSS}} ( g-r | L_r)}

\newcommand{\PNM}{\mathrm{P}(N \vert M)}
\newcommand{\philm}{\Phi(L|M)}

\newcommand{\wproj}{w_{\mathrm{p}}}
\newcommand{\rproj}{r_{\mathrm{p}}}

\newcommand{\fq}{F^{\mathrm{sat}}_{\mathrm{q}}}

\newcommand{\beq}{\begin{equation}}
\newcommand{\eeq}{\end{equation}}
\newcommand{\beqray}{\begin{eqnarray}}
\newcommand{\eeqray}{\end{eqnarray}}

\newcommand{\ben}{\begin{enumerate}}
\newcommand{\een}{\end{enumerate}}
\newcommand{\bit}{\begin{itemize}}
\newcommand{\eit}{\end{itemize}}

\begin{document}

\title[Assembly Bias and the Galaxy-Halo Relationship]{Galaxy Assembly Bias: A Significant Source of Systematic Error in the Galaxy-Halo Relationship}

\author[A.R.~Zentner, A.P.~Hearin, \& F.~C. van den Bosch]
{Andrew R. Zentner$^{1}$, 
Andrew P. Hearin$^{2}$, 
Frank C. van den Bosch$^{3}$ \\
  $^1$ Department of Physics and Astronomy \& \\
Pittsburgh Particle physics, Astrophysics and Cosmology Center (PITT PACC),\\ 
University of Pittsburgh, Pittsburgh, PA 15260;\\
$^2$ Fermilab Center for Particle Astrophysics, Fermi National Accelerator Laboratory, Batavia, IL, 60510-0500 \\
$^3$ Department of Astronomy, Yale University, P. O. Box 208101, New Haven, CT 06520-8101
}

\maketitle

%%%%%%%%%%%%%%%%%%%%%%%%%%%% ABSTRACT %%%%%%%%%%%%%%%%%%%%%%%%%%%%
\begin{abstract}

Methods that exploit galaxy clustering to constrain the galaxy-halo relationship, 
such as the halo occupation distribution (HOD) and conditional luminosity 
function (CLF), assume halo mass alone suffices to determine a halo's
galaxy content. Yet, halo clustering strength depends upon properties other than mass, 
such as formation time, an effect known as {\em assembly bias}. {\em If} galaxy 
characteristics are correlated with these auxiliary halo properties, the basic 
assumption of standard HOD/CLF methods is violated. We estimate the potential
for assembly bias to induce systematic errors in inferred 
halo occupation statistics. We construct realistic mock galaxy catalogues 
that exhibit assembly bias as well as companion mock catalogues 
with identical HODs, but with assembly bias removed. We fit 
HODs to the galaxy clustering in each catalogue. In the absence of 
assembly bias, the inferred HODs describe the true HODs well, validating 
the methodology. However, in all cases {\em with} assembly bias, the inferred 
HODs exhibit {\em significant systematic errors}. We conclude that the galaxy-halo 
relationship inferred from galaxy clustering is subject to significant 
systematic errors induced by assembly bias. Efforts 
to model and/or constrain assembly bias should be priorities as assembly 
bias is a threatening source of systematic error in galaxy evolution and 
precision cosmology studies.

\end{abstract}

%%%%%%%%%%%%%%%%%%%%%%%%%%%%%%%%%%%%%%%%%%%%%%%%%%%%%%%%%%%%%%%%%%


%%%%%%%%%%%%%%%%%%%%%%%%%%%%%% INTRO %%%%%%%%%%%%%%%%%%%%%%%%%%%%%%

\section{INTRODUCTION}
\label{sec:intro}

%%%%%%%%%%%%%%%%%%%%%%%%%%%%%%%%%%%%%%%%%%%%%%%%%%%%%%%%%%%%%%%%

Theoretical models connecting galaxies to dark matter halos unlock the predictive power of cosmological N-body simulations. 
The two most widely used models of the galaxy-halo connection are the Halo Occupation Distribution 
\citep[HOD, e.g.,][]{berlind02,zheng05} and the Conditional Luminosity Function 
\citep[CLF, e.g.,][]{yang03,vdBosch13}. The central quantity in the HOD is $\PNM,$ 
the probability that a halo of mass $M$ hosts $N$ galaxies brighter than some luminosity threshold. 
The CLF instead models $\philm,$ the mean abundance of galaxies of luminosity 
$L$ that reside in a dark matter halo of mass $M.$ 
These formalisms are closely related: an HOD can be derived by integrating the CLF against luminosity; 
a CLF can be derived by differentiating the HOD with respect to luminosity. Both formalisms have been 
studied extensively 
to constrain the galaxy-halo connection 
\citep[an incomplete list of recent examples includes][]{magliocchetti03,zehavi05a,yang_etal05,zheng_etal07,vdBosch07,zheng09,skibba_sheth09,simon_etal09,ross_etal10,zehavi11,geach_etal12,parejko_etal13} as well as cosmology 
\citep{tinker05,leauthaud11a,more_etal13,cacciato_etal13,mandelbaum_etal13}.

All conventional formulations of both the HOD and the CLF assume that galaxy occupation statistics are 
governed exclusively by the masses of the dark matter halos hosting the galaxies of interest. We will 
refer to any such formulation of the HOD or CLF formalisms as the "standard" approach henceforth. 
In this paper, we explore a class of simple, but well-motivated models for the galaxy-halo connection in which 
the assumption that halo statistics depend upon mass alone is violated, and demonstrate that the degree to which 
such violations lead to systematic errors in the inferred relationship between galaxies and halos can be significant.

That halo mass should be the halo property that most strongly influences the properties of 
the galaxies within them is now widely accepted and has significant and long-standing 
theoretical support \citep[e.g.,][]{whiterees78,blumenthal_etal84}. 
Related to this, it has also long been recognized that halo mass is the halo property that most 
strongly influences halo abundance and halo clustering. The theoretical underpinnings of this fact lie in the uncorrelated 
nature of the random walks describing halo assembly and clustering in the simplest implementations of the excursion set 
formalism \citep{press_schechter74,Bond91,LaceyCole93,mo_white96,zentner07}. 
Yet, the spatial distribution of dark matter halos in dissipationless N-body simulations depends 
not only on halo mass, but also on additional properties such as 
formation time \citep{gao_etal05,wechsler06,gao_white07,li_etal08}. Moreover, the lack of any correlation 
between halo environment and formation time is an assumption of the simplest excursion set models, rather than 
a derived property of halos. Relaxing this assumption, excursion set theory also predicts a correlation between 
halo formation time and halo environment \citep[e.g.,][]{zentner07,dalal_etal08}. 
The dependence of the spatial distribution of 
dark matter halos upon properties besides mass is generically 
referred to as ``halo assembly bias,'' and is typically quantified 
in terms of the halo two-point correlation function.

The character of assembly bias depends on halo mass relative to $\Mstar,$ 
the characteristic collapse mass at a given redshift: for halos at fixed mass 
$M \ll \Mstar,$ older halos cluster more strongly than young \cite{gao_white07}; 
for halos with $\Mvir \gg \Mstar,$ the reverse is true \citep{wechsler06}. 
The physical explanation for this trend was laid out in 
\citet{zentner07} and \citet{dalal_etal08}. At the high-mass end, assembly 
bias is expected purely from the statistics of peaks in the initial density field. At the low-mass end, 
halo assembly bias arises due to the cessation of mass accretion onto halos residing in dense environments 
\citep[See also][]{wang_etal09,lacerna_padilla11}. This correlation between halo formation time $\zform$, and 
environment suggests that other halo properties that are correlated with $\zform$ will also exhibit ``assembly bias'' 
trends, including concentration, triaxility, spin, and velocity anisotropy. Indeed, this is the case 
\citep{wechsler06,faltenbacher_white10,lacerna_padilla12}. If the properties of the galaxies that 
reside in a halo are correlated with any of these properties that are known to be correlated with halo 
assembly, then the standard HOD/CLF assumption will be violated and these models will not be able 
to predict the clustering statistics of galaxies correctly. Of course, this violation could be 
sufficiently weak as to be of little practical importance, 
but whether or not this is the case remains an open question.

\citet{croton_etal07} studied the effect of assembly bias in semi-analytic models 
of galaxy formation on three-dimensional galaxy clustering, finding that assembly 
bias is an important ingredient in determining the clustering strengths of galaxies. 
In their models, the effect of assembly bias depends non-trivially 
on galaxy luminosity and color. Disconcertingly, \citet{croton_etal07} 
find that the assembly bias effects they detect cannot be accounted for with 
either halo formation time or concentration alone. This suggests that 
galaxy properties may depend upon the assembly histories of halos in a 
sufficiently complicated manner as to make empirical modeling extremely challenging.

Observational investigations of assembly bias in the galaxy
distribution have produced mixed results. \citet{yang_etal06} studied
the cross-correlation between galaxies and galaxy groups in the Two
Degree Field Galaxy Redshift Survey \citep[2dFGRS,][]{colless01}, and
found that for groups of the same mass, the correlation strength
depends on the star formation rate (SFR) of the central galaxy: at
fixed mass, the clustering strength of galaxy groups decreases as the SFR of the
central galaxy increases. \citet{wang_etal08} and \citet{wang_etal13} confirmed these
findings using much larger data sets obtained from the SDSS, showing 
that the color dependence is more prominent in less massive groups,
and demonstrating that these results are consistent with predictions
from semi-analytical models. 
These studies suggest that assembly bias may be present in the 
observed galaxy distribution at a statistically significant level.\footnote{See also \citet{cooper_etal10} for a 
reported detection of assembly bias that does not employ a group-finder.} 
On the other hand,  \citet{blanton_berlind07} use a different technique as well as an alternative SDSS group
catalogue, and found little, if any, evidence for assembly bias on large
scales, and only a modest signal on small scales $(r \lesssim 300\, h^{-1}\mathrm{Mpc})$. 
Similarly, \citet{tinker_etal08b} has shown that HOD
models fit to galaxy clustering measurements predict void statistics
in good agreement with the data, providing independent support for the
standard HOD assumption that halo mass is the only relevant property that
determines galaxy occupation statistics.

Assembly bias is a generic prediction of  two related classes of models for the galaxy-halo connection that enjoy 
significant success in reproducing a wide variety of observed galaxy statistics. The first is the widely used {\em abundance matching} 
technique \citep{kravtsov04a, vale_ostriker04, tasitsiomi_etal04,vale_ostriker06, conroy_wechsler09, guo10, simha10,neistein11a, watson_etal12b, rod_puebla12,kravtsov13}. In this approach, one assumes that every (sub)halo in the universe hosts a single galaxy at its center, 
and that there is a monotonic relation between a halo property (usually something that can serve as a proxy for 
halo size or potential well depth, such as the maximum circular velocity $\vmax,$) and the luminosity 
(or stellar mass) of the galaxy it hosts. Abundance matching using $\vmax$ has been shown to predict accurately a wide range of 
statistics of the observed galaxy distribution, including the two-point projected correlation function of galaxies at both 
low- and high-redshift \citep{conroy06}, the conditional stellar mass function \citep{reddick12,hearin_etal13}, 
magnitude gap statistics \citep{hearin_etal12b}, and galaxy-galaxy lensing statistics \citep{hearin_etal13}. 
As we will demonstrate explicitly in this paper, {\em assembly bias is a generic prediction of any abundance 
matching technique predicated upon halo circular velocity.} Broadly speaking, 
this is because halo mass alone does not suffice to specify the halo velocity profile, and halo profiles are 
correlated with assembly \citep{gao_etal05,wechsler06,dalal_etal10}.

The second class of galaxy-halo models we study in this paper is the recently developed 
{\em age distribution matching} \citep{hearin_watson13a}. In age matching, galaxy color 
at fixed luminosity (or stellar mass) is assumed to be in monotonic correspondence with 
halo age at fixed $\vmax$. As was shown by \citet{hearin_watson13a}, age matching predicts 
the observed color-dependence of galaxy clustering as well as the scaling between galaxy color and 
host halo mass remarkably well. In a follow-up study, \citet{hearin_etal13} showed that age matching also 
provides a good description of the excess surface mass density about galaxies as a function of galaxy color, 
as measured from galaxy-galaxy lensing in the SDSS. Age matching explicitly correlates galaxy color with halo 
age, so it should not be surprising that age matching naturally predicts galaxy assembly bias. A variety of 
definitions of halo age exist in the literature and in the present paper, we use the same definition of halo 
age as \citet{hearin_watson13a}, which we reiterate in \S~\ref{subsub:agematching}.

In this paper, we address the following question: To what degree might assembly bias threaten 
the ability of standard HOD models to draw unbiased inferences and/or make unbiased predictions for the 
galaxy-halo connection? We focus on the statistic that is most often modeled using 
these techniques, namely, the projected two-point correlation function of galaxies. 
We take the empirical successes of abundance matching and age matching as 
motivation to use these models as bases for our assessment. 
To be clear, there are known weaknesses of these 
models \citep[e.g.,][]{hearin_etal12b,reddick12,hearin_watson13a,hearin_etal13} 
and they certainly do not provide a complete 
description of the galaxy-halo connection. However, these models are simple to use, 
contain assembly bias in a transparent manner, and describe observed galaxy 
clustering reasonably well. Thus our approach complements that taken in \citet{pujol_gaztanaga13}, 
who instead study a variety of semi-analytic models (SAMs) which violate the simple assumptions of the standard 
HOD due to the numerous complex baryonic processes that SAMs parameterize.

We proceed by fitting the projected two-point functions of abundance and age matching mock galaxy 
catalogues to a standard HOD model. We compare these to fits of the two-point function in mock 
galaxy catalogues with {\em identical} true HODs, but built to have {\em no assembly bias}. 
The degree to which the inferred HODs differ from the true HODs can be used to assess 
the potential importance of assembly bias (as well as provides an 
important validation exercise for HOD-based inferences).

We find that reasonable levels of assembly bias in the galaxy population can lead to statistically 
significant systematic errors in the galaxy-halo connection inferred using standard HOD techniques. This is 
true for luminosity threshold samples, and quite dramatic for color-selected subsamples. 
Moreover, we show that these biases induce systematic errors in 
predictions for independent quantities made using HOD parameters inferred from galaxy clustering. 
Finally, we demonstrate that an independent statistic used previously to diagnose assembly bias, 
namely the void probability function \citep{tinker06}, is relatively insensitive 
to the assembly bias present in abundance/age-matching mock galaxy catalogues. 
These results suggest that 
\ben
\item inferences drawn regarding the galaxy-halo connection from 
galaxy clustering should include a significant, and previously neglected, systematic error in their error budgets, 
and 
\item incorporating assembly bias effects into HOD/CLF-like models should be a priority henceforth, 
including for (re)analyses of existing datasets such as SDSS.
\een
Indeed, as we discuss below, the mock galaxy catalogues that we study necessarily contain fewer galaxies within smaller effective volumes 
than either existing or forthcoming observational samples. Therefore, the galaxy two-point functions from our mock catalogues exhibit 
larger errors than observational samples, suggesting that observational samples may be subject to a systematic error from assembly 
bias that is even more statistically significant than those that we present in this paper.

Our paper is organized as follows. In \S~\ref{sec:methods} we describe our methods. These include 
the construction of fiducial mock galaxy distributions based on abundance matching and age matching and 
our methods for fitting HODs to mock galaxy catalogues. In \S~\ref{sec:assembias} we demonstrate the 
importance of assembly bias in abundance matching and explain our algorithm for erasing the assembly 
bias from mock galaxy catalogues based on abundance/age matching models. We present our results in \S~\ref{sec:results}. 
In \S~\ref{sec:predictions}, we give examples of quantities that can be predicted (perhaps erroneously) using HODs while 
neglecting assembly bias and we demonstrate that assembly bias large enough to affect HOD inferences 
is not easy to diagnose using void statistics. We discuss the implications of our findings in \S~\ref{sec:discussion}. 
We conclude in \S~\ref{sec:summary} with a summary of our primary results.

%%%%%%%%%%%%%%%%%%%%%%%%%%%%%%%%%%%%%%%%%%%%%%%%%%%%%%%%%%%%%%%%%%%%%%%%%%
% METHODS
\section{Methods}
\label{sec:methods}

%%%%%%%%%%%%%%%%%%%%%%%%%%%%%% MOCK GALAXY SAMPLE %%%%%%%%%%%%%%%%%%%%%%%%%%%%%%

\subsection{Mock Galaxy Catalogues}
\label{sub:catalogues}

%%%%%%%%%%%%%%%%%%%%%%%%%%%%%%%%%%%%%%%%%%%%%%%%%%%%%%%%%%%%%%%%%%%%%

%---------Bolshoi
\subsubsection{The Bolshoi Simulation and Halos}
\label{subsub:bolshoi}

The bedrock of all of the mock galaxy catalogues used in this study is the high-resolution, 
collisionless $N-$body Bolshoi simulation \citep{bolshoi_11}. 
The simulation is based on a $\Lambda$CDM cosmological model with 
$\Omega_{\mathrm{m}}=0.27$, $\Omega_{\Lambda}=0.73$, 
$\Omega_{\mathrm{b}}=0.042$, $h=0.7$, $\sigma_{8}=0.82$, and $n_{\mathrm{s}}=0.95$. 
Bolshoi tracks $2048^{3}$ particles in a periodic box with side length $250\,\hmpc$, 
has a particle mass of $m_{\mathrm{p}} \simeq 1.9\times10^8\,\Msun$, 
and a force resolution of $\epsilon=1\,\hkpc$. The simulation was run with the 
Adaptive Refinement Tree Code 
(ART; \citealt{kravtsovART97, gottloeber_klypin08}). 
Snapshots and halo catalogues are available at 
{\tt http://www.multidark.org}. We refer the reader to \citet{riebe_etal11} 
for additional information.

Our mock galaxy sample is based on the ROCKSTAR merger trees and halo catalogues. ROCKSTAR is a phase-space, 
temporal halo finder capable of resolving Bolshoi halos and subhalos down to 
$\vmax\sim55\kms$ \citep{rockstar_trees,rockstar}. 
These catalogues are publicly available and can be found at {\tt http://hipacc.ucsc.edu/Bolshoi/MergerTrees.html}. 

The dark matter halos in the redshift-zero catalogues are defined to be regions within which the 
average density is $\Delta_{\mathrm{vir}} \simeq 360$ times the mean 
matter density of the Universe when centered on the local density peak. This 
is often called the ``virial'' criterion and 
the mass defined in this way is often referred to as the ``virial mass.'' The virial 
criterion implies that the relationship 
between the virial mass and the virial radius of a halo is 
$M=4\pi \Omega_{\mathrm{m}} \rho_{\mathrm{crit}} \Delta_{\mathrm{vir}} R_{\mathrm{vir}}^3/3$. 
``Subhalos'' are distinct, bound structures within the virial radii of still larger 
``parent'' or ``host'' halos. We consider a halo to 
be a subhalo of a larger host halo if the density peak on which the subhalo is 
centered resides within the virial radius of a 
more massive halo. The structures of subhalos are typically strongly affected by 
the potentials of their host halos. Consequently, 
the masses and radii of subhalos do not follow the virial definition given above.

%----------- Abundance Matching
\subsubsection{Luminosity-Only Mock Galaxies: Abundance Matching}
\label{subsub:shammocks}

We begin building mock catalogues by assigning galaxies of particular luminosities 
to halos and subhalos. In particular, absolute magnitudes 
in the r-band are assigned to Bolshoi (sub)halos using the prevalent abundance 
matching algorithm. We employ the same implementation 
of abundance matching described in detail in Appendix A of \citet{hearin_etal12b}. 
In this section, we merely sketch the basic features of this method.

The halo property $\vmax\equiv \mathrm{Max}\left\{ \sqrt{GM(<r) / r}\right\},$ where $M(<r)$ 
is the mass enclosed within a distance $r$ of the halo center, defines the maximum circular 
velocity of a test particle orbiting in the halo's gravitational potential well. The abundance 
matching technique requires that the cumulative abundance of SDSS galaxies brighter than 
luminosity\footnote{Typically in the $r$ band as in our study.} $L,$ $n_{g}(>L)$, is equal 
to the cumulative abundance of (sub)halos with circular velocities larger than 
$\vmax$, $n_{h}(>\vmax)$. This assumption specifies a monotonic relationship between 
luminosity and $\vmax$, enabling us to assign a unique r-band magnitude to every 
(sub)halo in the simulation.

In our implementation of abundance matching, we use the halo property $\vpeak,$ the peak 
value that $\vmax$ obtains throughout the entire assembly history of the halo 
\citep{reddick12}. Our model for the stochasticity in the brightnesses of mock 
galaxies results in uniform scatter in luminosity of $\sim 0.15$~dex at fixed $\vpeak;$
due to scatter between $M_{\mathrm{vir}}$ and $\vpeak,$ our model has $\sim0.18$~dex of scatter in luminosity at fixed $M_{\mathrm{vir}}.$
This amount of scatter is in accord with results from satellite kinematics \citep{more09b} and other 
abundance matching studies \citep{reddick12,hearin_etal12b}.

%----------- Age Matching
\subsubsection{Age Matching: Mock Galaxies with Luminosities and Colors}
\label{subsub:agematching}

In addition to assigning luminosities to mock galaxies,  
we also assign galaxies in our mock catalogues $g-r$ colors in order to 
study color-dependent clustering in mock galaxy catalogues. For our mock galaxy 
samples with both r-band luminosity and $g-r$ color, 
we use the publicly-available mock catalogue based on ``age distribution matching,'' which 
can be found at {\tt http://logrus.uchicago.edu/$\sim$aphearin}. We refer the interested reader 
to \citet{hearin_watson13a} for a detailed presentation of the age distribution matching technique and 
\citet{hearin_etal13} for additional applications. In what follows, we offer a 
brief review of the methods used to construct this catalogue.

Luminosities are first assigned to dark matter halos by abundance matching as 
described above in \S~\ref{subsub:shammocks}. 
After the luminosity assignments have been made for each galaxy, these galaxies 
are then assigned $g-r$ colors. This is accomplished 
by enforcing a monotonic relation between a proxy for halo age and galaxy color 
at fixed luminosity. To be more specific, we assign to each halo a formation redshift, 
$z_{\mathrm{starve}}$, which is designed to mimic the redshift at which the gas supply 
to the galaxy within the halo was cut off and new star formation began to be suppressed. 
Operationally, $z_{\mathrm{starve}}$ is the maximum of (1) the highest redshift at 
which the halo mass exceeded $10^{12}\, h^{-1}\mathrm{M}_{\odot}$, (2) the redshift 
at which the halo was accreted onto another, larger halo and thus became a subhalo, 
and (3) the halo formation redshift defined according to the fitting function of 
\citet{wechsler02}. Subsequently, all mock galaxies are placed in narrow bins 
of $r$-band luminosity and rank ordered by $z_{\mathrm{starve}}$, a proxy for 
halo age. Color assignments are made by drawing from the observed color 
distribution at fixed luminosity $\Psdss$, in such a way that the redder 
galaxies at fixed galaxy luminosity are placed within the older halos (higher $z_{\mathrm{starve}}$) 
and the observed color distribution is imposed upon the mock galaxy sample. 
As was shown in \citet{hearin_watson13a}, the resulting mock 
galaxy distribution exhibits good agreement with the luminosity 
{\em and} color-dependent two-point clustering measurements made by the SDSS \citet{zehavi11}. 
Moreover, the observed, color-dependent galaxy clustering and galaxy-galaxy lensing statistics are also describe well 
by mock galaxy catalogues constructed using this algorithm \citet{hearin_etal13}. In the interest of brevity, 
we will refer to this technique as {\em age matching} for the remainder of this manuscript.

%%%%%%%%%%%%%%%%%%%%%%%%%%%%% %%%%%%%%%%%%%%%%%%%%%%%%%%%%%%
% Halo Model and HOD fitting
\subsection{The Halo Model and the Halo Occupation Distribution}
\label{subsec:hodfitting}

The primary aim of this paper is to investigate the potential threat of assembly bias 
to standard HOD parameter inference when fitting models to observed clustering statistics. To do so, 
we treat the mock galaxy samples described in \S~\ref{sub:catalogues} as if they were the true universe. 
Mock catalogues are necessary because we must know the correct answers in order to assess our analysis methods. 
Moreover, we must be able to construct mock catalogues with varying levels of assembly bias in order 
to attribute systematic differences to assembly bias. 
There are a number of reasons that we utilize the abundance matching 
and age matching mock galaxy catalogues. First, the two-point functions 
of these catalogues faithfully represent the clustering observed in SDSS and we can have some reason to assume 
that such catalogues may exhibit at least some of the complexity of the true galaxy population. Second, abundance 
matching and age matching catalogues are easy to construct and manipulate. Lastly, and most importantly, these models 
are the simplest algorithms for assigning galaxies to halos in a way that exhibits assembly bias. The standard 
HOD formalism is predicated upon the assumption that assembly bias is zero, so by fitting our 
mock galaxy distributions with HOD models we can study how the inferred parameters 
must adjust to compensate for the presence of assembly bias in the mock catalogues.

We fit HOD parameters in a manner aimed at emulating the types of analyses that have been applied to 
data. In particular, we fit HODs to projected two-point correlation functions 
using the same halo model formulation described in \citet{tinker_etal12}. 
We use the halo mass function from \citet{tinker08a} and the halo bias of 
\citet{tinker_etal10} to describe the statistics of halos. This is 
augmented by the prescription for scale-dependent halo bias in \citet{tinker05}. 
This implementation of the HOD-formalism in conjunction with the halo model 
formalism, or slight modifications thereof, is relatively standard and similar 
models have been used in numerous studies including the completed SDSS 
galaxy clustering analysis of \citet{zehavi11} as well as \citet{zehavi05a},\citet{yang_etal05}, 
\citet{zheng_etal07}, \citet{vdBosch07}, \citet{zheng09}, \citet{simon_etal09}, \citet{abbas_etal10}, \citet{ross_etal10}, 
\citet{watson_etal10}, \citet{matsuoka_etal11}, \citet{miyaji_etal11}, \citet{leauthaud_etal11}, 
\citet{leauthaud_etal12}, \citet{tinker_etal12}, \citet{geach_etal12}, \citet{kayo_oguri12}, \citet{vdBosch13}, 
\citet{tinker_etal13}, \citet{parejko_etal13}, and \citet{cacciato_etal13}, to name some of the 
recent contributions to this extensive literature.

As described in the introductory section, a standard HOD model is defined by $\PNM,$ the probability for a halo of mass $M$ to 
contain $N$ galaxies of the type chosen for analysis (e.g., selected by luminosity, color, etc.). 
Keeping with the widely used convention, we describe central galaxies separately from satellite galaxies by assuming that 
$$\PNM =  \mathrm{P}(\Ncen|M) + \mathrm{P}(\Nsat|M).$$ 

For the first moment of $\mathrm{P}(\Ncen|M),$ we 
take the mean occupation of central galaxies in halos of mass $M$ to be 
\begin{equation}
\label{eq:ncen}
\langle N_{\mathrm{cen}}\rangle = \frac{1}{2} \left[ 1 + \mathrm{erf}\left(\frac{\log M - \log M_{\mathrm{min}}}{\sigma_{\log M}}\right)\right]. 
\end{equation}
The scale $M_{\mathrm{min}}$ describes the halo mass above which you are 
likely to have a central galaxy ($\langle N_{\mathrm{cen}}\rangle$ = 1/2 
at $M=M_{\mathrm{min}}$). Below $M_{\mathrm{min}}$ the average number of central galaxies 
approaches zero with decreasing mass while 
above $M_{\mathrm{min}}$ the number of central galaxies trends asymptotically toward unity. 
The parameter $\sigma_{\log M}$ determines the sharpness of the transition 
between $\langle N_{\mathrm{cen}}\rangle = 0$ at low mass and $\langle N_{\mathrm{cen}}\rangle=1$ at high mass. 
We describe the HOD of satellite galaxies as a Poisson distribution with mean 
\begin{equation}
\langle N_{\mathrm{sat}}\rangle = \left(\frac{M}{M_1}\right)^{\alpha} \, \exp \left(-\frac{M_{\mathrm{cut}}}{M}\right). 
\end{equation}
The mass scale $M_1$ is the halo mass scale at which halos have one satellite galaxy on average. 
At larger masses, the satellite number increases as a power-law of halo mass $M$, with index $\alpha$, 
and the satellite occupation power law is truncated below masses of $M_{\mathrm{cut}}$. In the interest of 
simplicity, we present results in which the satellites are distributed about central galaxies following 
a standard \citet{nfw97} profile with concentrations fixed to $60\%$ of the best-fitting 
average dark matter concentrations in the Bolshoi simulation \citep{bolshoi_11}. This 
value is a good description of the distributions of satellite galaxies in our mock catalogues 
and mimics the treatment of satellites in the majority of the preceding literature. We show examples of 
the manner in which our results change when the concentration parameters describing the satellite 
distributions are allowed to vary in Appendix~\ref{sec:appendix}.

In order to fit our mock galaxy data, we {\em fix} the cosmological parameters to be identical to the parameters 
of the Bolshoi simulation and vary the HOD parameters $\log(\sigma_{\log M}/\hMsun)$, 
$\log (M_1/\hMsun)$, $\log (M_{\mathrm{cut}}/\hMsun)$, and $\alpha$. 
The parameter $M_{\mathrm{min}}$ is a derived parameter that guarantees the 
galaxy sample has the correct mean number density, given the remaining HOD parameters. We limit 
$\sigma_{\log M}>10^{-3}$, and we caution the reader that marginalized posteriors on individual 
parameters are sensitive to the allowed range of $\sigma_{\log M}$, but none of our qualitative 
conclustions are sensitive to this choice. As suggested by 
\citet{tinker_etal12}, we also include a multiplicative parameter $f_b$, which is the ratio of the halo bias that we 
use to predict the galaxy clustering to the bias predicted by the formula of \citet{tinker_etal10}. The motivation 
for introducing $f_b$ is that it can partially account 
for the fact that the halo bias and the scale-dependence of the halo bias are only 
imperfectly calibrated from simulations. Following, \citet{tinker_etal12}, 
we place a Gaussian prior on the halo bias parameter with $f_b=1.0\pm 0.15$. In the case of 
fitting color-selected samples, we fit both red and blue samples {\em simultaneously} and require that a halo 
have no more than one central galaxy. We enforce this constraint by allowing the red samples to have a 
central galaxy HOD described by Eq.~(\ref{eq:ncen}) above and restricting the blue central galaxy HOD to be 
the minimum of Eq.~(\ref{eq:ncen}) and $1-\langle N_{\mathrm{cen}}^{\mathrm{red}}\rangle$. 
Thus the blue and red sub-samples have {\em distinct} values for each of the halo model parameters, 
but these parameters are related through the above constraint. We sample the parameter space using a 
standard Metropolis-Hastings Monte Carlo Markov Chain (MCMC) procedure. We compute the errors from 
the mock galaxy samples themselves using jackknife resampling with $25$ sub-samples and compute 
$\chi^2$ in the standard manner using the full covariance matrix derived from the jackknife procedure.

We have explored numerous modifications to our baseline fits. In particular, we have tried dropping 
the assumption of Poisson statistics for the satellite galaxies and varying 
separately the concentration of the parameter of the spatial distributions of satellites 
and find that our qualitative conclusions are insensitive to these assumptions. 
This is in large part because these parameters alter only small-scale clustering (projected separations 
below $r_p \lesssim 400\, h^{-1}\mathrm{kpc}$), while the differences that we describe are on large scales, and 
because the projected two-point correlation function receives a significant contribution from galaxies with relatively large 
three-dimensional separations (see \S~\ref{subsec:abimportance}). We have tested to ensure that neither 
anisotropy in the distribution of satellite galaxies around their hosts nor the peculiar velocities of 
satellite galaxies alter our conclusions in any significant manner. To make contact with previous 
literature, we address briefly both the influences of the parameter $f_b$ and 
of varying the satellite galaxy spatial distribution in Appendix~\ref{sec:appendix}. 
Finally, we have verified that our halo model implementation agrees with implementation used 
to generate halo model predictions of galaxy clustering in the recent publications of \citet{vdBosch13} 
as well as \citet{tinker_etal12} and \citet{reddick_etal13} to better than $\sim 1\%$ 
(private communication with J. Tinker and R. Reddick) and, furthermore, that we recover 
consistent HOD parameters (R. Reddick, private communication).

Generally, we expect some extant and a a number of forthcoming data sets to be significantly 
{\em more} sensitive to assembly bias than the HOD fits that we present in Section~\ref{sec:results} suggest. 
Consequently, results derived from observational data may exhibit markedly more significant systematic 
errors than those that we quote later in this paper. The published SDSS data from \citet{zehavi11} have errors that 
are more than a factor of four smaller than the errors that we derive from the Bolshoi simulation on scales 
$\gtrsim 1\, h^{-1}\mathrm{Mpc}$ for the $M_r < -21$ samples 
(this factor is approximately $\sim 2-3$ for $M_r < -20$ and $\sim 1.15-2$ for $M_r < -19$). 
Moreover, in most previous analyses, no nuisance parameter analogous to $f_{\mathrm{b}}$ 
was marginalized over in order to account for the limited calibration of halo clustering formulas. Taken together, 
our results in \S~\ref{sec:results} strongly suggest that extant inferences drawn from observational 
data are subject to a significant systematic error associated with the unknown true level of assembly bias.

%%%%%%%%%%%%
\section{Assembly Bias}
\label{sec:assembias}

The simple HODs described in \S~\ref{subsec:hodfitting} presume that halo mass is the only halo property 
that influences the number of galaxies residing in a (host) halo. If this is the case, then it is 
only necessary to enumerate the properties (e.g., abundance, structure, clustering, etc.) of halos 
as a function of their masses, averaging over all other halo properties, and compute galaxy clustering 
using the standard halo model. The term {\em assembly bias} is often used broadly to refer to the 
dependence of host halo clustering on a property other than halo mass. It may be useful to 
consider this the {\em assembly bias of halos}. Insofar as the HOD is independent of these other 
properties, galaxy clustering statistics are unaltered by 
the assembly bias of halos.\footnote{Note that this may not be the case for matter clustering, 
as may be probed by gravitational lensing or peculiar velocity statistics, for example. Matter 
clustering depends upon halo properties such as concentration and shape directly.}

What is relevant for galaxy clustering studies is the 
dependence of the HOD on any property $x \neq M$ upon which halo clustering 
{\em also} depends. It may be useful to refer to this as the {\em assembly bias of galaxies}. Mathematically, 
galaxy assembly bias is non-zero if and only if there exists some halo property $x$ such that 
$\mathrm{P}(N \vert M,x) \neq \PNM$ {\em and} the clustering of host halos depends upon 
$x$ at fixed halo mass. If there exists such a property, then $\mathrm{P}(N \vert M,x)$ may place galaxies in halos 
that are more or less strongly clustered than the average halo of mass $M$, and the standard halo 
model will fail to describe the galaxy clustering correctly\footnote{Strictly speaking, knowledge of 
the $x-$dependence of only the first two moments of 
$\mathrm{P}(N \vert M,x)$ is sufficient to determine the 
impact of galaxy assembly bias on two-point galaxy clustering.}. 
This definition of assembly bias includes the case in which the galaxy population within 
a halo depends only on a single halo property $y$ other than mass, $\mathrm{P}(N \vert y)$ 
($y$ may be a function of basic halo properties such as mass, concentration, formation time, etc.). 
In this paper, we estimate how poorly the 
standard halo model will do for a reasonable example of a mock galaxy sample that exhibits 
galaxy assembly bias explicitly.

%%%%%%%%%%%%%%%%%%%%%%%%%%%%%
% Assembly Bias in Abundance Matching
%
\subsection{Assembly Bias in Abundance Matching}
\label{subsec:assembiassham}

While this has not been emphasized in any of the previous work on the subject, 
non-zero assembly bias is a generic prediction of all contemporary subhalo abundance matching 
techniques. In fact, there are at least two distinct effects that are important to consider:
\ben
\item At fixed mass, halo clustering is known to depend upon halo concentration, $c$ 
\citep[e.g.,][]{gao_etal05,wechsler06,gao_white07,zentner07,li_etal08,wang_etal09,lacerna_padilla11}. Halos 
with higher $c$ will have larger circular velocities, $\vmax$. Therefore abundance matching 
is more likely to place galaxies of a given luminosity in higher-concentration, rather than lower-concentration, 
halos of the same mass. 
\item At fixed mass, host halos with larger $c$ contain {\em fewer} subhalos \citep[e.g.][]{zentner05} and 
therefore these hosts contain fewer satellites than host halos of the same mass with lower concentrations. 
\een
Other effects may also be important as well, such as the correlation of the spatial distributions of satellite 
galaxies within host halos and host halo environment, including the triaxiality of the satellite galaxy distribution 
and the alignment of this triaxiality among nearby host halo pairs.

Effect (i) is a consequence of the variety of halo profiles at fixed mass and 
results in halos of higher concentration hosting brighter galaxies at fixed mass. 
For any mock galaxy distribution defined by a luminosity threshold determined by a $\vmax$-based abundance matching model, 
 it follows that the sample will be biased to contain more 
highly concentrated halos relative to a mass threshold sample.

Effect (ii) can be understood in terms of merging and tidal disruption of subhalos \citep[see, for example, Figure 14 of ][]{zentner05}. 
Halo concentration correlates with formation time so the sense of this effect is intuitive: more highly concentrated halos assembled 
their masses earlier, thereby leaving more time for processes such as dynamical friction to deplete their subhalo populations 
\citep[see also][]{vdbosch_etal05}. Moreover, the magnitude 
of this effect is not small. For example, in our fiducial $\mathrm{M}_{r}<-19$ mock galaxy catalogue, 
if the population of central galaxies in halos with virial masses $M \sim 10^{12}\Msun$ is split on halo concentration, 
central galaxies that reside in halos in the bottom half of the concentration distribution have 
{\em over twice as many satellites} as their higher-concentration counterparts. We emphasize that this is a genuine 
feature of substructure content, and not simply due to the effect of subhalos on NFW fits: trends of similar magnitude 
obtain when centrals are instead split on other host halo properties such as halo formation time and environment density.

%---------------------------------------------------------
\subsection{Assembly Bias in Age Matching}
\label{subsec:assembiasadm}

Age matching is the name of the algorithm that we use to assign $g-r$ colors to galaxies as described briefly in \S~\ref{sec:methods}. 
Age matching is predicated upon a very simple assumption: older halos host galaxies with older stellar populations. 
In \citet{hearin_watson13a}, halo age is quantified by the property $\zstarve,$ so that in the language of the HOD, 
$\mathrm{P}(N_{\mathrm{red}} \vert M,\zstarve) \neq \mathrm{P}(N_{\mathrm{red}} \vert M)$, and likewise for 
$N_{\mathrm{blue}}.$ Halo clustering is known to depend upon halo age, 
so again age matching explicitly introduces galaxy assembly bias into mock catalogues. 

%-----------------------------------------------------------
% How to erase assembly bias
%
\subsection{Erasing Assembly Bias}
\label{subsec:assembiaerase}

In order to assess the significance of assembly bias, 
it is necessary to construct mock catalogues that do not exhibit assembly bias effects, 
but which have the same $\mathrm{P}(N \vert M)$ as our fiducial catalogues. 
This isolates the effects of assembly bias from effects 
due to changes in the HOD. For any model of the galaxy-halo connection, 
it is possible to construct new mock catalogues without assembly bias while preserving the exact HOD. 
We describe our algorithm for erasing assembly bias presently. 

First, we divide the {\em central} galaxies in a given sample into bins of halo mass $M$. 
We use fifty logarithmically-spaced mass bins spanning 
$11.5 \leq \log M/\Msun \leq15$, corresponding to a bin width of $0.07$~dex, 
to minimize effects due to finite binning.\footnote{We have performed a 
variety of explicit tests to ensure that our results are insensitive to our choice for bin width.} 
We then  assign each central galaxy to a new, randomly-selected host halo in the same mass bin, 
including as candidates those halos that did not originally host a central galaxy. 
This erases the memory our mock central galaxies have of all halo properties 
besides mass, while leaving $\langle\Ncen(M)\rangle$ fixed by construction. 
Since this randomization does not alter satellite populations, this step also leaves the satellite 
occupation $\mathrm{P}(\Nsat|M)$ unchanged for all luminosity thresholds.

Next, we assign each {\em system} of satellite galaxies to a new, randomly-selected host halo of the same mass, 
keeping fixed each satellite's host-centric spatial position. This reassignment preserves $\langle\Nsat(M)\rangle$ 
as well as the higher order moments of $\mathrm{P}(\Nsat \vert M)$ because all of satellites in each 
system are assigned to the same new host halo. Thus $\mathrm{P}(\Nsat \vert M)$ is identical in 
the original and randomized catalogues, but any correlation between satellite occupation and host halo 
properties besides $M$ is erased. Likewise, the intra-host spatial distribution of the randomized satellites 
has no memory of the host halo assembly history or environment.

Note that our assembly bias-erasing algorithm differs from the procedure adopted in \citet{croton_etal07} 
in a subtle but important way. \citet{croton_etal07} assign the 
entire galaxy population of each host halo to a new, 
randomly-selected host. In particular, the central galaxy {\em and} 
its satellites are relocated together as an ensemble. 
\citet{croton_etal07} correctly pointed out that this procedure 
exactly preserves the 1-halo term in the original mock catalogue. 
Thus if there is assembly bias present in the 1-halo term, their algorithm does not erase it. 
Effect (ii) discussed in \S~\ref{subsec:assembiassham} is an example of assembly bias that impacts 
the 1-halo term, and for the purposes of this paper, this effect must be erased because we wish 
to construct a counterpart mock catalogue that has zero assembly bias. For this reason, we separately 
assign centrals and satellites to new host halos, in accord with the standard HOD assumption that 
central and satellite galaxy occupation is independent, embodied by the equation 
$$\mathrm{P}(N_{\mathrm{gal}}|M) = \mathrm{P}(N_{\mathrm{cen}}|M) + \mathrm{P}(N_{\mathrm{sat}}|M).$$

We employ the exact same procedure described above when erasing assembly bias in the 
age matching mock catalogues. In particular, we apply this procedure {\em independently} to the 
red and blue samples of mock galaxies. 
This appropriately mimics the assumption of our 
HOD model that the clustering of red galaxies is independent from the clustering of blue galaxies, 
and conversely. By performing the above two-step procedure 
separately on blue and red galaxy populations, we preserve both red 
and blue HODs, and leave no trace of assembly bias on either population. 

Upon erasing assembly bias using these procedures, 
we are left with galaxy catalogues with {\em identical} HODs; however, one set 
of galaxy catalogues, our fiducial catalogues, have explicit 
galaxy assembly bias, while the other set of galaxy catalogues {\em cannot} 
exhibit galaxy assembly bias. This enables us to estimate 
how large an affect assembly bias may have on galaxy clustering 
statistics independent of any HOD fitting. 

%-----------------------------------------------------------
% The importance of assembly bias
%
\subsection{The Importance of Assembly Bias}
\label{subsec:abimportance}

\begin{figure}
\includegraphics[width=8cm]{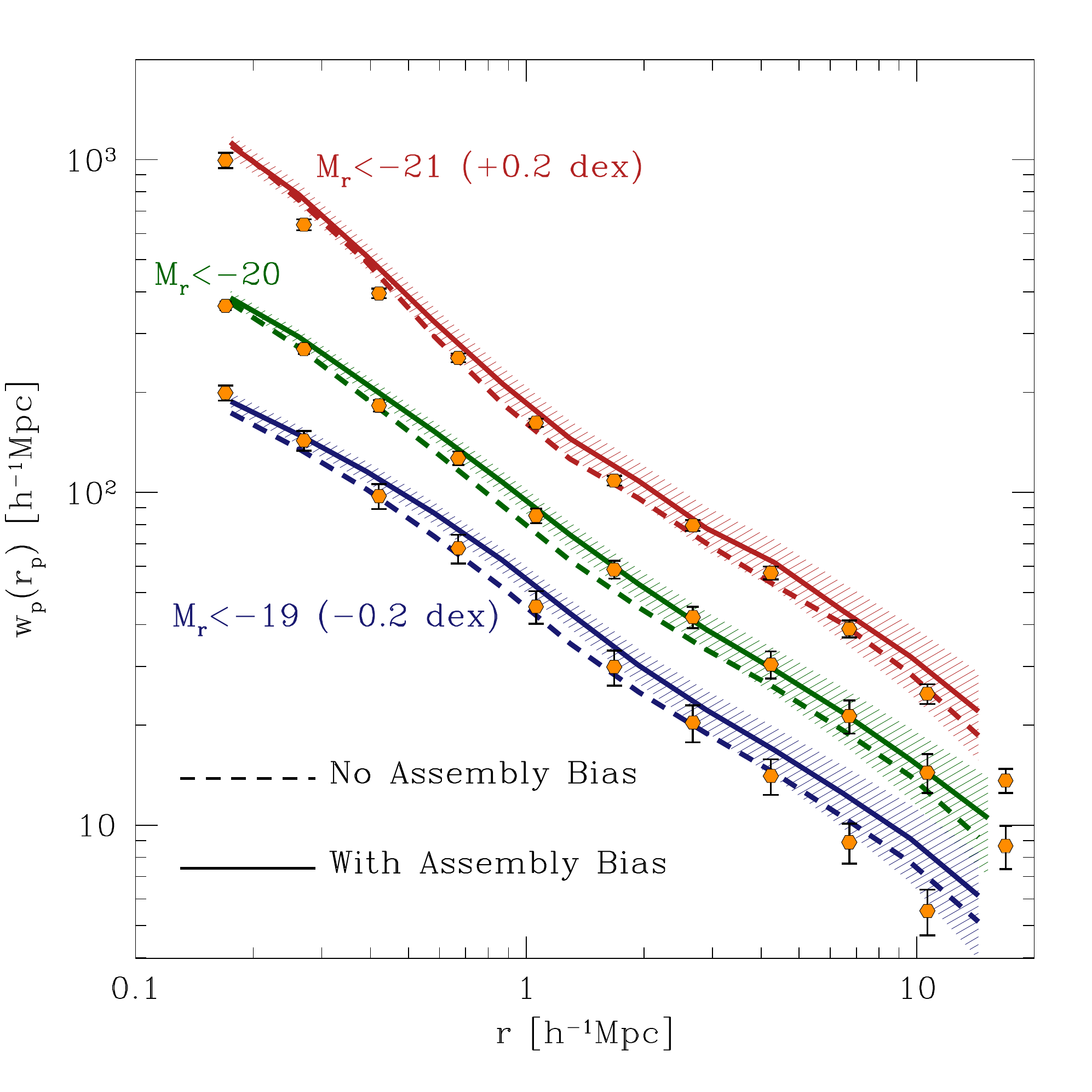}
\caption{
Assembly bias in abundance matching models. The panel shows the projected two-point correlation functions 
for several galaxy luminosity threshold samples with $M_r<-21$ (top, offset +0.2 dex for clarity), 
$M_r<-20$ (middle), and $M_r<-19$ (bottom, offset -0.2 dex). The points with error bars 
represent measurements of $\wprp$ from SDSS DR7 \citep{zehavi11}. The lines represent the predicted values 
of $\wprp$ from abundance matching mock catalogues based on the Bolshoi simulation ({\em solid}) and 
mock catalogues with precisely the same HODs, but with assembly bias erased ({\em dashed}). 
The hatched regions about the abundance matching mock catalogue measurements represent the errors on 
the predicted $\wprp$ estimated from jackknife resampling of the simulation volume. We show errors only 
for the assembly bias mock catalogues in the interest of clarity; however, the errors on the $\wprp$ in models with 
assembly bias erased are similar.
}
\label{fig:wpassem}
\end{figure}

Figure~\ref{fig:wpassem} compares the projected two-point clustering 
of galaxies in our fiducial abundance matching catalogues, which exhibit assembly bias, 
and in our catalogues with assembly bias erased for three different magnitude threshold samples. 
The effects of assembly bias are not insignificant compared to the errors on the simulation measurements (the 
hatched regions), and are large compared to the precision of the SDSS measurements. The relative effect of 
assembly bias is largest on large scales and ranges from approximately $\sim 15\%$ on large scales for the 
$M_r < -19$ threshold sample to $\sim 6\%$ for the $M_r < -21$ sample. That the effect is most prominent for 
the lower luminosity thresholds is consistent with the dependence of halo clustering on formation time, which  is 
more prominent for lower-mass halos \citep{gao_etal05,wechsler06}. The relative effect of assembly bias 
in our abundance-matching mock catalogues is grossly similar to that in the semi-analytic models of 
\citet{croton_etal07}. However, in detail \citet{croton_etal07} find assembly bias to have a 
more complex dependence upon luminosity. In particular, their red galaxy sub-sample is consisent 
with no clustering enhancement due to assembly bias at the highest luminosities, so that the 
clustering of their brightest luminosity threshold samples exhibit diminished, 
rather than enhanced, clustering as a result of assembly bias.

Neither set of our mock catalogues suffice for a detailed 
description of the SDSS clustering data; however, these predictions are 
broadly similar to SDSS clustering, so it is reasonable to suppose that these catalogues exhibit some of the 
richness of the observed galaxy data and may yield insight into galaxy clustering. For the purposes of this paper, 
the salient point is that the clustering 
differences shown in Figure~\ref{fig:wpassem} between the fiducial and assembly bias-erased catalogues 
will drive our halo model fits to to recover (erroneously) {\em distinct} HODs.

%%%%%%%%%%%%%%%%%%%%%%%%% FIGURE %%%%%%%%%%%%%%%%%%%%%%%%%%%%%

%%%%%%%%%%%%%%%%%%%%%%%%%%%%%%%%%%%%%%%%%%%%%%%%%%%%%%%%%%%%%%%%%%%%

As Figure~\ref{fig:wpassem} shows, the relative size of the 
effect of galaxy assembly bias on galaxy clustering statistics in these 
catalogues is large. The clustering is most altered on relatively large scales ($r_p \gtrsim 1\, h^{-1}\mathrm{Mpc}$), 
suggesting that the effect is primarily due to the occupation statistics of central galaxies. This is 
indeed the case, so it is useful to examine the differences in host halo clustering among our mock 
catalogues. 

\begin{figure}
\begin{center}
\includegraphics[width=8.1cm]{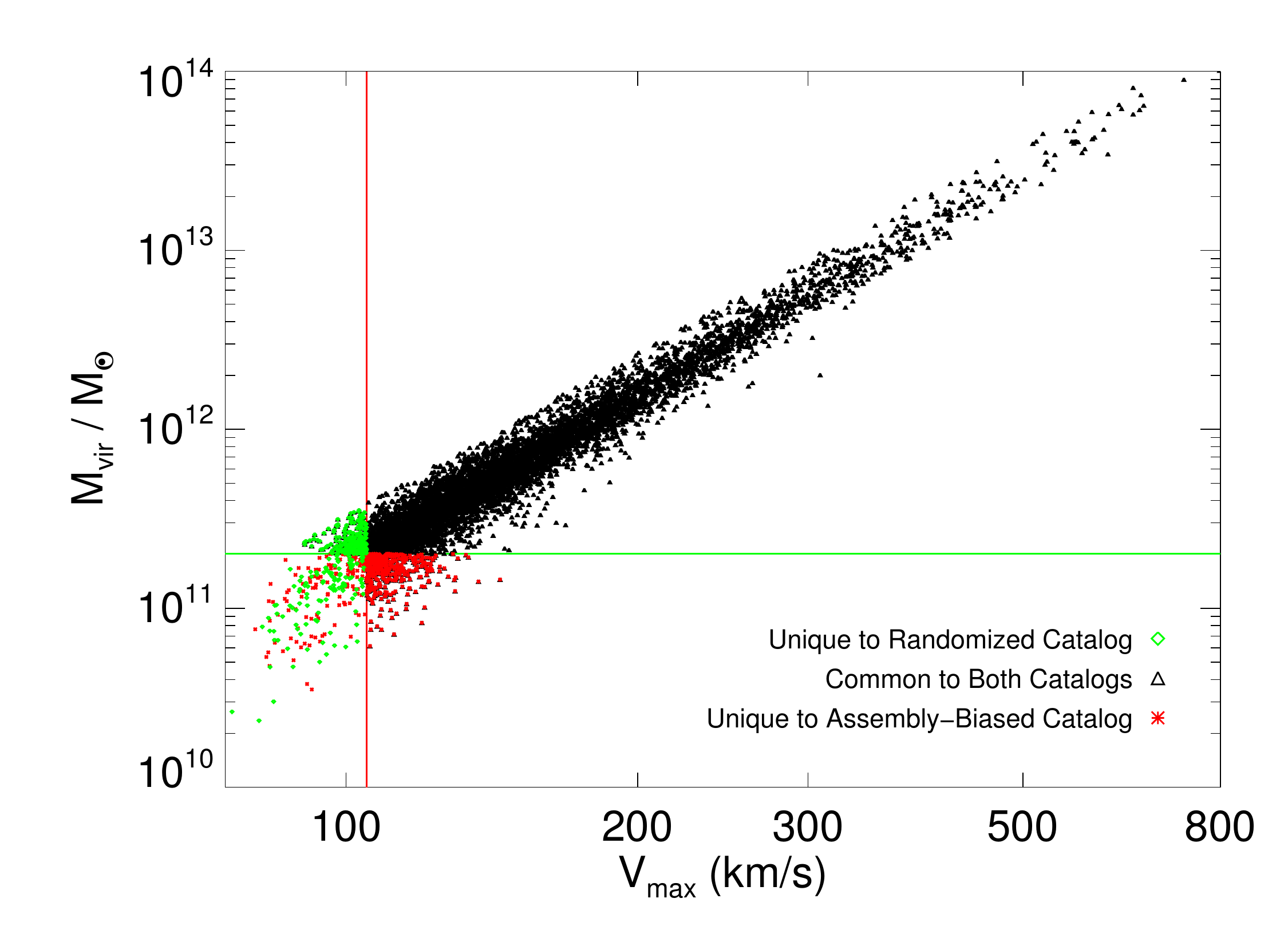}
\includegraphics[width=8cm]{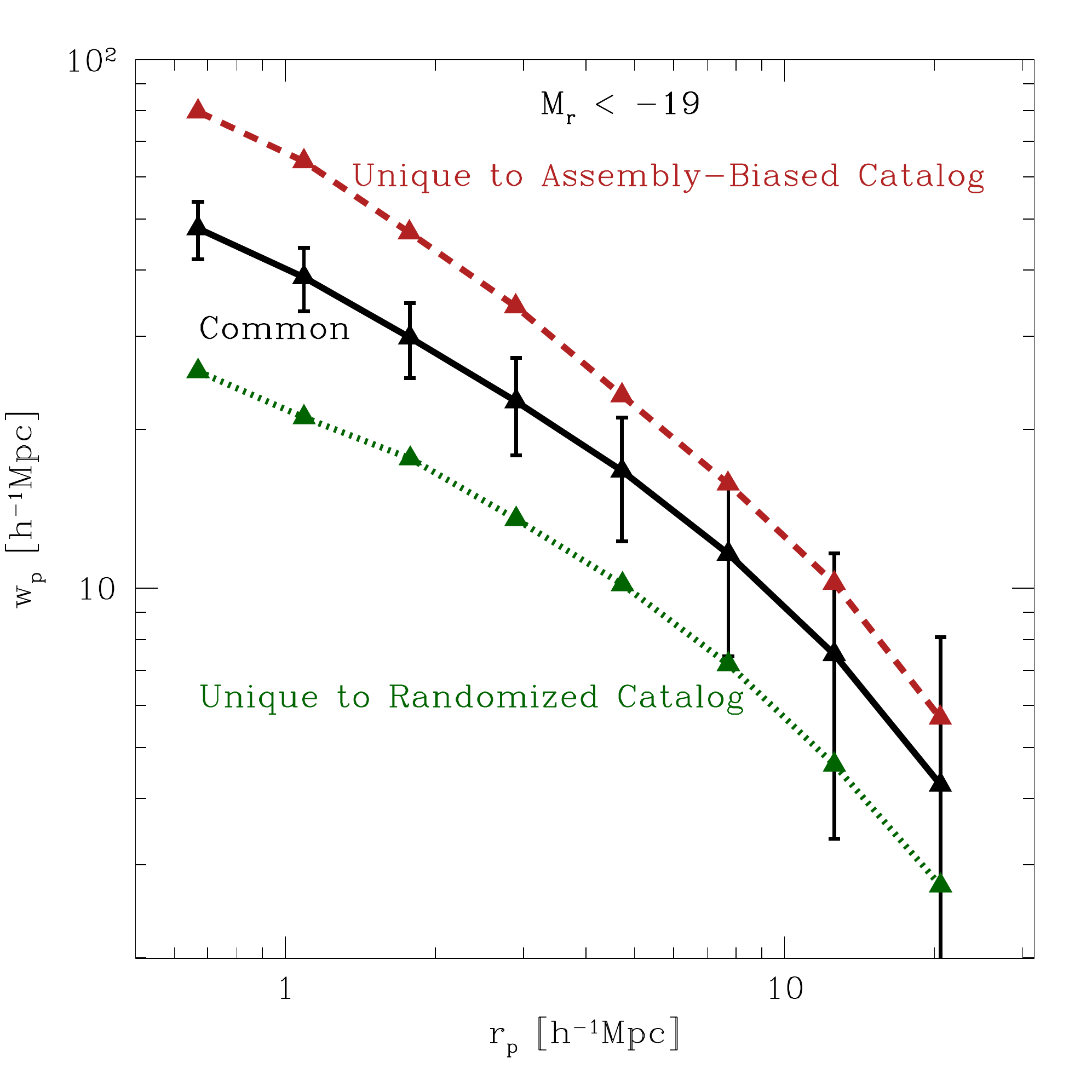}
\caption{
Comparing the halos that host central galaxies in mock catalogues with and without 
galaxy assembly bias. Each point in the {\em top} panel 
represents a central galaxy in one or both of our luminosity-only $M_r<-19$ catalogues. 
In the {\em bottom} panel, the black triangles 
connected by the solid line shows the two-point clustering of halos selected to have central galaxies 
in {\em both} catalogues. The {\em upper, dashed} line shows the two-point clustering of host halos 
that are assigned central galaxies only in our fiducial galaxy catalogue that exhibits assembly bias. 
The {\em lower, dotted} line shows the two-point clustering of host halos that are assigned central 
galaxies only in the galaxy catalogue in which assembly bias has been erased. All data in this 
plot refer to the $M_r < -19$ sample with a mean galaxy density of 
$n_{\mathrm{g}} \simeq 1.57 \times 10^{-2}\, h^{-1}\mathrm{Mpc}$. Errors are 
shown only for the two-point function of halos common to all catalogues in the interest of 
clarity.
}
\label{fig:wpdemo}
\end{center}
\end{figure}

In Figure~\ref{fig:wpdemo}, we compare the host halo populations in our mock catalogues. 
The top panel of Fig.~\ref{fig:wpdemo} shows the masses and maximum circular velocities of objects in our 
catalogues with and without assembly bias. The bottom panel of Fig.~\ref{fig:wpdemo} compares 
the clustering of halos that are selected to have central galaxies in our fiducial catalogues, 
with assembly bias, to the clustering of host halos in our catalogues in which assembly 
bias has been erased. For demonstration purposes we choose the 
$M_r < -19$ threshold sample for this example because the galaxy assembly bias is largest for this 
sample (Fig.~\ref{fig:wpassem}), and because halo assembly bias is largest in low-mass 
host halos \citep{gao_etal05,wechsler06}. Recall that the mean occupation statistics of 
central galaxies in these catalogues are {\em identical} by construction.

%%%%%%%%%% Halo Comparison FIGURE %%%%%%%%%%%%%%%%%%%%%%%%%%%%%
%
%
%%%%%%%%%%%%%%%%%%%%%%%%%%%%%%%%%%%%%%%%%%%%%%%%%%%%%%%%%%%%%%%%%%%%

Each point in the {\em top} panel 
represents a central galaxy in one or both of our luminosity-only $M_r<-19$ catalogues. 
Centrals that are common to both catalogues appear as black triangles. The vertical red line illustrates the $\vmax$ 
cut corresponding to $M_r<-19;$ the horizontal green line illustrates the cut on halo mass $M$, that produces the same 
corresponding number density of halos. As discussed above, when randomizing central galaxy occupation we 
include halos that did not necessarily host a central galaxy in the fiducial catalogue. Thus there is no guarantee 
that a halo hosting a central galaxy in the fiducial catalogue will host a central in the no-assembly-bias counterpart 
catalogue, and conversely. With red asterisks (green diamonds) we show those host halos in the $M_r<-19$ fiducial 
catalogue (erased assembly bias catalogue) that do not appear in the catalogue without (with) assembly bias. The halos 
that are common to both catalogues  represent approximately $\approx 74\%$ of the host halo population. The 
remaining $\approx 26\%$ of halos differ between the two catalogues.

Now we turn attention to the {\em bottom} panel of 
Fig.~\ref{fig:wpdemo}. The solid, black line shows the projected correlation function of 
the halos that are selected to have central galaxies in both our fiducial mock catalogue (with assembly bias) 
and in our mock catalogue in which assembly bias has been erased. 
The dashed and dotted lines show the clustering of the halos that are unique to the 
fiducial catalogue and the catalogue with assembly bias erased, respectively. Fig.~\ref{fig:wpdemo} 
shows that the halos that are distinct to the galaxy populations with and without assembly 
bias are clustered significantly differently. The halos unique to the fiducial catalogues are a factor of 
$\sim 3$ more strongly clustered on small scales ($r_p \lesssim 1\, h^{-1}\mathrm{Mpc}$) 
and a factor of $\sim 2$ more strongly clustered on large scales ($r_p \gtrsim 10\, h^{-1}\mathrm{Mpc}$) 
than the halos unique to the galaxy populations with no assembly bias. The difference in host halo clustering 
shown in Fig.~\ref{fig:wpdemo} is nearly sufficient to account for the entirety of the differences between 
the two-point clustering in the $M_r < -19$ samples, even on scales $r_p \lesssim 1\, h^{-1}\mathrm{Mpc}$.

%%%%%%%%%%%%%%%%%%%%%%%%%%%%%% RESULTS %%%%%%%%%%
%%%%%%%%%%%%%%%%%%%%

\section{RESULTS}
\label{sec:results}

%%%%%%%%%%%%%%%%%%%%%%%%%%%%%%%%%%%%%%%%%%%%%%%%%%%%%%%%%%%%%%%%%%%%%

Using the methods described in \S~\ref{subsec:hodfitting}, 
we fit the clustering of galaxies in luminosity threshold samples 
as well as red and blue galaxy subsamples, in both our fiducial mock galaxy catalogues 
with assembly bias and our galaxy samples in which assembly bias 
has been removed. In this section, we describe our results with an emphasis on how these 
fits differ between the mock galaxy samples with and without assembly bias 
(but with {\em identical} true HODs). 

%-----------------------------------------------------
\subsection{Fits to Projected Clustering}
\label{subsec:wpfits}

We begin the discussion of our results by stating that, 
with a single exception, every sample of mock galaxies we study passes a naive ``goodness-of-fit'' 
test based on the best-fitting $\chi^2$, including those galaxy samples exhibiting assembly bias. 
An immediate and unavoidable conclusions is that the ability to achieve an acceptable fit to projected 
galaxy clustering data with a standard HOD model has little bearing on the question of whether or 
not galaxy assembly bias is present in the real Universe.

%%%%%%%%% Example Fits Including our Worst-Case-Scenario %%%%%%%%%%%%%%%%%%
\begin{figure*}
\includegraphics[width=8cm]{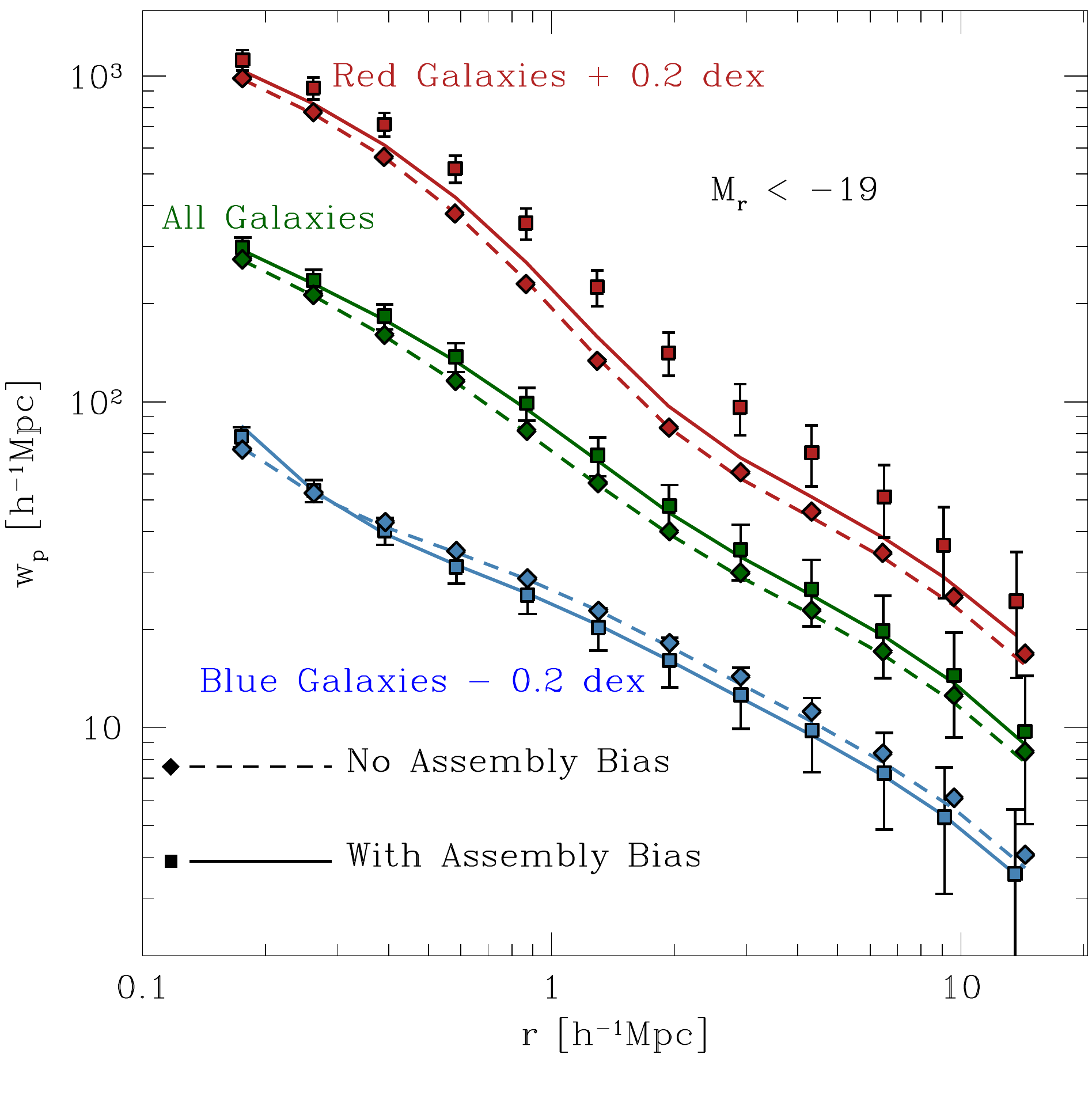}
\includegraphics[width=8cm]{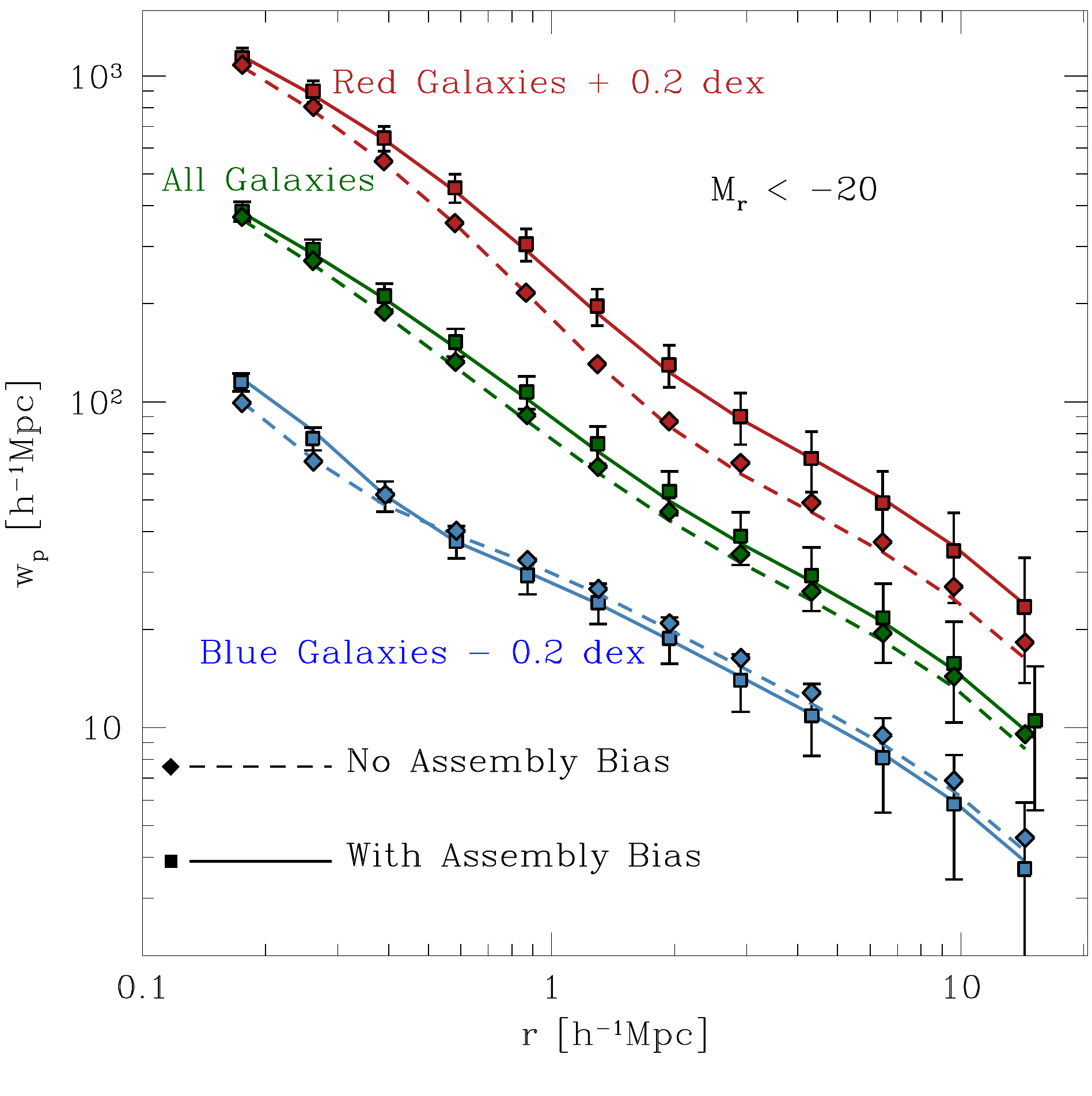}
\includegraphics[width=8cm]{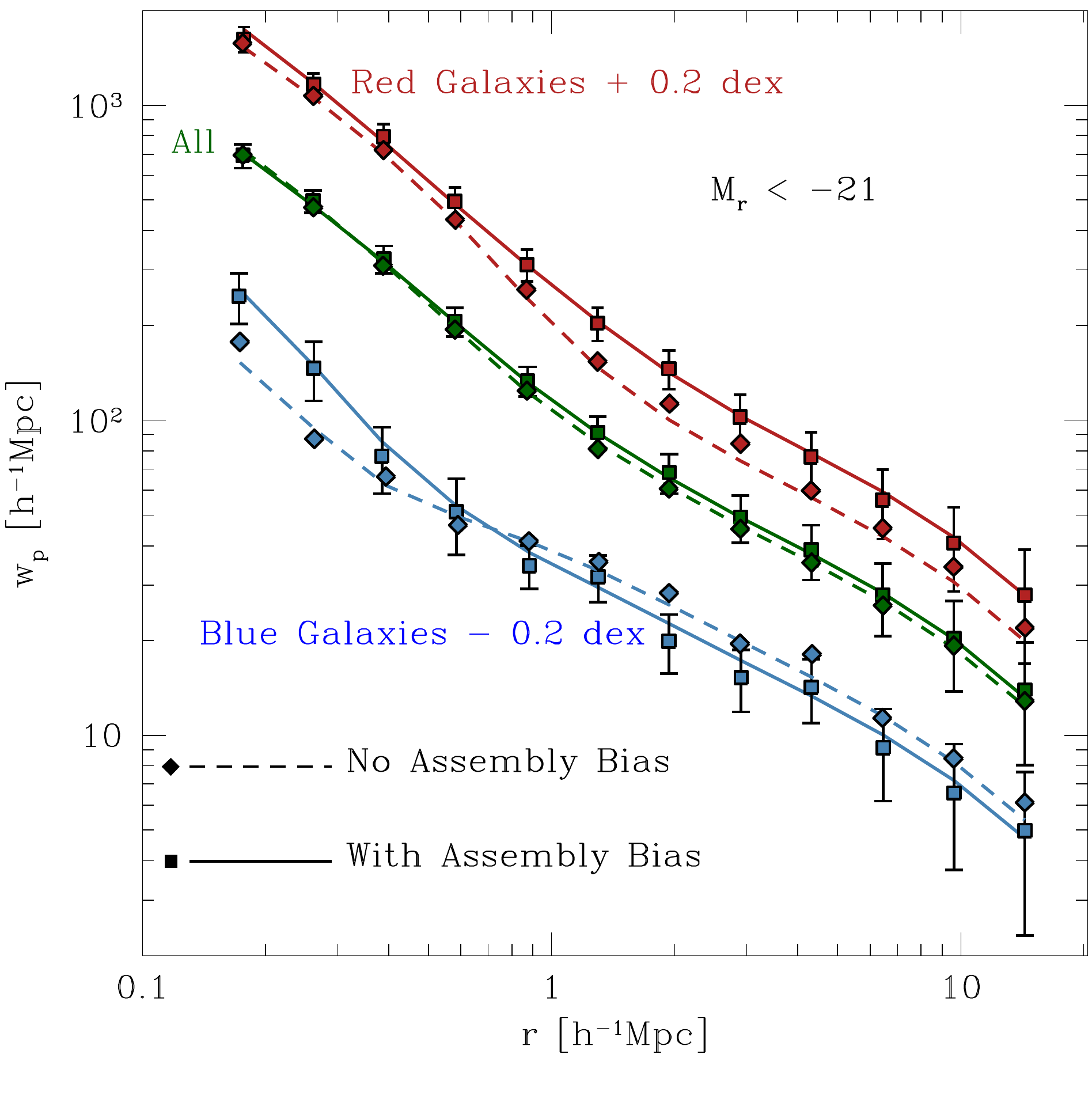}
\caption{
Halo model fits to the $\wprp$ measured in the Bolshoi simulations. Different panels show fits 
to samples with different luminosity thresholds, where the brightness cut is indicated in each panel. 
{\em Solid} lines are the best fits to our fiducial models that exhibit assembly bias while {\em dashed} 
lines are the best fits to our models with {\em no assembly bias}. The fiducial galaxy samples 
and their counterparts in which assembly bias has been erased have {\em identical} HODs. Mock catalogue 
data is represented by {\em squares} for our fiducial models and by {\em diamonds} in the case of 
erased assembly bias. The topmost pair of mock catalogue data points and curves (in red) correspond to the subsamples 
of red galaxies; the bottom pair (in blue) correspond to the subsamples 
of blue galaxies; the central pair (in green) correspond to all galaxies in the luminosity threshold sample 
with no color selection. For visual clarity, the projected correlations for the red (blue) samples have been offset in 
the positive (negative) direction by 0.2 dex, and errors are shown only for the fiducial mock catalogues. 
We emphasize that it is not trivial to determine $\chi^2$ by visual inspection of 
this plot because the $w_{\mathrm{p}}(r_{\mathrm{p}})$ data are highly correlated. 
}
\label{fig:wp19}
\end{figure*}
%%%%%%%%%%%%%%%%%%%%%%%%%%%%%%%%%%%%%%%%%%%%%%%%%%%%

As we showed in \S~\ref{subsec:abimportance}, 
assembly bias in abundance matching mock catalogues is strong. The strength of assembly bias predicted by 
the age matching assignments of galaxy clustering is, perhaps, extreme because galaxy color 
is in monotonic correspondence with halo age at fixed halo $\vmax$ in this model. Yet nearly all 
of these mock galaxy samples may be fit by a galaxy-halo model in which assembly bias 
is (incorrectly) assumed to be zero. 

In Figure~\ref{fig:wp19}, we show our fits to each of the samples we study in this paper. 
We remind the reader that the correlation function data at different spatial separations 
are highly correlated, so it is not trivial to estimate $\chi^2$ by inspection of the lines and 
points in Fig.~\ref{fig:wp19}. This is particularly true for the fits to the color-split samples because 
both red galaxy and blue galaxy clustering are fit simultaneously, 
subject to the constraint that there can be only one central galaxy per halo.

With this caveat in mind, Fig.~\ref{fig:wp19} suggests that our fitting procedure 
is generally quite successful in that it recovers the correct two-point galaxy clustering. In particular, 
the fits to the samples with no assembly bias are of high quality in all cases. This is, perhaps, not 
particularly surprising because the models with no assembly bias are consistent with the 
premises on which the halo model is based and halo clustering statistics have now been 
calibrated very accurately from cosmological simulations \citep[e.g.,][]{tinker08a,tinker_etal10}. 
Fig.~\ref{fig:wp19} also demonstrates that our fits to the projected clustering of samples {\em with} 
assembly bias generally describe the mock catalogue data quite well.
With only a single exception (see below), each of our fits results in a chance probability for attaining 
larger best-fit $\chi^2$ values that exceeds $\gtrsim 0.05$.

Let us now discuss the only mock galaxy sample whose best-fit HOD fails the goodness-of-fit test: 
the $M_r < -19$, color-split fiducial catalogue {\em with} assembly bias. Recall that we 
fit the red and blue samples simultaneously subject only to the constraint that there can 
be no more than one central galaxy per halo. 
The simultaneous fit to the $M_r<-19$ red and blue galaxy samples results in a best-fitting 
$\chi^2 \simeq 74.4$; for 16 degrees of freedom, a $\chi^2$ this large or larger would occur by 
chance with a probability of $\simeq 2 \times 10^{-9}$. This suggests strongly 
that our halo model description of color-dependent clustering cannot describe the distribution of 
galaxies with $M_r < -19$ assigned to Bolshoi halos through age matching.

Intriguingly, the poor fit to the color-selected $M_r<-19$ sample occurs in a 
situation similar to the unacceptable fit to SDSS data for color-split samples with $-20 \le M_r \le -19$ in 
\citet{zehavi11}. However, the \citet{zehavi11} result is driven by the fact that the blue galaxies are 
more weakly clustered than their model can accommodate, while in the case of our fits, the model 
fails because it cannot accommodate the strength of the red galaxy clustering in the age matching 
mock galaxy distribution. In particular, we find that the large-scale bias of the red galaxies in this 
sample is $\sim 20\%$ higher than in our best fit model. In order to accommodate this, it would be 
necessary to place galaxies in halos that are $\sim 5$ times more massive because the 
halo bias function is a shallow function of halo mass in the relevant mass range, near  
$M_{\mathrm{min}} \sim 7 \times 10^{11}\, h^{-1}\mathrm{M}_{\odot}$. 
However, this increase in mass is impossible because a model 
with a significantly larger value of $M_{\mathrm{min}}$ does not yield the 
correct average number density. Moreover, the parameter $f_b$ alone cannot accommodate 
the strength of the red galaxy clustering for this sample without simultaneously over predicting 
the clustering of the blue galaxies. For the remainder of this paper, 
we proceed by giving the results for each of our halo model fits with the 
caveat that, according to a $\chi^2$ goodness-of-fit test, the fit to the color-split sample with 
$M_r < -19$ is unacceptable. 

%-----------------------------------------------------
\subsection{The Character of Color-Dependent Assembly Bias}
\label{subsec:colorab}

Before taking a detailed look at our best-fit HOD parameters in the following section, 
let us first use Fig.~\ref{fig:wp19} to consider the qualitative imprint that assembly bias 
leaves on color-selected galaxy samples.
Notice that for both the luminosity 
threshold samples and the red galaxy subsamples, assembly bias tends to drive galaxy 
clustering higher on all scales, just as in Fig.~\ref{fig:wpassem} and Fig.~\ref{fig:wpdemo}. 
This is because the same physical mechanism gives rise to assembly bias in both cases. 
Abundance matching-generated luminosity threshold samples are preferentially populated 
with centrals living in halos that have large values of $\vmax$ for their masses; such halos are 
more strongly clustered, as discussed in \S~\ref{subsec:assembiassham}. Age matching exaggerates 
this effect in red galaxy samples, because red centrals are explicitly chosen to reside in the earliest forming 
halos, with the highest values of $\vmax$, at a given mass. 

On the other hand, the blue galaxy samples in Fig.~\ref{fig:wp19} exhibit distinctly different behavior 
in two respects. First, assembly bias drives these galaxies to be more weakly clustered 
on large scales ($r_{\mathrm{p}} \gtrsim 200-700\, h^{-1}\mathrm{Mpc}$ depending upon 
the sample under consideration). This is because age matching assigns blue galaxies to halos that acquired 
their mass relatively more recently, and it is now well-known that 
relatively-later forming halos tend to be more weakly clustered than their earlier-forming 
counterparts at fixed mass. 

Second, on small scales ($r_{\mathrm{p}} \lesssim 200-700\, h^{-1}\mathrm{Mpc}$), blue galaxy clustering 
is {\em strengthened} by assembly bias. This is primarily caused by Effect (ii) discussed 
in \S~\ref{subsec:assembiassham}: 
at fixed mass, later-forming host halos have a larger than average number of satellites. 
In age matching, halos with a blue central galaxy are 
 the latest-forming halos of a given mass. Therefore, at fixed mass, age matching predicts 
 that the presence of a blue central is correlated 
 with having a larger than average number of satellites. Mathematically, this is represented 
 as $$\langle N_{\mathrm{cen}}^{\mathrm{blue}}\Nsat\rangle>\langle N_{\mathrm{cen}}^{\mathrm{blue}}\rangle\langle\Nsat\rangle.$$
 
 This small-scale effect is further enhanced by the phenomenon of ``galactic conformity'' \citep[e.g.][]{weinmann06b}, 
 a term referring to the observed tendency for a blue (red) central galaxy to host a preferentially blue (red) satellite population.
 Age matching naturally predicts galactic conformity because the formation time of a host halo correlates with the formation 
 time of its subhalos, as both the host and its subhalos collapse from the same region of the cosmic density field. 
 Mathematically, this second effect is represented as 
 $$\langle N_{\mathrm{cen}}^{\mathrm{blue}}N_{\mathrm{sat}}^{\mathrm{blue}}\rangle>\langle N_{\mathrm{cen}}^{\mathrm{blue}}\rangle\langle N_{\mathrm{sat}}^{\mathrm{blue}}\rangle.$$
 Note that both of these effects violate the common HOD assumption that the 
 satellite and central HODs are uncorrelated: 
 $\mathrm{P}(N_{\mathrm{gal}}|M) = \mathrm{P}(N_{\mathrm{cen}}|M) + \mathrm{P}(N_{\mathrm{sat}}|M).$

These two small-scale effects work together to boost the central-satellite pair counts in 
blue galaxy samples with assembly bias relative to blue samples with no assembly bias. This 
boost strengthens clustering on small scales where the 1-halo term dominates. That the impact on two-point 
clustering is significant primarily for blue samples derives from two facts. First, it is only for the 
blue samples that $P(N_{\mathrm{cen}}=1 \vert M) < 1$ for an appreciable portion of the halo mass range of 
interest (see \S~\ref{subsec:hodmodels} below). Furthermore, early-forming halos host the reddest galaxies 
in the age matching procedure, yet early-forming halos have fewer satellites because processes have 
operated for a greater number of dynamical times in such systems \citep[e.g.,][]{zentner05,watson_powerlaw11}, with the consequence that the 
boost in small-scale clustering that may be provided by conformity may be partially canceled by dynamical evolution.

%%%%%%%%%%%%%%%%%%%%%%%%% FIGURE %%%%%%%%%%%%%%%%%%%%%%%%%%%%%

\begin{figure}
\begin{center}
\includegraphics[width=8.5cm]{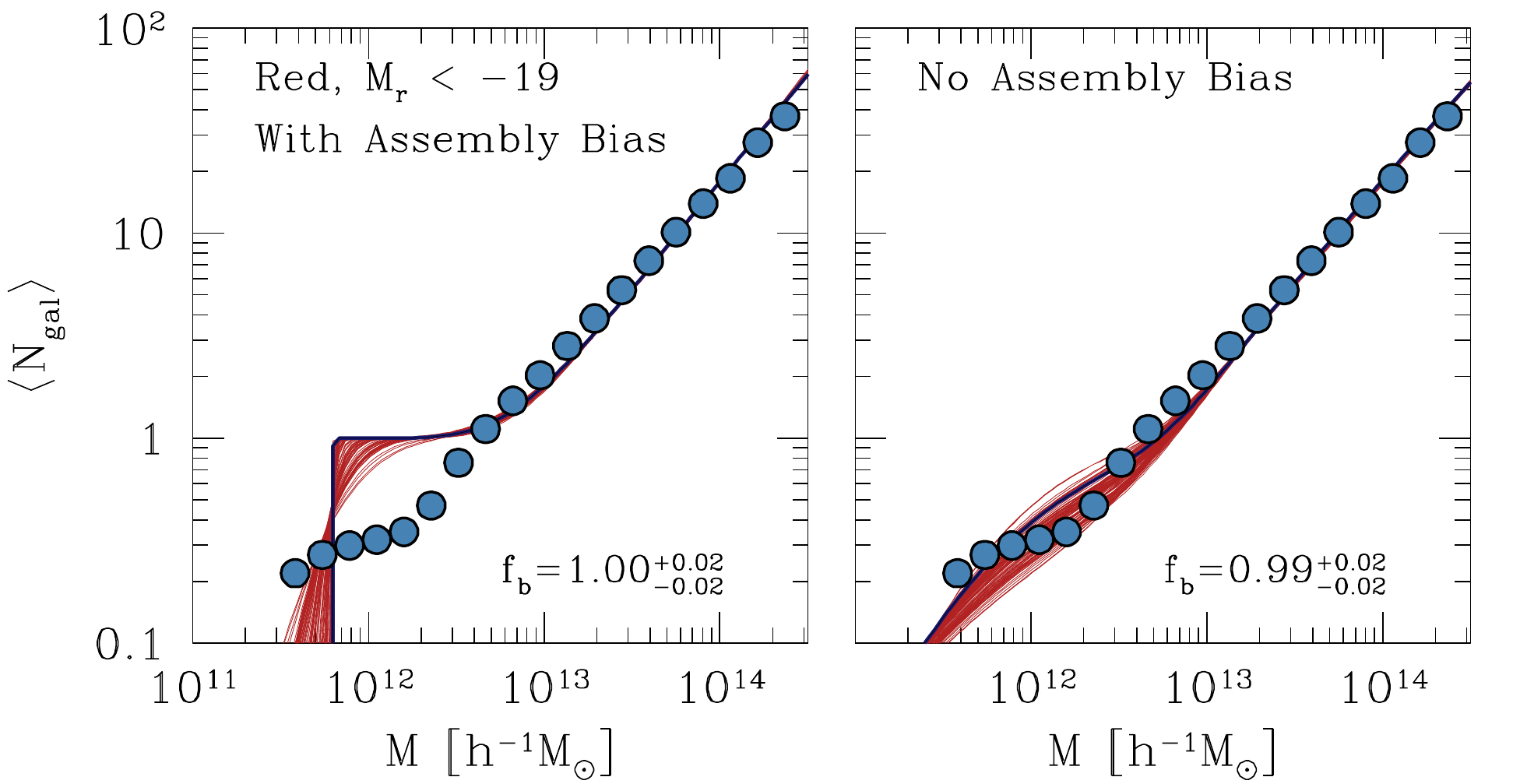}
\includegraphics[width=8.5cm]{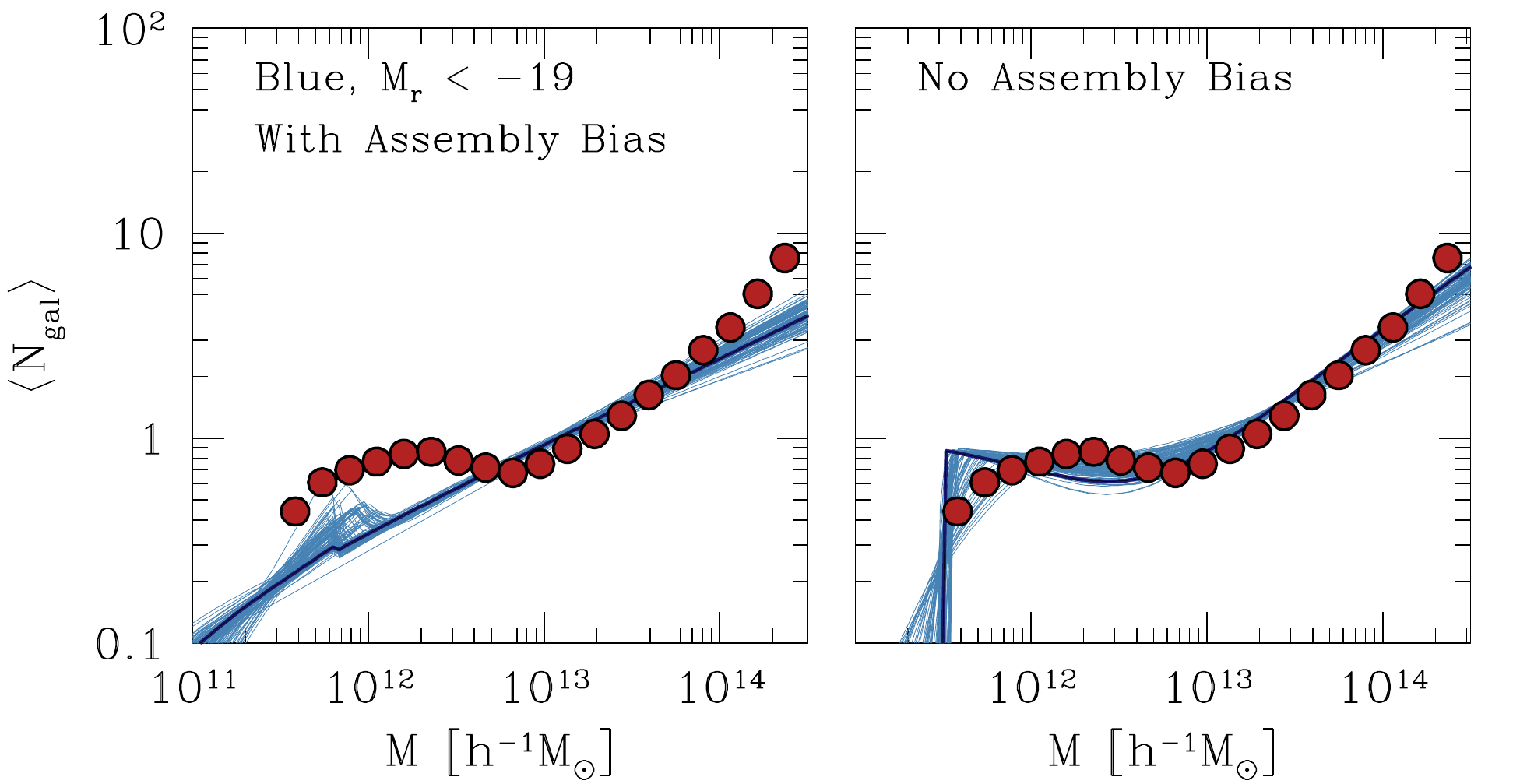}
\includegraphics[width=8.5cm]{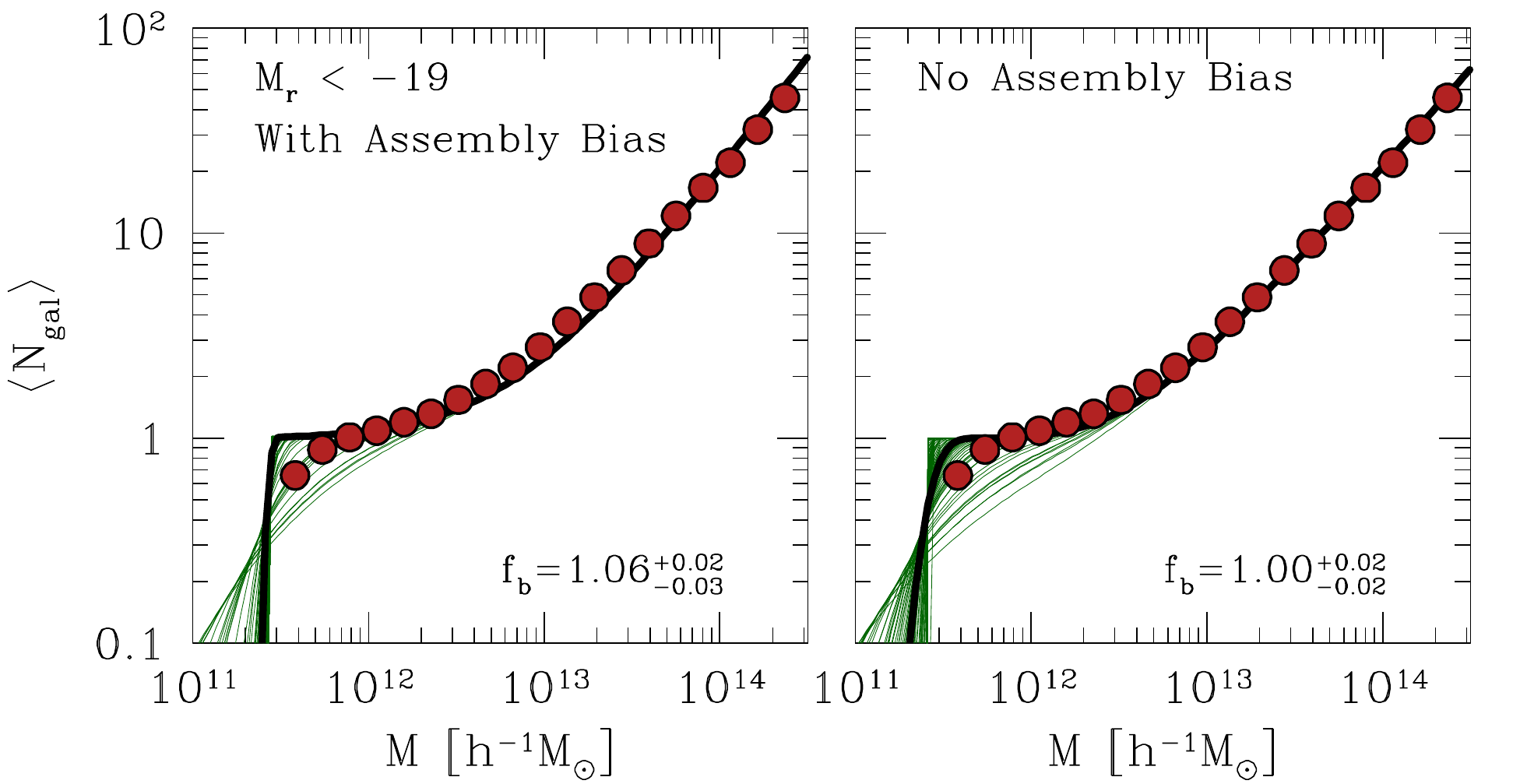}
\caption{
Comparison of the best-fit HODs for galaxies in the 
$M_r < -19$ sample with the true HODs in the mock galaxy catalogues (points). 
In each case, the {\em thick, solid} lines represent the best fits and the numerous, 
{\em thin, colored} lines represent 100 randomly-selected HODs from the MCMC chain 
that are within $\Delta \chi^2 \le 1$ relative to the best-fit model. The marginalized constraint 
on the halo bias nuisance parameter $f_b$ is shown in each of the panels for the red, color-selected 
samples and the luminosity threshold samples. Note that the red and blue samples are fit simultaneously, 
so they are fit with a common value for $f_b$.
}
\label{fig:hod19}
\end{center}
\end{figure}

%%%%%%%%%%%%%%%%%%%%%%%%%%%%%%%%%%%%%%%%%%%%%%%%%%%%%%%%%%%%%%%%%%%%

%------------------------------------------------------
\subsection{Best-Fit HOD Models}
\label{subsec:hodmodels}

The primary aim of this paper is to emphasize the significance of assembly bias in abundance matching and age matching, 
and to exploit these differences to estimate the potential for assembly bias to introduce systematic errors in the HOD parameters 
inferred in galaxy clustering analyses. Therefore, we now turn to examining the HODs inferred 
from our fits. Before turning to constraints on individual HOD parameters in the following section, 
in this section, we begin by examining $\Ngal(\Mvir),$ the average halo occupation in our acceptable 
models. We consider this to be the most useful manner in which to represent our 
results because the posterior distribution of HOD parameters that results from our fits 
exhibits strong correlations. 

%------------------------------------------------------
\subsubsection{HOD fits to $M_r<-19$ samples}
\label{subsubsec:mr19fits}
 
In this section we compare our inferred HODs for the $M_r < -19$ samples to the actual 
HODs measured in the Bolshoi simulation. 
We remind the reader that according to a $\chi^2$ goodness-of-fit test,
we do not achieve a good fit to the clustering of red galaxies for this luminosity threshold. This is due to the strength 
of assembly bias in our chosen fiducial model. However, the considerations driving systematic biases in HOD fits 
to assembly-biased samples are the same for all our luminosity thresholds, and so we consider it instructive to begin 
discussion of our fits with the $M_r < -19$ samples. 

Figure~\ref{fig:hod19} illustrates $\Ngal(M)$ for both luminosity- and color-selected $M_r<-19$ galaxy samples. 
First examine the fits to the luminosity threshold samples in 
the bottom panels of Fig.~\ref{fig:hod19}. In both cases, the inferred HODs are similar to the 
true HODs. However, in the fits to the fiducial catalogues with assembly bias, the fits are slightly more strongly 
biased toward a rapid transition from $\langle N_{\mathrm{cen}}\rangle=0\rightarrow\langle N_{\mathrm{cen}}\rangle=1$. 
The character of this bias will be encountered in many of our fits. It is caused by the stronger clustering 
of galaxies in the fiducial catalogues, and can be understood as follows. In general, a long tail of non-negligible 
$\Ncen$ extending to low mass drives clustering strength down because low-mass halos are more weakly 
clustered than high-mass halos. Our HOD model is unable to exploit halo assembly bias, so our HOD-fit to 
the fiducial catalogue with assembly bias achieves adequate clustering strength, in part, by making the central 
galaxy transition sharp, rather than extended; this avoids populating a low-mass tail in $\Ncen,$ thereby boosting the clustering. 

Furthermore, it is also apparent that the average number of satellite galaxies is a steeper function of halo mass in the 
fits to the fiducial catalogues with assembly bias. In the HOD language, this is reflected in a larger power-law index $\alpha$, 
describing the average number of satellites as a function of halo mass. Fits to samples with assembly bias yield 
significantly larger values of the parameter $\alpha$ because the large scale clustering of galaxies can be increased 
by packing relatively more galaxies into more strongly clustered higher-mass halos and fewer galaxies into 
less strongly clustered, lower-mass halos. This bias in the number of satellites in large halos is another feature 
of our inferred HODs that will be encountered repeatedly in this section.

These trends are more strongly in evidence for the red galaxy subsamples in the top row of panels in Fig.~\ref{fig:hod19}. 
In particular, in our best-fit HOD to the red sample with assembly bias, the central galaxy 
transition is far sharper than the true transition. 
The effect is of the same character, and more pronounced, in this case because assembly bias in red subsamples is largely 
an extreme version of assembly bias in luminosity threshold samples, as discussed in \S~\ref{subsec:colorab}. 

An additional distinguishing feature of fits to the color-selected samples is driven by 
the need to simultaneously reproduce the blue and red galaxy 
clustering. A consequence of this is that the bias parameter $f_{\mathrm{b}}$ cannot vary to accommodate the large-scale clustering of 
the red sample without predicting blue galaxy clustering in excess of the mock catalogue data. Recall that assembly bias in these models increases 
red galaxy clustering strength and decreases blue galaxy clustering strength. 
The marginalized constraints on $f_{\mathrm{b}}$ cleanly demonstrate this point. 
In the threshold samples, the strong clustering of the samples 
with assembly bias is taken up, in part, by $f_{\mathrm{b}}$ so that the best-fit value of $f_{\mathrm{b}}=1.06$ sets halo 
clustering to be $6\%$ stronger than predicted by the \citet{tinker08a} formula. On the contrary, the best-fit bias 
parameter from the fits to the color-split samples is $f_{\mathrm{b}}=1$. In Appendix~\ref{sec:appendix}, 
we show how eliminating the extra freedom afforded by the parameter $f_{\mathrm{b}}$ alters inferred 
HODs. In short, allowing $f_{\mathrm{b}}$ to vary along with the HOD increases the 
errors in the inferred HOD parameters and mitigates the biases in the inferred HODs. 

Continuing our discussion of the red galaxy samples in the top panels of Fig.~\ref{fig:hod19}, 
it is interesting to note that the true HOD has a feature at $M \approx 2 \times 10^{12}\, \hMsun$ 
and $\Ngal \simeq 0.4$ that cannot be accommodated by the standard functional form for the 
central galaxy HOD, yet the acceptable HODs go broadly through this feature in a smooth fashion. 
In contrast, the failure of the fit to recover the correct HOD in the case of the fiducial catalogues with 
assembly bias is dramatic. 
%%%%%%%%%%%%%%%%%%%%%%%%% FIGURE %%%%%%%%%%%%%%%%%%%%%%%%%%%%%

\begin{figure}
\begin{center}
\includegraphics[width=8.5cm]{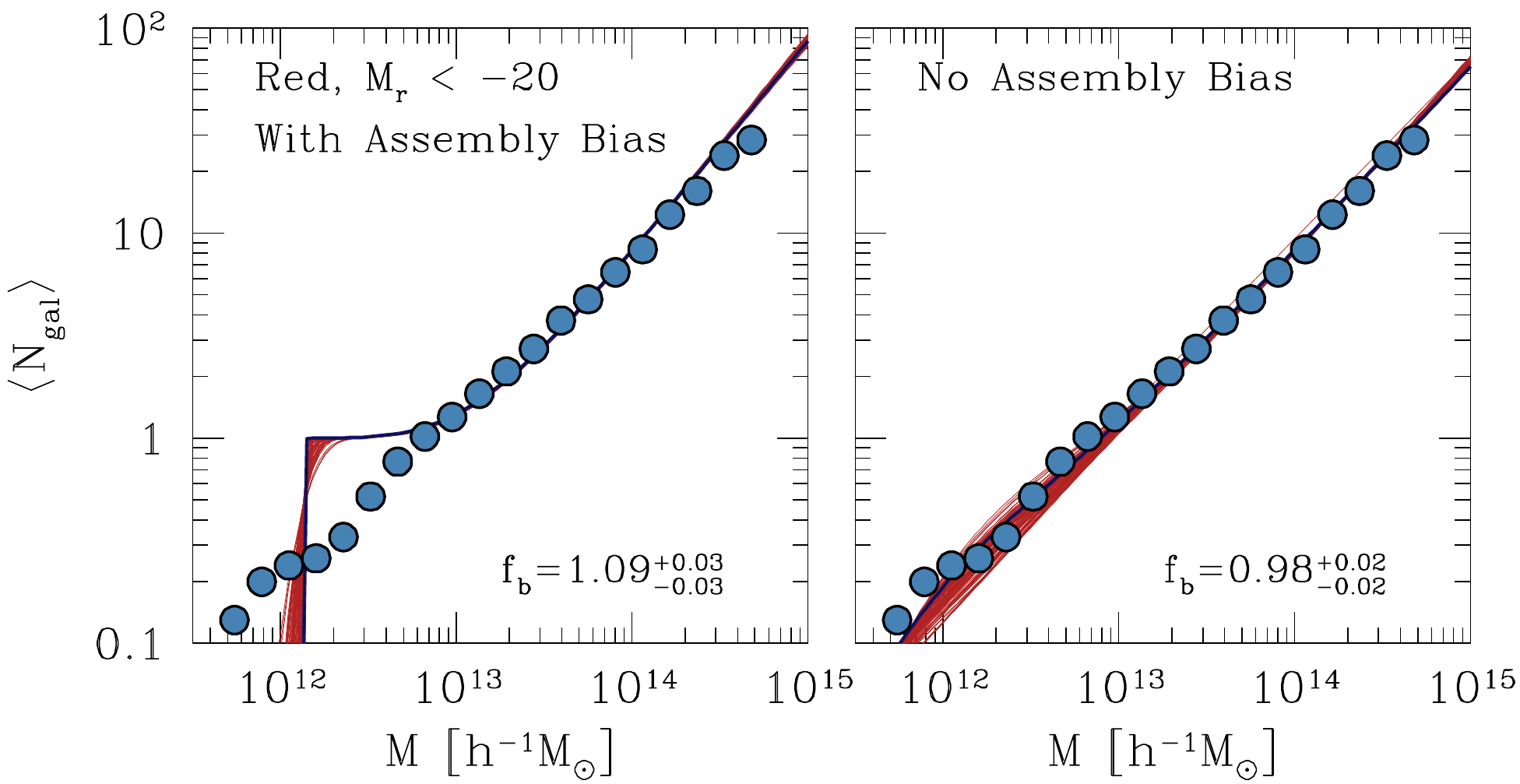}
\includegraphics[width=8.5cm]{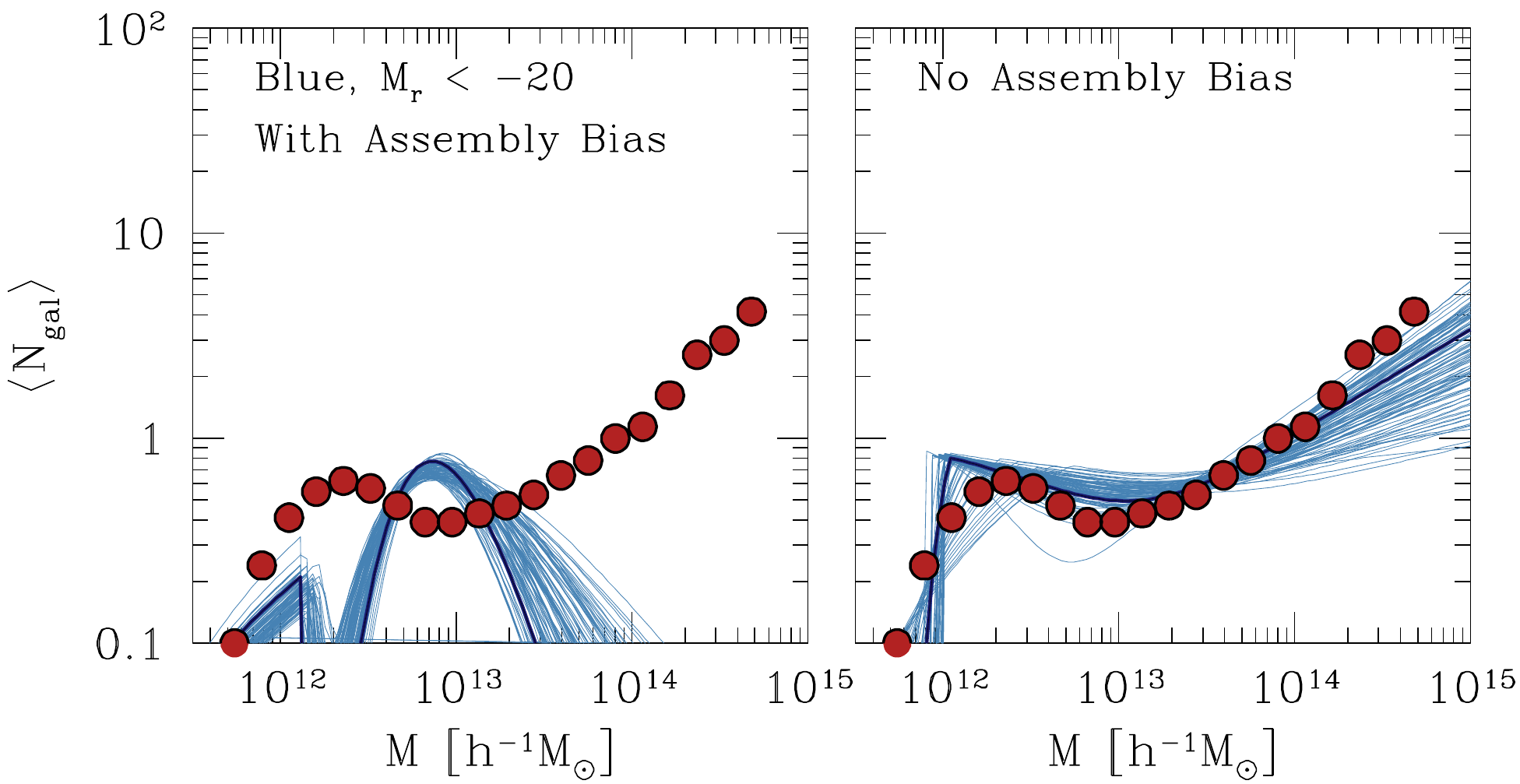}
\includegraphics[width=8.5cm]{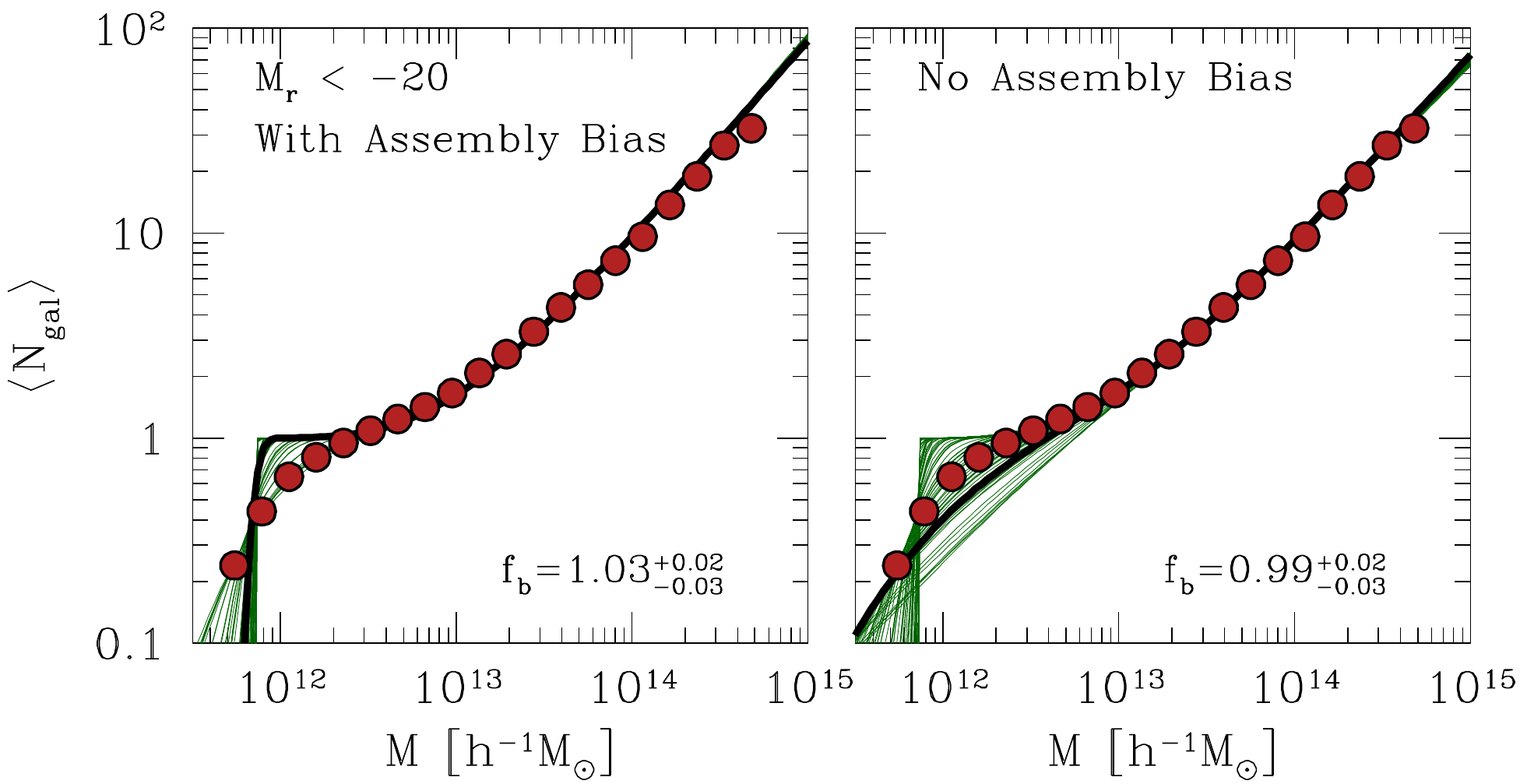}
\caption{
Comparison of the best-fit HODs for galaxies in the 
$M_r < -20$ sample with the true HOD in the simulation (points). 
This figure is the same as Fig.~\ref{fig:hod19}, but addresses the 
higher-luminosity samples with $M_r < -20$.  
}
\label{fig:hod20}
\end{center}
\end{figure}

%%%%%%%%%%%%%%%%%%%%%%%%%%%%%%%%%%%%%%%%%%%%%%%%%%%%%%%%%%%%%%%%%%%%

Consider now the blue galaxy samples in Fig.~\ref{fig:hod19}. Again, the HOD is recovered 
comparably well from the sample in which assembly bias has been eradicated, whereas the 
recovery of the true HOD is significantly poorer in the fiducial sample. However, in this case 
the sense of the difference is opposite to that in full the luminosity threshold sample. To be 
specific, the fits are driven to more gradual (rather than more rapid) central galaxy transitions 
and lower values of $\alpha$. The reason is because assembly bias {\em weakens} the large-scale 
clustering of blue galaxies, as discussed in \S~\ref{subsec:colorab}.  
Halo model fits attempt to compensate for this by placing as many galaxies as possible in 
lower-mass halos, thereby broadening the central galaxy transition and reducing $\alpha$. 

To conclude this section, we note that our fits to the clustering of $M_r<-19$ galaxies in the mock catalogues in 
which assembly bias has been eliminated are significantly less biased than fits to our fiducial samples. 
In general, the fits to the galaxy catalogues with no assembly bias generally recover the true, input HOD well. 
This is promising because it suggests that halo model methods have been sufficiently well developed 
such that when the data are consistent with the premises of the standard HOD implementation 
(i.e., no assembly bias), they do correctly model the clustering of galaxies in a wide range of reasonable 
models, and can be used to interpret galaxy clustering data. This is important because there exist 
few validation exercises that demonstrate this fact in the literature (we are aware of only 
\citet{reddick_etal13}, but even in this case the focus is on cosmological parameters, and 
HOD parameters are not treated in any detail). Moreover, the strength of assembly bias in the 
real universe remains an open question.

%------------------------------------------------------
\subsubsection{HOD fits to brighter samples}
\label{subsubsec:brightfits}
 
We now consider results pertaining to galaxy samples with brighter luminosity thresholds. 
We remind the reader that the clustering of {\em all} samples we consider in this 
section is adequately fit by an HOD model, as determined by a $\chi^2$ goodness-of-fit test.

Figure~\ref{fig:hod20} is analogous to Figure~\ref{fig:hod19}, but here we show the results from fits to $M_r < -20$
mock galaxy samples. In general, the 
true HOD is recovered relatively well for samples with no assembly bias, particularly for the luminosity threshold sample. 
On the other hand, the HODs inferred by fitting the clustering to the samples with 
assembly bias are poorer representations of the true HODs, even for the luminosity-only sample.

The sense of the biases are familiar from our discussion in \S~\ref{subsubsec:mr19fits}. 
In particular, in the luminosity threshold sample the central galaxy transition is biased to be significantly 
sharper than the true transition and the power-law index $\alpha$ is biased higher in order to increase 
the clustering strength to mimic the effect of assembly bias. More dramatic biases 
of the same sense are evident for the red galaxy subsample. Conversely, the inferred 
HODs of the blue galaxies are biased toward placing galaxies in low mass halos in 
order to {\em reduce} the clustering in blue samples (and mimic the effect of assembly 
bias on blue galaxies). This manifests in two ways. First, the best-fit HOD model places essentially all 
blue centrals in low-mass halos, rather than over a broad range of halo masses. Second, blue 
satellites are inferred to be significantly less abundant within high-mass 
($M_{\mathrm{halo}} \gtrsim 10^{13}\, \hMsun$) than they truly are. In this case, the 
failure is so dramatic that an analyst performing such a fit would recognize it as 
incorrect. For example, this fit implies no blue satellite galaxies in large clusters. 
This is manifestly false. In this case, adding additional data, such as galaxy-galaxy lensing 
\citep[as in][]{leauthaud_etal12,vdBosch13}, or galaxy number-to-halo mass 
ratios \citep[as in][]{tinker_etal12}, or additional priors should greatly mitigate against 
this failure mode.

Fig.~\ref{fig:hod20} makes it apparent that the best-fit HOD models fail 
to recover the true halo occupation statistics exhibited by any of our fiducial 
$M_r < -20$ galaxy samples. In contrast, the HODs of the $M_r < -20$ without assembly bias 
are recovered comparably well. However, both samples have identical true HODs. 
We conclude that the failure of the standard HOD model to accurately describe our fiducial galaxy distributions is 
due in large part to the presence of assembly bias in these samples.

Figure~\ref{fig:hod20nofb} in the Appendix shows inferred HODs for the same 
samples as in Fig.~\ref{fig:hod20} but in in an alternative 
case in which $f_{\mathrm{b}}$ has been fixed to unity. In this case, the biases 
are more statistically significant, indicating that including 
such a nuisance parameter is important for mitigating, in part, biases in inferred 
HODs due to assembly bias. We have included the $f_{\mathrm{b}}$ bias parameter 
in order to be conservative, and minimize the systematic errors in HODs induced by assembly 
bias. However, it is worth noting that our conservative choice to include such a parameter is 
not standard practice.

%%%%%%%%%%%%%%%%%%%%%%%%% FIGURE %%%%%%%%%%%%%%%%%%%%%%%%%%%%%

\begin{figure}
\begin{center}
\includegraphics[width=8.5cm]{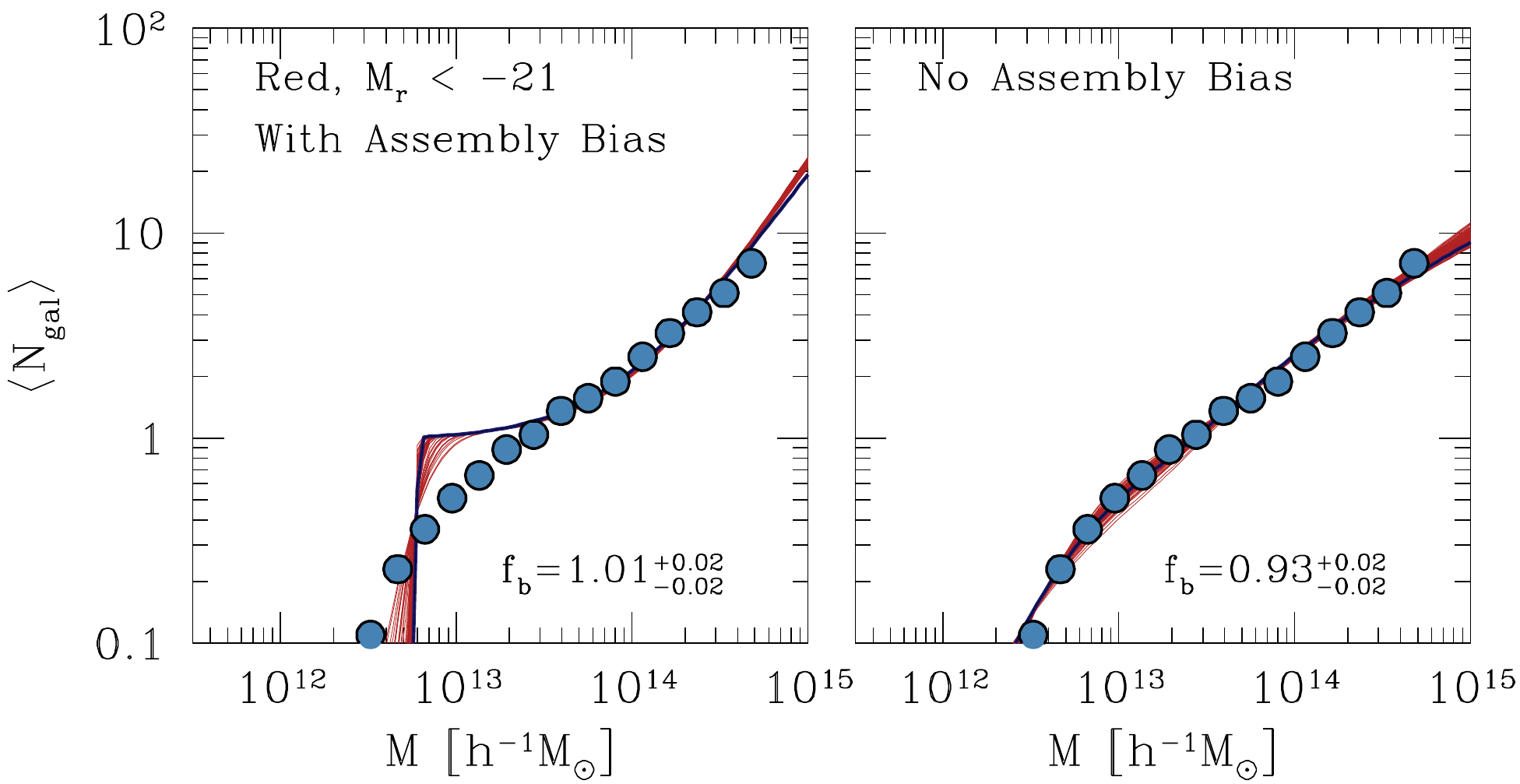}
\includegraphics[width=8.5cm]{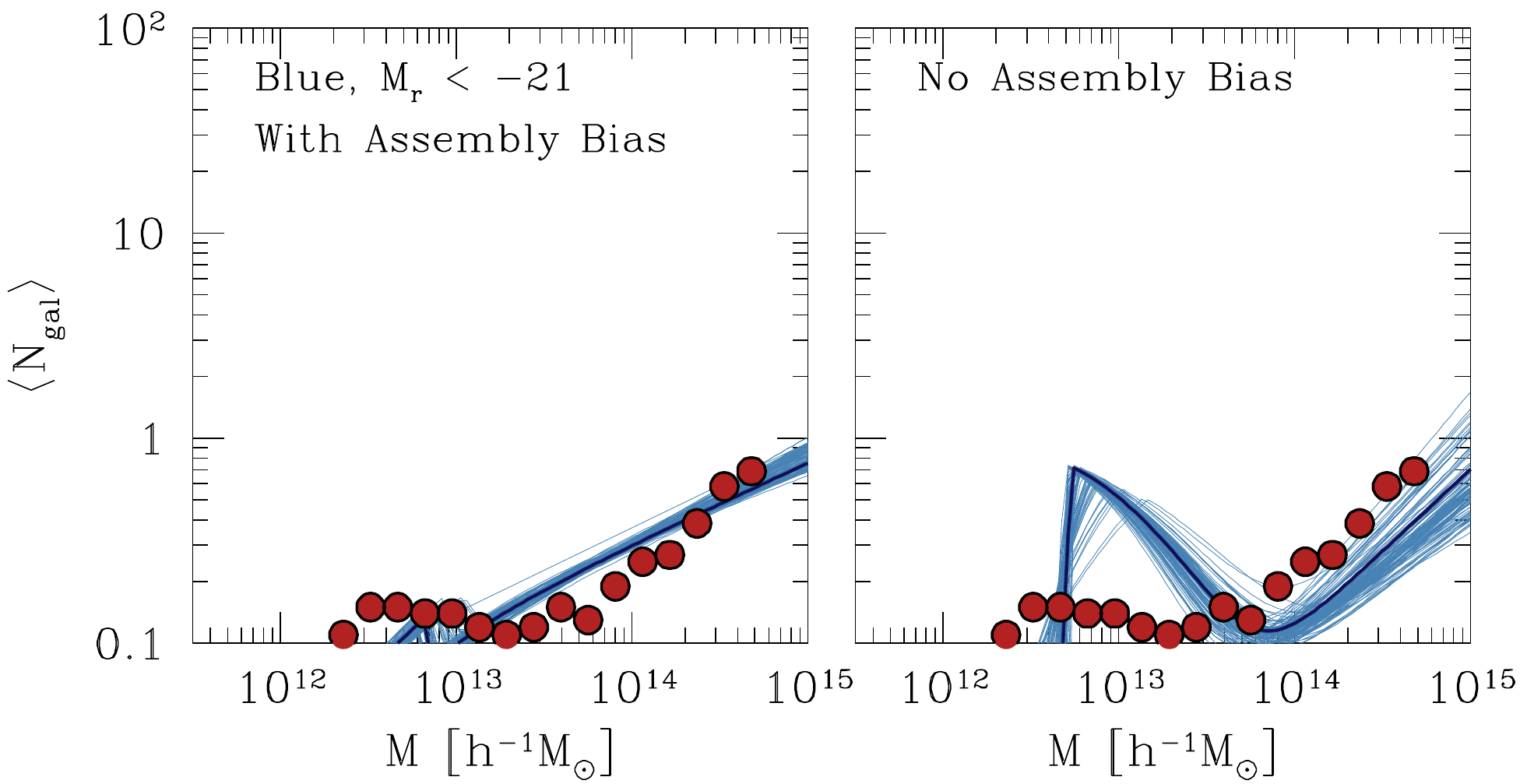}
\includegraphics[width=8.5cm]{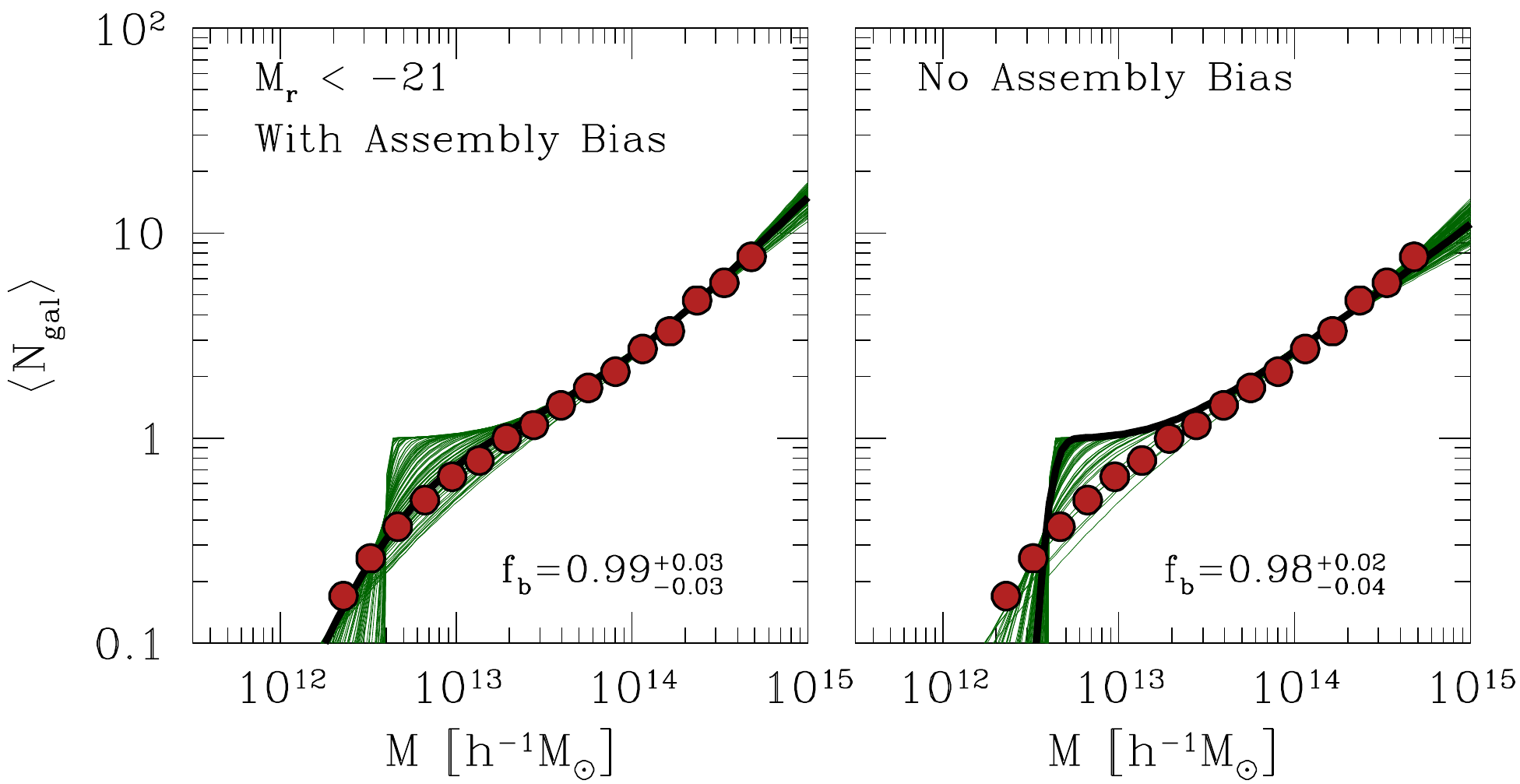}
\caption{
Comparison of the best-fit HODs for galaxies in the 
$M_{\mathrm{r}}<-21$ sample with the true HODs in the simulation (points). 
This is the same as Fig.~\ref{fig:hod19}, but addresses this higher-luminosity 
$M_r < -21$ samples.
}
\label{fig:hod21}
\end{center}
\end{figure}

%%%%%%%%%%%%%%%%%%%%%%%%%%%%%%%%%%%%%%%%%%%%%%%%%%%%%%%%%%%%%%%%%%%%

Figure~\ref{fig:hod21} depicts the inferred HODs for the $M_r < -21$ mock galaxies. 
The same broad features that we expounded upon above are evident for the case of red subsamples. 
However, the other cases do not 
qualitatively resemble their counterparts in the lower luminosity samples. In the luminosity 
threshold samples shown in the bottom row of panels in 
Fig.~\ref{fig:hod21}, the true underlying HOD is recovered more faithfully in the presence 
of assembly bias, despite the fact that assembly bias is {\em not} included in the modeling. 
The bias in the HOD inferred from the mock galaxy sample with no assembly bias is, 
in part, due to the extra parameter freedom from the halo bias parameter, $f_{\mathrm{b}}$. 
The increased clustering due to the erroneously sharp central galaxy transition can 
be compensated by decreasing $f_{\mathrm{b}};$ these parameter shifts result 
in a slightly better $\chi^2$ despite the offset in the inferred HOD. Additionally, 
the blue mock galaxies with $M_r < -21$ are an interesting exception 
to the general trends observed in our other sample analyses. The inferred HODs in these 
cases certainly differ from each other; however, in neither case are the inferred blue HODs 
representative of the true, underlying blue galaxy HODs. These results suggest one or more 
failures of the halo model to describe the clustering of the halos in these samples.

We summarize these results by reiterating the salient point of this section: 
fits to the clustering exhibited by mock galaxies with assembly bias typically yield 
inferred HODs that are biased. Thus in the absence of independent evidence 
justifying the assumption that assembly bias is zero, 
we conclude that conventional implementations of the HOD are ill-equipped to 
robustly constrain galaxy-halo models with two-point clustering measurements alone. 
We reach this conclusion even though we have made the conservative choice 
to marginalize over $f_{\mathrm{b}},$ whereas what is almost universally done is 
to assume that halo bias is perfectly calibrated and hold $f_{\mathrm{b}}$ fixed to unity.

%-----------------------------------------------------------------------
\subsection{Constraints on Individual HOD Parameters}
\label{subsec:hodparams}

The preceding figures make clear the trends in the systematic errors on inferred HOD parameters 
to be expected when fitting galaxy samples in which galaxy properties are correlated with halo properties other 
than mass. They also give a reasonable representation of how different the inferred and 
true HODs can be when analyzing real galaxy samples, where assembly bias may or may not 
be present. Furthermore, the HODs represented in Fig.~\ref{fig:hod19} through Fig.~\ref{fig:hod21} 
include the covariance among the inferred HOD parameters in each case.

Nonetheless, there is considerable interest in the constraints on individual HOD parameters, and estimates of these 
parameters can often be of practical use. We turn now to the marginalized constraints on  
particular HOD parameters in each of our samples. We focus on the three parameters that tend to 
garner the broadest interest: (1) the mass scale at which 
the average number of central galaxies per halo is $1/2$, 
$\log (M_{\mathrm{min}}/\hMsun)$ (an inferred parameter); (2) 
the mass scale at which the average number of satellite galaxies is $1$, $\log (M_1/\hMsun)$; 
and (3) the power-law index of the satellite portion of the HOD, $\alpha$.

%%%%%%%%%%%%%%%%%%%%%%%%%%%%%%%%%%%%%%%%%%%%%%%%%%%%%%%%%%%
%%%%%%%%%%%%%%%%%%%%%%%%% FIGURE %%%%%%%%%%%%%%%%%%%%%%%%%%%%%
\begin{figure}
\includegraphics[width=7.5cm]{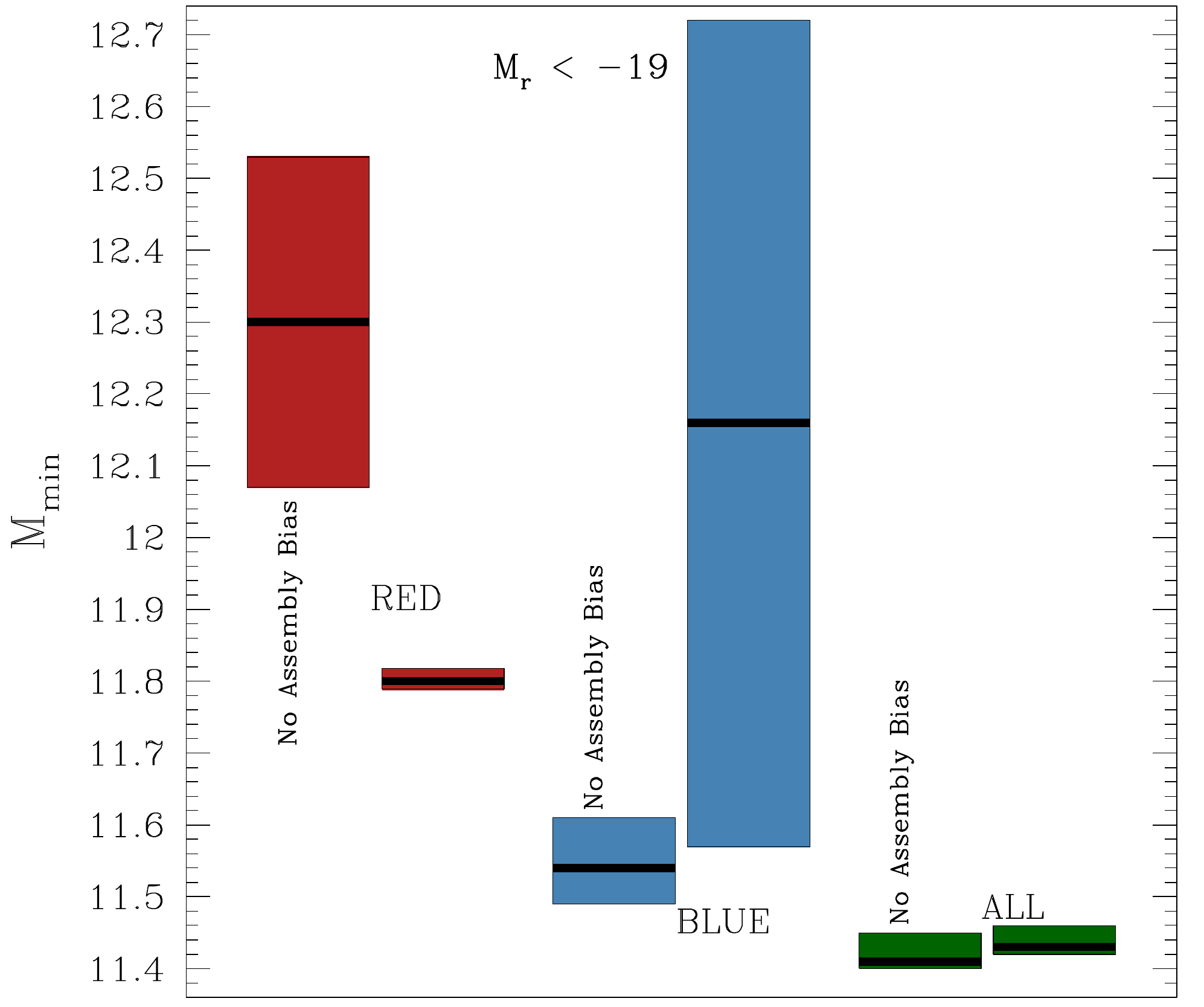}
\includegraphics[width=7.5cm]{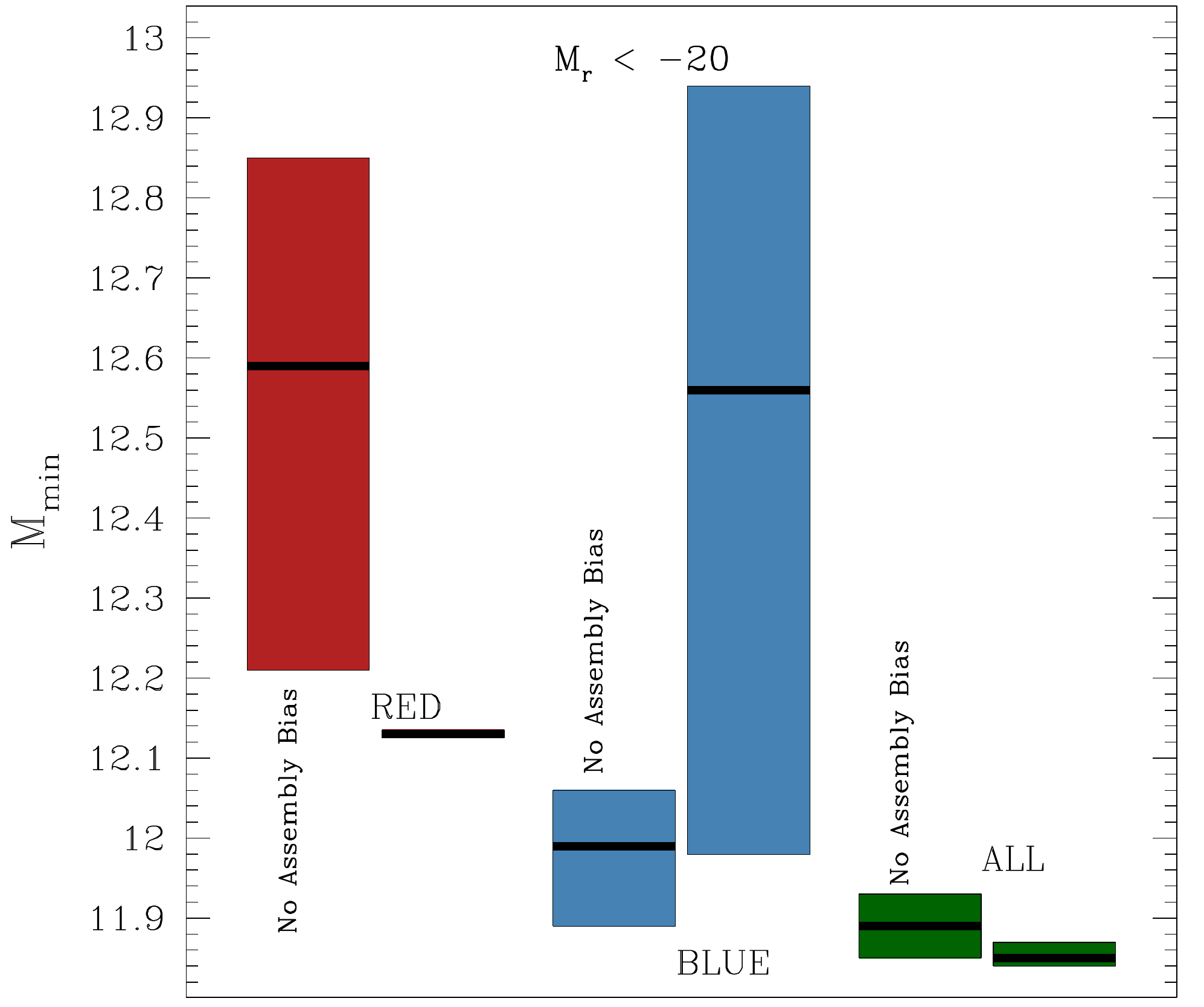}
\includegraphics[width=7.5cm]{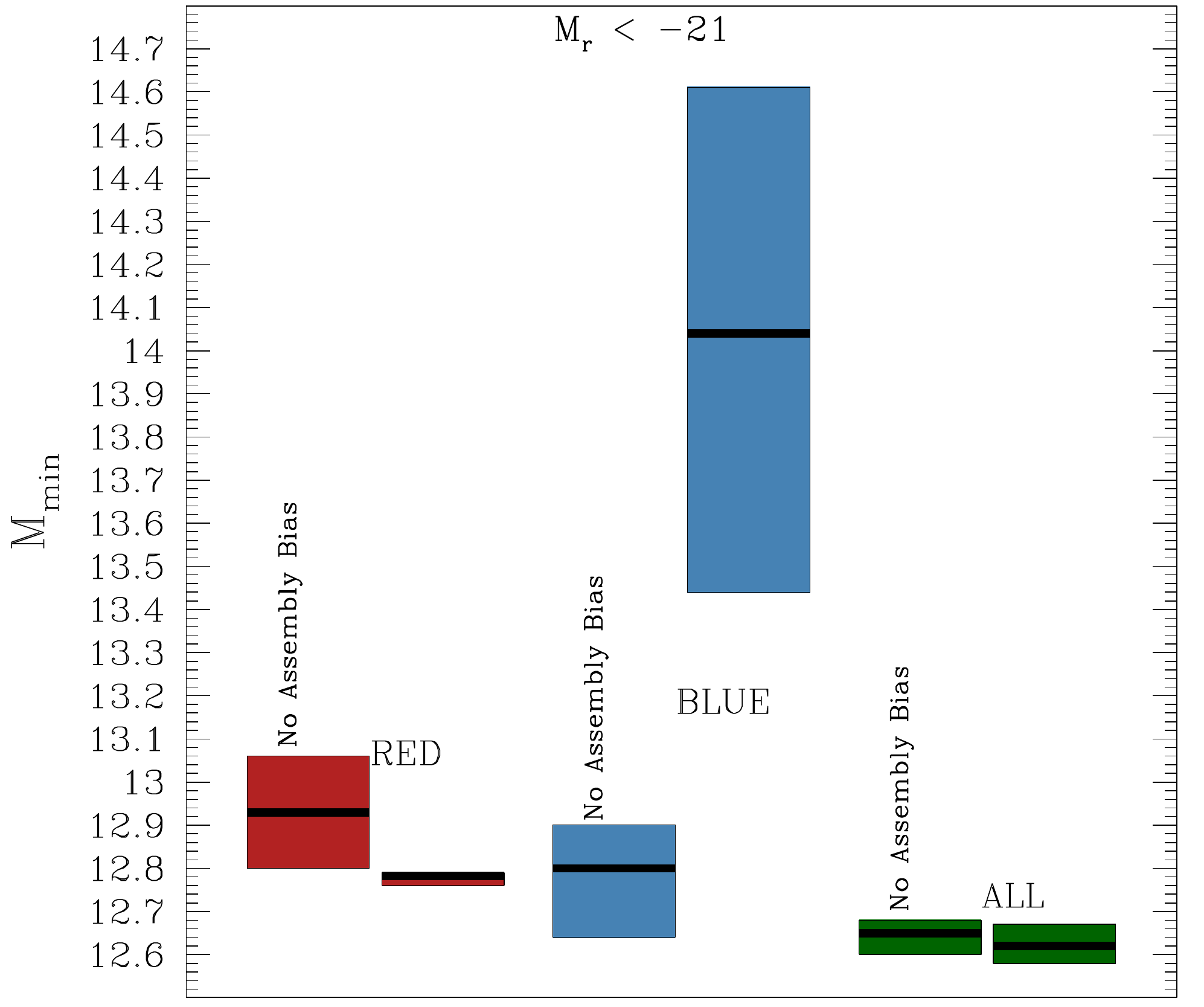}
\caption{
Constraints on $\log \Mmin/\hMsun$ inferred by fitting $\wprp$ from the mock galaxy catalogues. 
In each case the bars span the $1\sigma$ marginalized constraint 
on $\Mmin$ defined to be symmetric in the sense that the posterior integrates 
to $68.3\%$ over the domain delineated by the bands while the high-$\Mmin$ and 
low-$\Mmin$ tails outside of the bands each integrate to $(100-68.3)\%/2 = 15.85\%$ 
of the posterior. The {\em thick, solid} lines show the median values of $\Mmin$. 
From top to bottom, the panels show the results of the 
$M_r < -19$, $M_r < -20$, and $M_r < -21$ samples. 
Within each panel, we show the results for the luminosity threshold samples 
(labelled "ALL" and at the far right), as well as the color-split samples separately.
}
\label{fig:mmins}
\end{figure}
%%%%%%%%%%%%%%%%%%%%%%%%%%%%%%%%%%%%%%%%%%%%%%%%%%%%%%%%%%%

Figure~\ref{fig:mmins} shows the $1\sigma$ marginalized constraints on the derived parameter 
$\log (M_{\mathrm{min}}/\hMsun)$ for each of the samples we have fit with an HOD. While we 
have derived constraints on the logarithm $\log (M_{\mathrm{min}}/\hMsun)$, 
we will refer to these as ``$\Mmin$ constraints'' in the interest of brevity.

For the luminosity threshold samples and the red sub-samples, 
the constraints on $\Mmin$ are typically tighter when derived 
from the fiducial mock galaxy populations with assembly bias. This is largely because these 
models are driven to have very low $\sigma_{\log M}$ values in order to place galaxies in the 
most massive halos possible, thereby boosting clustering. When $\sigma_{\log M}$ is limited to 
a very narrow range, $\Mmin$ is also limited to a very narrow range in order to guarantee 
that the HOD describes a model with the correct average number density of galaxies. 
Counter-intuitively, this also explains why the inferred values of $\Mmin$ are 
generally {\em smaller} in the samples with assembly bias. When $\sigma_{\log M}$ is 
relatively large, a large fraction of all galaxies in the sample reside in halos with 
masses below $\Mmin$ because the halo mass function increases rapidly as 
halo mass is decreased. The blue sub-samples run counter 
to this general trend because their clustering is diminished by assembly bias 
rather than enhanced (see \S~\ref{subsec:colorab}).

It is clear from Figure~\ref{fig:mmins} that the differences in the inferred values of $\Mmin$ between samples with and 
without assembly bias can be quite significant. Moreover, it is also significant that the precision of the 
inferred constraints on HOD parameters varies significantly between the models with and without 
assembly bias. For perspective on this, consider the particular case of the 
luminosity threshold sample with $M_r < -20$ (the green bands in the middle panel of Fig.~\ref{fig:mmins}),  
in which the systematic difference may not seem egregious. In this case, a fit to the 
assembly-biased mock galaxy data would rule out the median $\Mmin$ for the case with no assembly 
bias by more than $\sim 2\sigma$, despite the fact that the inferred HOD in the case with no assembly 
bias is an excellent description of the true HOD (Fig.~\ref{fig:hod20}). Differences of this sort 
are particularly pronounced for red-selected samples for which the inferred values 
of $\Mmin$ differ by significantly more than the statistical errors on $\Mmin$. 
This strongly suggests that the error budgets of HOD analyses of galaxy clustering require a 
substantial, previously neglected contribution from the unknown strength of galaxy assembly 
bias in the real Universe. We will return to this point below, and in \S~\ref{sec:discussion}.

\begin{figure}
\includegraphics[width=7.5cm]{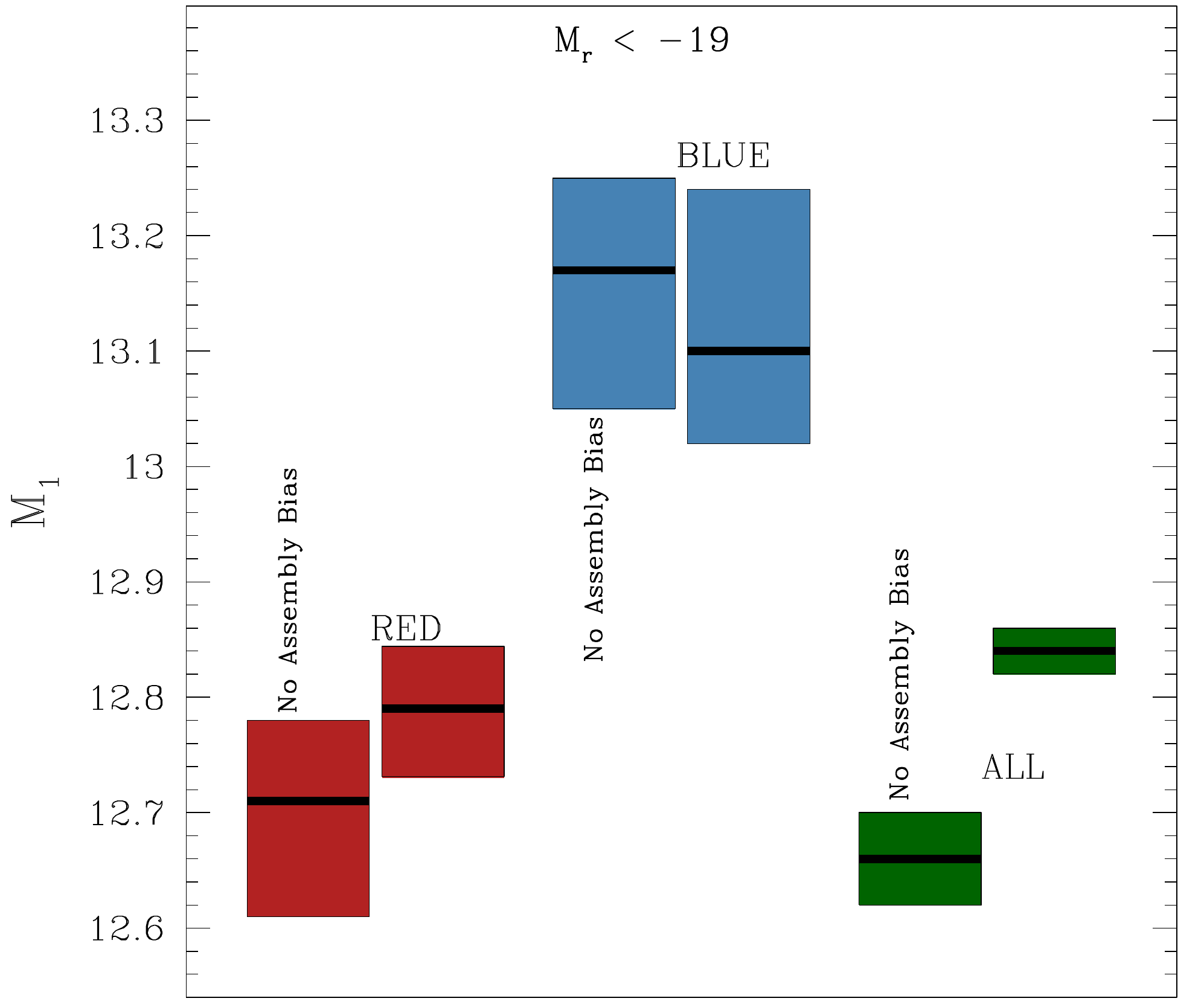}
\includegraphics[width=7.5cm]{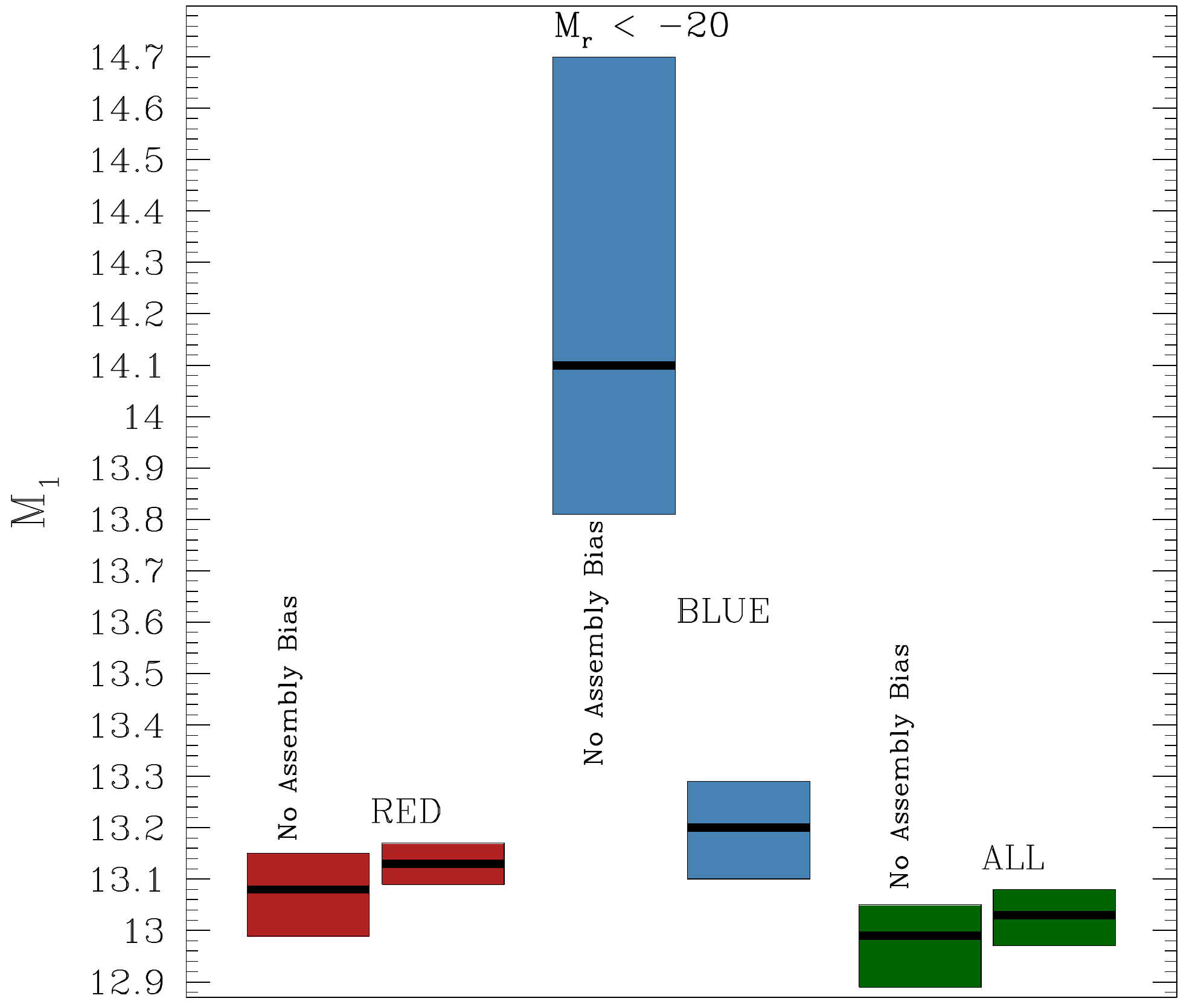}
\includegraphics[width=7.5cm]{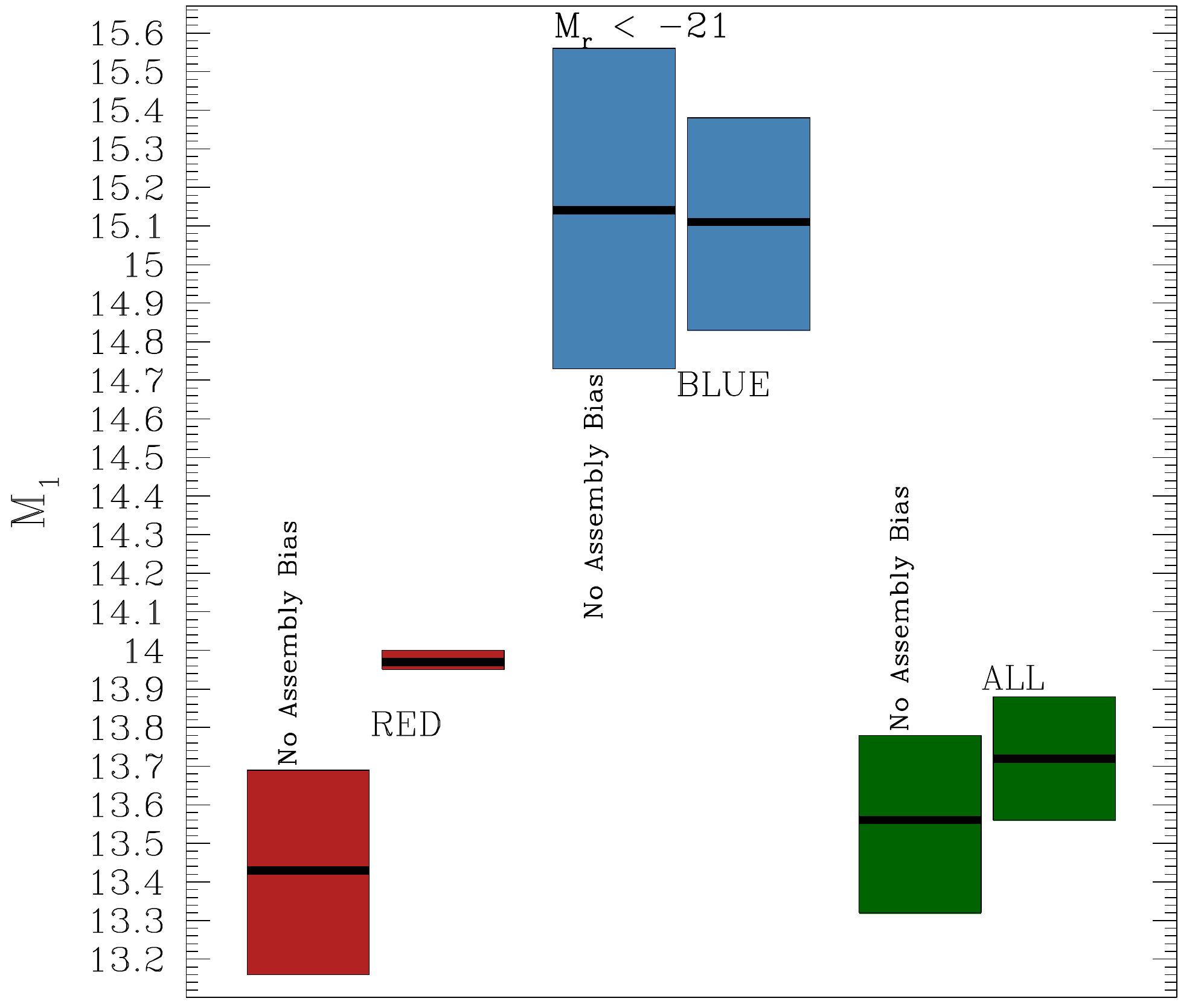}
\caption{
Constraints on $\log M_1/\hMsun$ inferred by fitting $\wprp$ from the mock galaxy catalogues. 
In each case the bars span the $1\sigma$ marginalized constraint on $M_1$ defined as 
described in Figure~\ref{fig:mmins}. The {\em thick, solid} lines show 
the median values of $M_1$. From top to bottom, the panels show the 
results of the $M_r < -19$, $M_r < -20$, and $M_r < -21$ samples. Within each panel, we show the 
results for the luminosity threshold samples (labelled "ALL" and at the far right), as well as the color-split 
samples separately. 
}
\label{fig:mones}
\end{figure}

Figure~\ref{fig:mones} depicts the inferred values of the parameter $M_1.$ 
As with $\Mmin$, the offsets in inferred $M_1$ values 
can be significant. In most cases, the systematic offsets in the inferred values 
of $M_1$ are comparable to, or larger than, the statistical errors on these 
inferences. Even for the luminosity threshold samples, the offsets in 
the constraints on $M_1$ are significant at all luminosities. 
Again, there is a relatively general pattern to these systematic offsets. 
For the luminosity threshold samples and the red galaxy samples, the trend is for the 
inferred $M_1$ to be larger in the fiducial models with assembly bias, whereas for the 
blue mock galaxy samples the inferred $M_1$ is typically lower in the 
fiducial mock catalogues. Again, this is driven by the HOD adjusting to 
increase clustering, by packing galaxies into the most massive halos possible, 
in the former cases and to reduce clustering in the latter case.

%%%%%%%%%%%%%%%%%%%%%%%%%%%%%%%%%%%%%%%%%%%%%%%%%%%%%%%%%%%
%%%%%%%%%%%%%%%%%%%%%%%%% FIGURE %%%%%%%%%%%%%%%%%%%%%%%%%%%%%
%%%%%%%%%%%%%%%%%%%%%%%%%%%%%%%%%%%%%%%%%%%%%%%%%%%%%%%%%%%

\begin{figure}
\includegraphics[width=7.5cm]{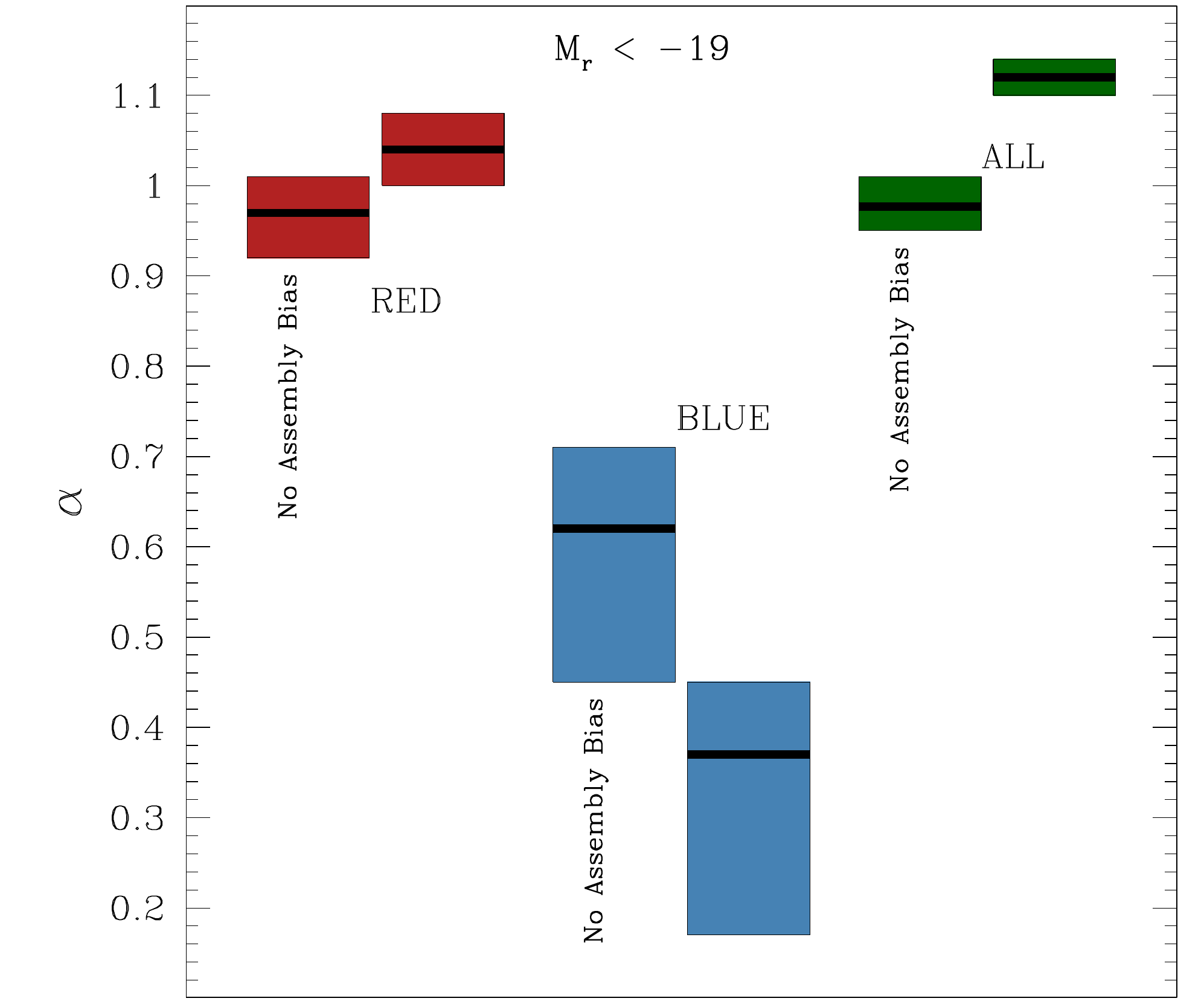}
\includegraphics[width=7.5cm]{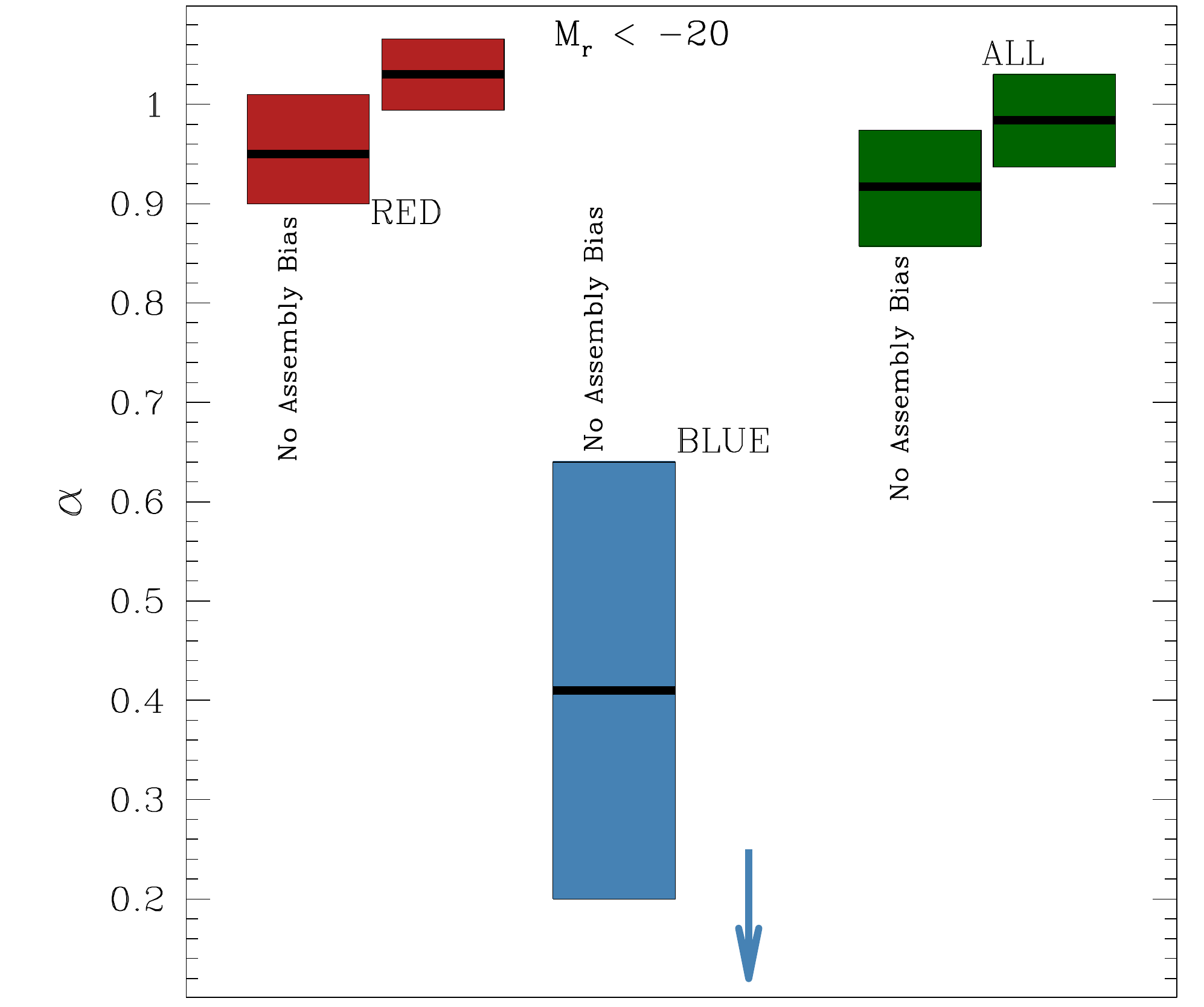}
\includegraphics[width=7.5cm]{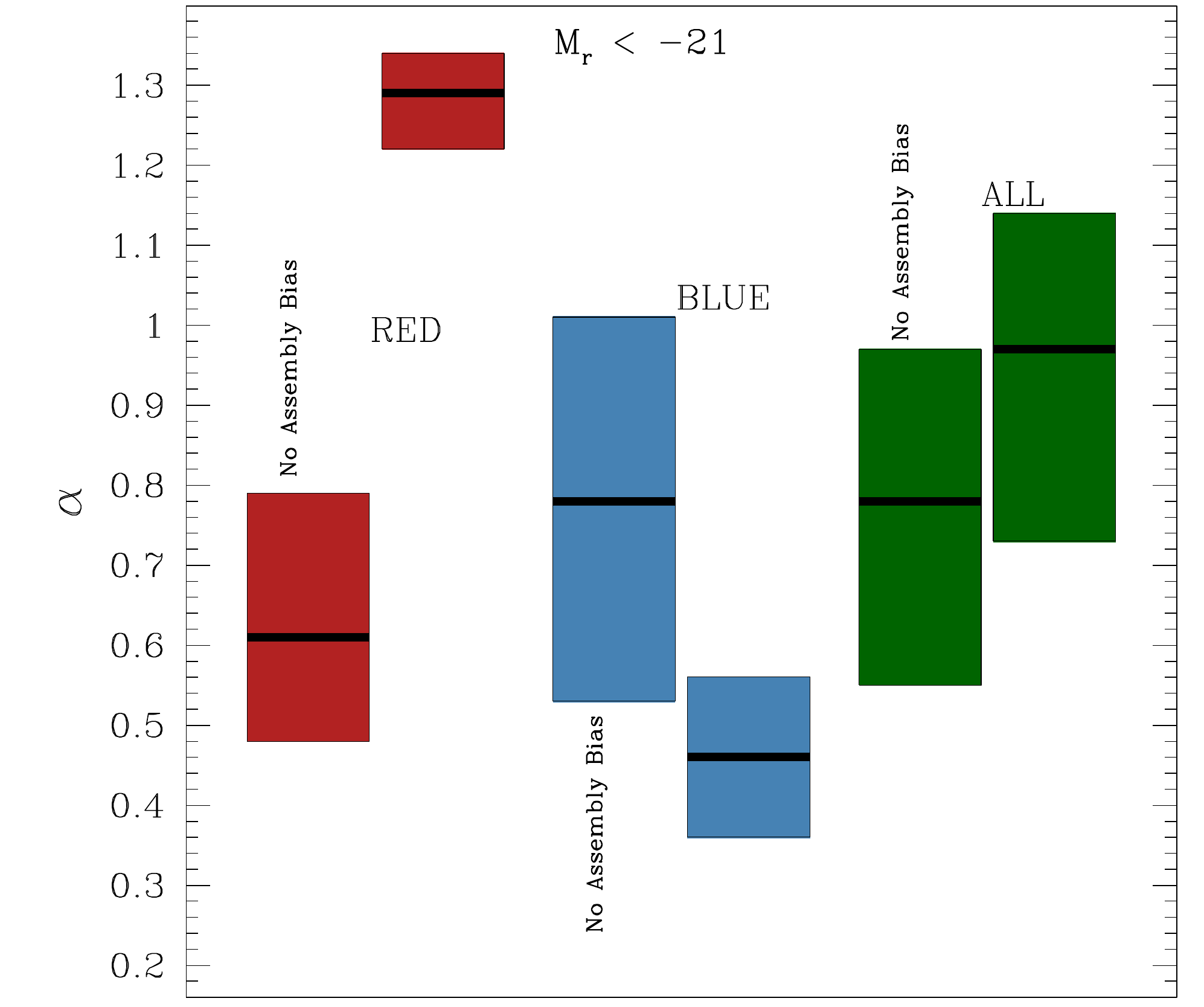}
\caption{
Constraints on $\alpha$ inferred by fitting $\wprp$ from the mock galaxy catalogues. 
In each case the bars span the $1\sigma$ marginalized constraint on $\alpha$ defined as 
described in Figure~\ref{fig:mmins}. 
The {\em thick, solid} lines show the median values of $\alpha$. From top to bottom, the panels show the 
results of the $M_r < -19$, $M_r < -20$, and $M_r < -21$ samples. Within each panel, we show the 
results for the luminosity threshold samples (labelled "ALL" and at the far right), as well as the color-split 
samples separately. The constraint for the blue mock galaxies in the fiducial, assembly-biased 
$M_r < -20$ sample is not shown because it is off the scale of the figure. The marginalized constraint 
in this case is $\alpha=-2.04^{+1.04}_{-2.24}$. 
}
\label{fig:alphas}
\end{figure}

Following our discussion of $\Mmin$ and $M_1$, Figure~\ref{fig:alphas} gives the inferred values 
of the satellite galaxy power-law index $\alpha$ in each of our fits. 
In the case of $\alpha$, the systematic offsets induced by assembly bias are comparable to, 
or larger than, the statistical errors in all cases, including the threshold samples. In the case of 
the blue galaxies in the fiducial catalogue with $M_r < -20$, the constraint on $\alpha$ is not shown 
in Fig.~\ref{fig:alphas} because it is significantly negative and depicting this constraint 
would alter the scale of the figure to the degree that clarity would suffer. The marginalized 
$1\sigma$ constraint on $\alpha$ in this case is $\alpha=-2.04^{+1.04}_{-2.24}$. 
As we mentioned above, an analyst would likely identify this value to be unphysical, 
as it predicts essentially zero blue galaxies in large clusters. In our analysis, 
we did not include any prior on $\alpha$. For the luminosity threshold samples and for the red galaxies, 
the trend is for the inferred values of $\alpha$ to be larger in the fiducial samples than 
in the samples with assembly bias removed. The fits are driven to larger values of 
$\alpha$ in order to place galaxies preferentially in relatively rare, highly-biased halos, 
thereby boosting clustering. The values of $M_1$ inferred from these samples are larger as well 
in order to keep the total galaxy number density fixed and to mitigate the increase in 
clustering strength on small scales ($r_p \lesssim 1\, h^{-1}\mathrm{Mpc}$), where pairs of galaxies 
within common host halos dominate the signal. We refer the reader to \citet{watson_powerlaw11} 
for a more in depth discussion of the factors that determine the relative strength of the 
small-scale (one-halo) and large-scale (two-halo) clustering of galaxies. As with $M_1$, and 
for the reasons discussed in \S~\ref{subsec:colorab}, the systematic error in $\alpha$ is 
of the opposite sense for blue galaxies as compared to red galaxies.

%%%%%%%%%%%%%%%%%%%%%%%%%%%%%%%%%%%%%%%%%%%%%%%%%%%%%%%%%%%
%%%%%%%%%%%%%%%%%%%%%%%%% FIGURE %%%%%%%%%%%%%%%%%%%%%%%%%%%%%
%%%%%%%%%%%%%%%%%%%%%%%%%%%%%%%%%%%%%%%%%%%%%%%%%%%%%%%%%%%

The results in this section suggest that galaxy assembly bias at levels that can not easily be ruled out 
may have a statistically significant effect on inferences made about the relationship between galaxies and the halos 
in which they reside. These biases are best represented in the complete HOD representations 
shown in Fig.~\ref{fig:hod19} through Fig.~\ref{fig:hod21} because the systematic errors in the 
conventional HOD parameters (e.g., $\Mmin$, $M_1$, $\alpha$) are strongly correlated. Nevertheless, 
the marginalized constraints on individual HOD parameters exhibit systematic errors that are statistically 
significant (Fig.~\ref{fig:mmins} through Fig.~\ref{fig:alphas}). In the Appendix, we demonstrate that 
removing the freedom provided by the $f_{\mathrm{b}}$ nuisance parameter renders these systematic 
errors slightly more significant, compared to the statistical errors on the inferred HOD parameters. Consequently, 
biases will generally be larger if no such nuisance parameter is introduced. We also show in the Appendix 
that marginalizing over the internal spatial distributions of satellite galaxies does not mitigate the systematic 
errors in HODs induced by assembly bias (and can even increase the systematic offset). These 
demonstrations indicate that nuisance parameters introduced in previous HOD-based studies 
do not necessarily alleviate the effects of assembly bias on inferred HODs.

At minimum, these results suggest that assembly bias is an observationally relevant effect that is 
not accounted for in standard HOD/CLF analyses. In fact, it is worth reiterating that we should expect 
actual data to be significantly {\em more} sensitive to assembly bias than the mock catalogues we have analyzed, with the 
consequence that inferences based on observational data may contain systematic errors of great statistical significance 
than those that we quote.

%---------------------------------------------------------------------------------------------------------------------------------
% Predictions of our catalogues and models
\section{Additional Predictions of the Halo Model and HOD}
\label{sec:predictions}

The results presented in \S~\ref{sec:results} demonstrate that traditional HOD fits to 
measurements of $\wproj(\rproj)$ and $\bar{n}_{g}$ can be significantly altered 
by the presence of assembly bias. Evidently, when galaxy assembly bias is present in 
the data there is sufficient parametric freedom in the HOD to compensate for incorrectly 
assuming assembly bias to be absent. In light of these results, we argue that it is necessary  
to search for statistics other than $w_{\mathrm{p}}(r_{\mathrm{p}})$ 
that may be sensitive to assembly bias so that these degeneracies may be broken. 
In \S~\ref{subsec:vpf} we investigate the potential for the void probability 
function to detect the presence of galaxy assembly bias. 

Once a set of acceptable HOD parameters has been determined by fitting galaxy clustering data, 
the HOD formalism enables new predictions about galaxy evolution to be made. We present an 
example of such a prediction in \S~\ref{subsec:quenching}, in which we study the inferred host mass-dependence 
of satellite quenching, with particular attention to the threat that unknown levels of assembly bias pose 
for such inferences.

%%%%%%%%%%%%%%%%%0%%%%%%%%%%%%
\subsection{The Void Probability Function}
\label{subsec:vpf}

The amount of galaxy assembly bias in the universe is unknown and cannot be revealed 
by galaxy clustering statistics alone. Consequently, a natural question to ask is whether some 
additional statistic describing the galaxy distribution can be used to test for assembly 
bias. One natural candidate for such a statistic is the {\em void probability function} (VPF), 
defined as the probability that a spherical region of some radius will be devoid of galaxies. 
The VPF has been studied previously for precisely this purpose. In \citet{tinker06,tinker_etal08b}, the 
authors scrutinized particular models with assembly bias. The models they studied 
had VPFs that are inconsistent with SDSS measurements, even though the clustering in those models 
{\em is} consistent with the data, effectively falsifying those models.\footnote{See also \citet{conroy_etal05} for 
an investigation of the information content in the VPF that is independent from the two-point function.}

In this section, we present  the VPF predicted by the mock catalogues explored in \S~\ref{sec:results}, 
focusing on the $M_r<-20$ sample for brevity\footnote{Recall that the standard HOD interpretation of the 
clustering in the $M_r < -19$ color-selected samples with assembly bias is already inconsistent, so the 
additional value of the VPF is limited in this case.}. We compute the VPF in our mock catalogues by randomly placing 
$10^6$ spheres of a given radius within the simulation box and counting the fraction that are empty. We estimate 
errors by jackknife resampling over the eight octants of the cubical simulation volume.

\begin{figure}
\includegraphics[width=6.3cm]{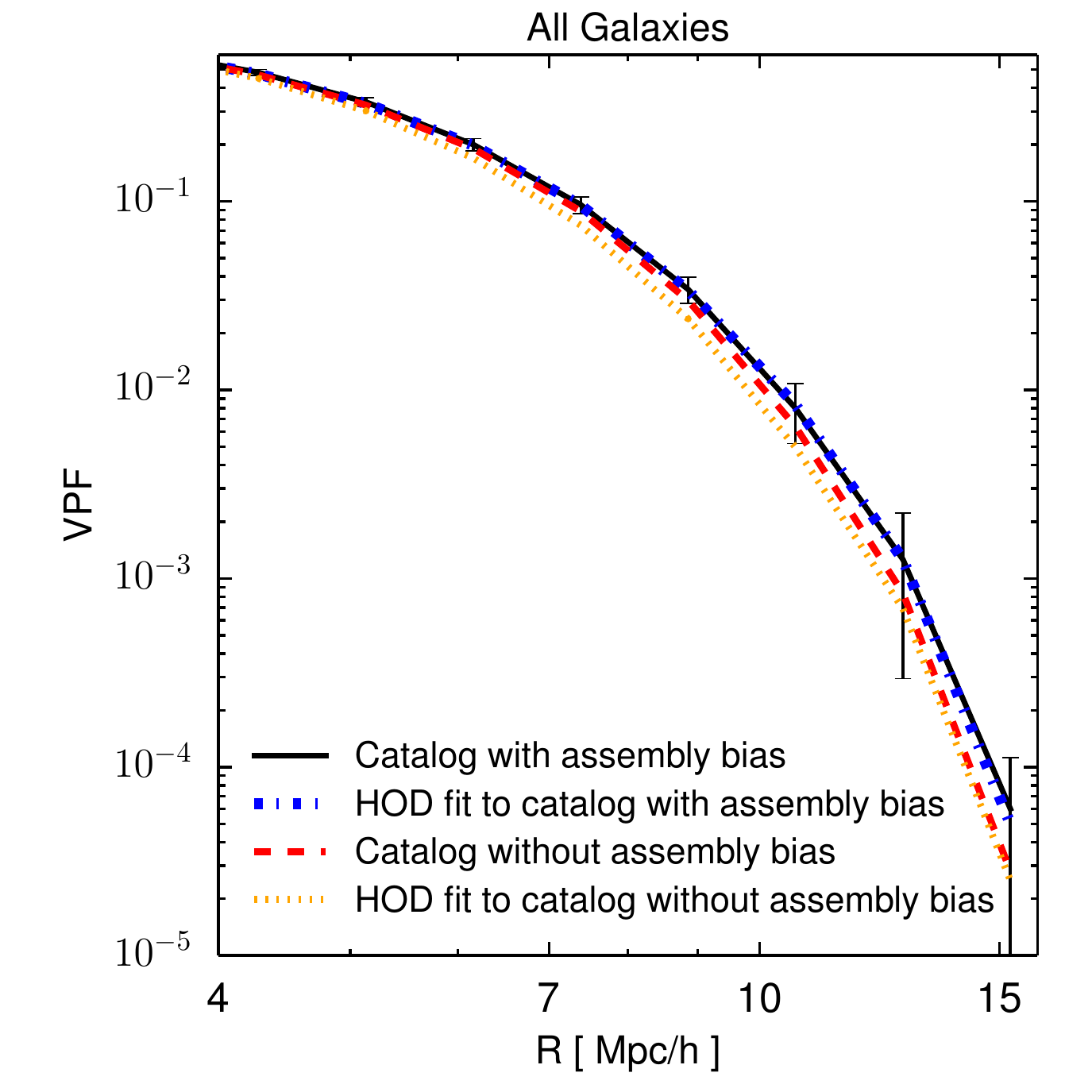}
\includegraphics[width=6.3cm]{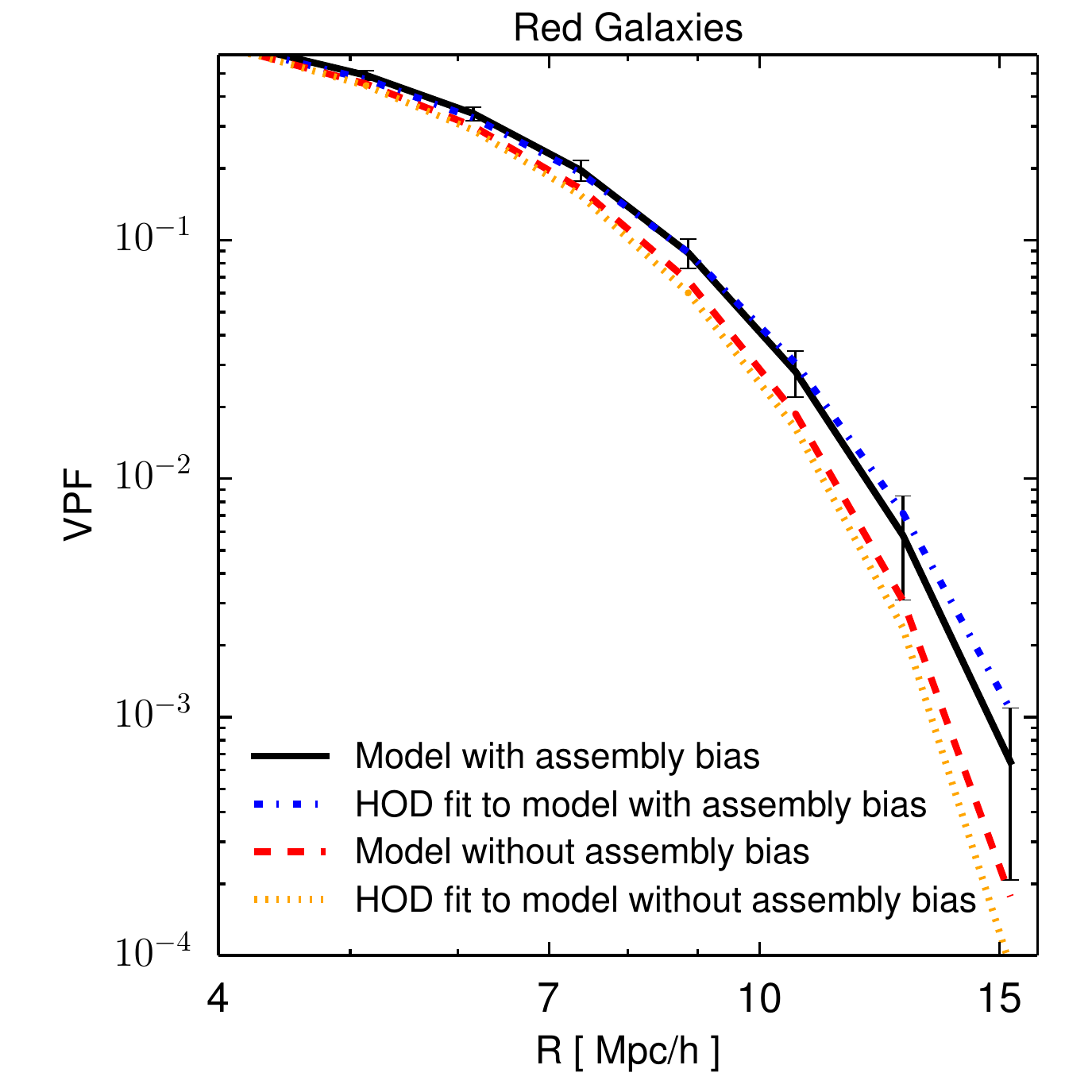}
\includegraphics[width=6.3cm]{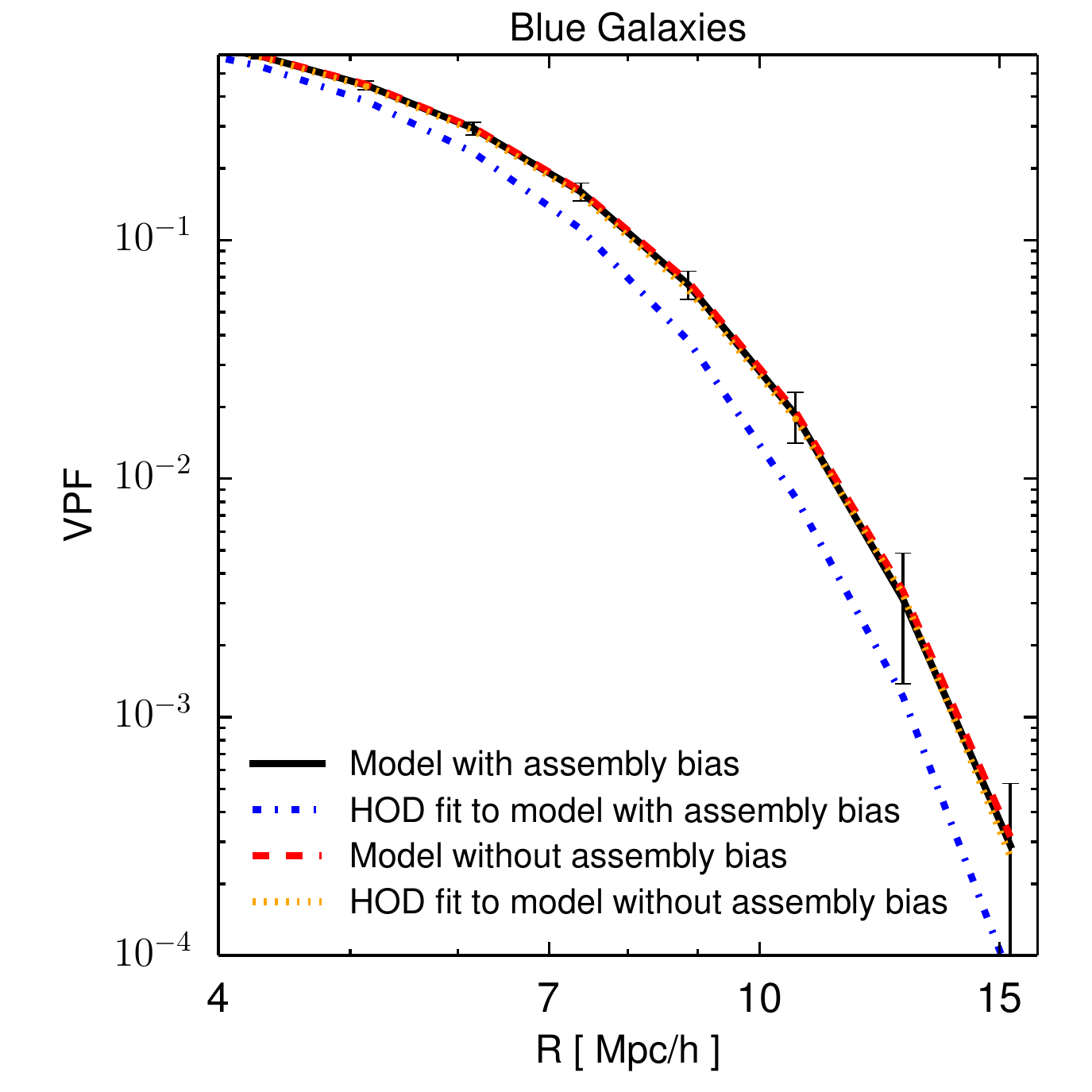}
\caption{
Void Probability Function (VPF) for four of our $M_r<-20$ threshold samples. 
Results for full luminosity-threshold samples appear in the {\em top} panel. 
The two {\em bottom} panels are analogous to the top panel, but show VPFs for our color-selected 
subsamples of the $M_r<-20$ galaxy catalogues. 
In each panel, the {\em points with error bars} represent the VPF in our fiducial mock galaxy 
catalogue exhibiting galaxy assembly bias. The {\em dashed, red} lines give the VPF 
predicted by our mock galaxy catalogues with assembly bias erased. The {\em dot-dashed, blue} lines 
represent the VPFs predicted by the HOD that best fits the two-point clustering of the 
fiducial catalogue with assembly bias and, finally, the {\em dotted, orange} lines pertain 
to HOD models fit to the mock galaxy catalogues in which assembly bias has been erased. 
}
\label{fig:vpf}
\end{figure}

Figure~\ref{fig:vpf} shows the VPFs for four mock galaxy samples with $M_r<-20$.
The top panel of Fig.~\ref{fig:vpf} shows results for our luminosity-only mock galaxy catalogues; VPFs in 
the color-selected samples appear in the lower panels. 
Four curves appear in each panel. The {\em solid black} curve pertains to our fiducial model with 
assembly bias, the {\em dashed red} curve to the mock catalogue in which we have 
erased the assembly bias, but preserved the HOD. To make the remaining two curves, 
we have populated host halos in the Bolshoi simulation with an HOD using the parameters of 
the best-fit models determined in \S~\ref{sec:results}. The {\em blue dot-dashed} curves correspond to 
HODs fit to the clustering in our fiducial mock galaxy catalogues, the {\em orange dotted} 
curves to HOD fits to our mock catalogues without assembly bias. Jackknife-estimated error 
bars appear for the VPFs in our fiducial models.

In all cases, there is good agreement between the VPFs measured directly from the mock galaxy catalogues 
without assembly bias ({\em red dashed} curves) and the VPFs predicted by the HODs inferred from the clustering 
in these mock catalogues ({\em orange dotted} curves). This is a reassuring result. A significant discrepancy 
between these VPFs would only be due to an inadequacy of the halo model that is {\em not} related to assembly 
bias because neither of these predictions includes the assembly bias effect. Therefore, this agreement serves as 
an important validation exercise for halo model predictions of void statistics in the absence of assembly bias.

Next, consider the VPFs for the luminosity threshold samples in the top panel of Fig.~\ref{fig:vpf}. 
The catalogues with assembly bias and with assembly bias erased (the {\em black solid} and 
{\em red dashed} curves, respectively) predict VPFs that are consistent with each other given 
statistical uncertainties. This comparison isolates the impact of assembly bias in these catalogues 
on the VPF and the similarity of the two predictions suggests that assembly bias of the strength and 
character predicted by abundance matching would require volumes significantly larger than that of the 
Bolshoi simulation to detected. The impact of assembly bias on the VPFs is relatively minor.

Moving on, consider comparing the VPFs in the luminosity threshold samples to the VPFs predicted from 
the HODs that best-fit the clustering of these samples. The VPF predicted by the mock catalogue with 
abundance matching ({\em solid, black} curve) is in excellent agreement with the VPF predicted by the 
HOD that best fits the two-point clustering in this sample ({\em blue, dot-dashed} curve). This excellent 
agreement leads to a somewhat unsettling conclusion, namely, that good agreement between the 
VPF measured in data and the VPF predicted by an HOD model that has been fit to the clustering of the same 
sample {\em cannot} be used to rule out the presence of significant assembly bias in the underlying galaxy 
sample.

Consider the middle and bottom panels of Figure~\ref{fig:vpf}, in which we study the 
VPF in color-selected subsamples of our $M_r<-20$ age matching mock catalogue. The line types 
are analogous to those in the top panel that we have discussed in detail in the preceding paragraphs. 
Again, comparing the {\em red, dashed} and {\em solid, black} curves yields the true effect of 
assembly bias on the VPF. For the red galaxy subsamples, the impact of assembly bias on the VPF is 
substantial, and the sense of the effect is easy to understand. Age matching preferentially places red galaxies 
into denser environments, rendering large regions devoid of red galaxies comparably common. Erasing the assembly 
bias from the catalogues mixes some of these red galaxies into regions of lower density and makes large voids relatively 
less abundant. The converse is true for blue samples, although in this case we see that the difference 
this produces in the VPF is negligible.

Finally, compare the VPFs predicted by the HODs fit to the catalogues with assembly bias ({\em blue, dot-dashed} curves)
and the VPFs measured directly from the catalogues with assembly bias ({\em black, solid} curves) for our color-selected 
samples. For red galaxies, the difference is statistically insignificant. Evidently, the systematic shift in 
the HOD parameters away from their true values has counter-balanced the effect of assembly bias on the VPF, 
so that the assembly bias is nearly entirely disguised. This provides another explicit example 
of a standard HOD model which accurately describes galaxy clustering as well as the VPF, 
but which has parameters that are significantly biased. However, for the blue subsamples, 
this is not true. The VPF predicted by the HOD that best fits the clustering of the catalogue with 
assembly bias differs significantly from the VPF measured directly from the catalogue with assembly bias. 
Recall from \S~\ref{subsubsec:brightfits} that the 
HOD model that best-fit the clustering of blue galaxies was grossly incorrect 
(see the middle left panel of Fig.~\ref{fig:hod20}). This biased 
HOD results in a significantly incorrect VPF, so that in this case, the VPF {\em does} 
provide some indication that the true HOD has {\em not} been correctly recovered.

We conclude that the assembly bias predicted by age matching may be strong enough to 
be detectable in VPF measurements. However, the VPF signal is neither a smoking gun signature 
of assembly bias nor is it necessarily an effective cross check for assembly bias. Indeed, 
with Figure~\ref{fig:vpf} we have given explicit examples in which assembly bias can be 
quite strong and yet go entirely undetected in an HOD analysis of galaxy clustering 
and void statistics. This is unfortunate, as it implies that the VPF alone cannot be relied upon 
to uncover reasonable levels of galaxy assembly bias in the data. 

Our conclusion comes with the caveat that the VPF 
may be measured to higher precision in both existing and forthcoming dotted
survey data and so, may be somewhat more discriminating than our results suggest. The errors 
are likely only moderately smaller for SDSS DR 7, such as those discussed above, because the 
effective volumes of the samples are only moderately larger than the volume of the Bolshoi 
simulation \citep[][see the discussion toward the end of \S~\ref{subsec:hodparams}]{zehavi11}. 
As such, our conclusions are likely to be relevant to the vast majority of extant data analyses. 
However, spectroscopic surveys of comparably bright galaxies over much larger volumes could result in statistical 
errors significantly smaller than those in Fig.~\ref{fig:vpf}. Conspicuous examples would be the 
baryon oscillation spectroscopic survey (BOSS)\footnote{URL {\tt http://www.sdss3.org/surveys/boss.php}}, 
the bigBOSS extension\footnote{URL {\tt https://bigboss.lbl.gov/}}, and the survey to be carried out by the 
Dark Energy Spectroscopic Instrument (DESI)\footnote{URL {\tt http://desi.lbl.gov/}}.

%%%%%%%%%%%%%%%%%%%%%%%%% FIGURE %%%%%%%%%%%%%%%%%%%%%%%%%%%%%
%%%%%%%%%%%%%%%%%%%%%%%%%%%%%%%%%%%%%%%%%%%%%%%%%%%%%%%%%%%

%%%%%%%%%%%%%%%%%%%%%%%%%%%%%%
\subsection{The Quenching of Satellite Galaxies}
\label{subsec:quenching}

Modeling the galaxy distribution with the HOD is useful beyond the interpretation of observed statistics 
such as the two-point function. Once a set of parameters has been constrained by fitting to some set of observations, 
the HOD makes definite predictions for the imprint galaxy evolution physics leaves on halo occupation. 

Consider the recent example of \citet{tinker_etal13}. After fitting COSMOS observations with a halo model,
the authors demonstrate how their best-fit HOD parameters can be used to make 
numerous predictions about galaxy evolution, such as the rate at which central galaxies 
migrate to the red sequence, and the evolution of the characteristic timescale of satellite quenching. 
In principle, predictions such as these can directly inform our understanding of galaxy 
evolution, as well as our modeling of the physical processes that govern star formation and quenching. 
In this section, we study another example of such a prediction: the host halo mass-dependence 
of satellite galaxy quenching. In particular, we assess the degree to which assembly bias may threaten the 
program to use HOD fits to projected galaxy clustering to study the impact of the parent halo on star 
formation in satellite galaxies.

To quantify satellite quenching we use the {\em quenched fraction} of satellites, 
defined as the fraction of satellite galaxies that are red, $\fq = N^{\mathrm{red}}_{\mathrm{sat}} / \Nsat.$ 
Although dust obscuration complicates the use of $\fq$ to quantify star formation activity 
\citep[see, e.g.,][]{wetzel_etal11}, this statistic is used widely throughout the literature for this purpose 
\citep{vdBosch08,kovac_etal13,tinker_etal13}. Moreover, $\fq(\Mhost)$ can be readily computed from our 
HOD fits to color-selected samples, permitting a direct comparison between the halo model prediction 
and the true quenching fraction in the mock catalogues.

\begin{figure}
\includegraphics[width=8cm]{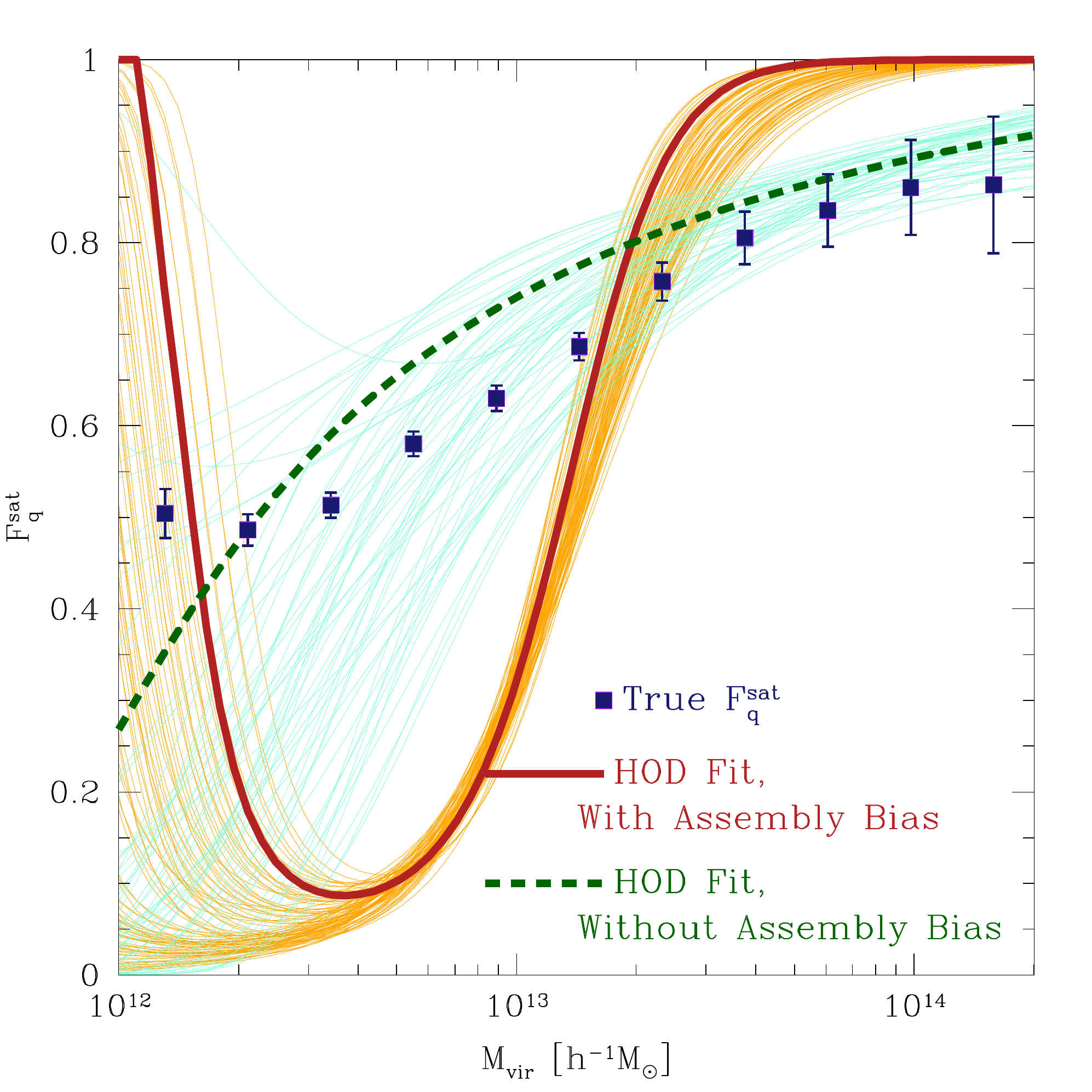}
\caption{
Host halo mass dependence of satellite galaxy quenching efficiency. 
The quantity $\fq$ is defined as the fraction of satellite galaxies that are red, 
shown here as a function of the virial mass of the host halo $\Mvir.$ {\em Points with error bars} 
show satellite quenching in our age-matching catalogues with $M_r<-20$. 
The {\em solid} curve shows the satellite quenching prediction that would be made using 
the best-fit HOD to the clustering of red and blue galaxies in our 
fiducial age-matching mock galaxy populations with assembly bias. 
The {\em dashed} curve shows $\fq$ predicted by the best-fit HOD of the 
erased-assembly-bias counterpart of the fiducial model. 
In each case, the {\em thin} lines in the background represent the 
100 randomly-selected samples from HODs with 
$\Delta \chi^2 < 1$ relative to the best-fit, 
analogous to the samples shown in Figures~\ref{fig:hod19} through \ref{fig:hod21}.
}
\label{fig:quenching}
\end{figure}

Figure~\ref{fig:quenching} shows the predictions for satellite quenching made by fits to our $M_r<-20$ samples. 
The {\em points with error bars} show $\fq$ predicted by our mock galaxy catalogues. 
Note that the catalogues with and without 
assembly bias have {\em identical} satellite populations on average 
(they have exactly the same HODs by construction), 
so both of these catalogues make {\em identical} predictions for $\fq$. 
The {\em thick, solid, red} curve in Fig.~\ref{fig:quenching} shows the 
$\fq$ predictions from the HODs fit to the clustering of our 
fiducial mock galaxy catalogues with assembly bias, while the 
{\em thick, dashed, green} curve shows the $\fq$ predicted by the best-fit 
HODs to our mock galaxy catalogues with assembly bias erased. In each 
case, the {\em thinner} lines show $\fq$ values predicted by the 100 
randomly-selected HODs with $\Delta \chi^2 < 1$ in the fits to clustering 
shown in Figure 4 through Figure 6.

Comparing the HOD predictions for the fiducial catalogues with assembly bias to the 
actual quenching fractions in Fig.~\ref{fig:quenching} 
illustrates clearly the following point: {\em if} there is significant galaxy assembly 
bias in the real universe that is neglected in HOD 
fits to galaxy clustering, then the conclusions that can be drawn about satellite 
quenching from such fits may be wildly incorrect. The sense of the 
discrepancy at high host halo masses follows directly from the HOD fits of 
\S~\ref{sec:results}. The catalogues with assembly bias 
predict stronger red galaxy clustering and weaker blue galaxy clustering on large scales. 
Consequently, HOD fits to those samples tend to prefer parameters that place 
a greater number of red satellites in strongly clustered, high-mass halos and 
few blue satellite galaxies in massive halos. The large quenching fractions in the 
assembly bias fits at low masses ($M_{\mathrm{vir}} \lesssim 3 \times 10^{12}\, h^{-1}M_{\odot}$) 
occurs at values of $M_{\mathrm{vir}}$ for which $\langle N_{\mathrm{sat}}\rangle \ll 1$, and so 
is less relevant to measurements of quenching fractions in groups and clusters and its cause 
is significantly more subtle. In short, the slight increase in small-scale, one-halo clustering in the 
blue subsamples with assembly bias (Fig.~\ref{fig:wp19}) drives a preference for HODs that 
place satellites in relatively rarer, higher-mass halos, contrary to the demands of the large-scale clustering \citep[for a 
more complete discussion of the manner in which HOD parameters affect small- and large-scale 
clustering, see][]{watson_powerlaw11}. This results in a relative deficit of blue satellites in low-mass 
host halos and an over-estimate of the quenching fraction at low host halo masses.

The quenching fraction discrepancy for the galaxy sample with assembly bias 
in Fig.~\ref{fig:quenching} is a particularly dramatic consequence of 
the systematic uncertainty in HOD fits to clustering. Indeed, in the 
case of the $M_r<-20$, the actual offset in quenching fraction realized in any study 
would likely not be as dramatic as shown in Fig.~\ref{fig:quenching}. 
Recall that the HOD fit to the assembly biased mock galaxy catalogues 
in this case give a blue galaxy HOD that is clearly incorrect (Fig.~\ref{fig:hod20}). 
An analyst confronted with this data would likely introduce additional data 
and/or a prior in order to bring the blue satellite description into closer 
agreement with expectations. However, in this case, the prediction of the 
quenching fraction would rely entirely on the reliability of the extra 
data to constrain the satellite population and/or the prior being 
truly informative.

In contrast, the {\em dashed} curves in Fig.~\ref{fig:quenching} give 
comparably good descriptions of the true, underlying quenching 
fractions, suggesting that HOD fits to clustering may perform fairly well 
in predicting similar quantities in the absence of galaxy 
assembly bias. The exception to this is at low values of host mass 
$\Mvir \lesssim 10^{13}\, h^{-1}\mathrm{M}_{\odot}$, particularly 
for the brighter sample, where the expected number of satellite galaxies is 
also small ($\langle N_{\mathrm{sat}} \rangle \ll 1$, see Fig.~\ref{fig:hod20} and Fig.~\ref{fig:hod21}). 
The residual offsets between the mock galaxy data points and the 
best-fitting HOD models in this case provide an estimate of the systematic 
errors on quenching fractions induced by using HOD/CLF-based methods as 
currently implemented. The influence of halo environment on star 
formation activity is of central interest in the physical interpretation of 
astronomical observations of the galaxy distribution. Our results should therefore 
provide strong motivation for a comprehensive effort to model and constrain 
the color-dependence of galaxy assembly bias.

%%%%%%%%%%%%%%%%%%%%%%%%% FIGURE %%%%%%%%%%%%%%%%%%%%%%%%%%%%%
%%%%%%%%%%%%%%%%%%%%%%%%%%%%%%%%%%%%%%%%%%%%%%%%%%%%%%%%%%%

%%%%%%%%%%%%%%%%%%%%%%%%%%%%%% DISCUSSION %%%%%%%%%%%%%%%%%%%%%%%%%%%%%%

\section{DISCUSSION}
\label{sec:discussion}

%%%%%%%%%%%%%%%%%%%%%%%%%%%%%%%%%%%%%%%%%%%%%%%%%%%%%%%%%%%%%%%%%%%%%

Assembly bias, the potential for galaxies to preferentially reside in halos in a manner that 
is correlated with halo clustering at fixed halo mass, has received significant attention in the recent 
literature both for its potential to impede our ability to draw reliable inferences  
from standard HOD/CLF-based statistical studies of galaxy clustering and, quite to the contrary, 
for its potential to be used as a signal from which new inferences about galaxy formation and evolution may be drawn. 
Nonetheless, it remains unclear (1) whether or not assembly bias at detectable levels is present in the actual, 
observed galaxy population, and (2) the degree to which reasonable levels of assembly bias in data may affect the inferences 
of statistical models of the galaxy-dark matter connection that neglect assembly bias. 
This is a significant omission because numerous studies infer the statistical relationships 
between galaxies and the halos in which they live using standard HOD/CLF-based methods that neglect 
assembly bias effects \citep[e.g.][]{zehavi05a,yang_etal05,zheng_etal07,vdBosch07,zheng09,simon_etal09,
abbas_etal10,watson_etal10,matsuoka_etal11,miyaji_etal11,leauthaud_etal11,leauthaud_etal12,
tinker_etal12,geach_etal12,kayo_oguri12,vdBosch13,tinker_etal13,parejko_etal13,cacciato_etal13}. 
The aim of this paper is to improve upon this situation by studying 
mock catalogues of galaxies that have been constructed to 
be broadly representative of the observed galaxy population, yet at the same time 
exhibit significant levels of assembly bias.

We chose to use halo abundance matching and age matching to build our mock galaxy catalogues to 
exhibit assembly bias. In \S~\ref{sec:assembias}, we discussed why modern abundance matching 
techniques all include assembly bias, and we demonstrated that this assembly bias is, indeed, 
significant by showing that the projected two-point correlation functions predicted using abundance 
matching differs markedly from the two-point clustering predicted from mock galaxy 
populations with identical HODs, but no assembly bias. By itself this is an interesting  point because 
abundance matching and age matching yield galaxy populations with clustering statistics that are 
broadly similar to those observed in large-scale galaxy surveys 
\citep[e.g.][]{kravtsov04a, vale_ostriker04,tasitsiomi_etal04,vale_ostriker06,conroy_wechsler09,guo10,simha10,neistein11a,watson_etal12b,rod_puebla12,kravtsov13,hearin_watson13a,hearin_etal13}. 
This suggests that the assembly bias predicted 
by abundance matching may not be wildly different from 
what is observationally permissible.

Assembly bias has been shown to be an important effect in other theoretical 
contexts as well. Of particular relevance to the present paper is the work 
of \citet{croton_etal07}. These authors studied the relative effect of 
assembly bias on the three-dimensional correlation functions of galaxies 
in semi-analytic models of galaxy formation. They found that 
halo properties other than mass do influence the properties of galaxies 
in their models and that these effects lead to significantly altered clustering 
that is of roughly the same size as the assembly bias effect in 
abundance matching and age matching. This bolsters our case for using 
abundance matching to construct simple galaxy catalogues exhibiting assembly 
bias. Interestingly, \citet{croton_etal07} also found that assembly bias 
could {\em not} be attributed only to concentration- or formation time-dependent 
halo clustering, suggesting that the relationship between galaxies and their 
halos may be complicated enough to make empirical modeling with 
sufficient precision to address extant data challenging. In abundance matching 
and age matching, the effects of assembly bias arise solely due to the 
concentrations and formation times of halos, and this is one of the great 
benefits of using abundance matching as a sandbox to study assembly bias, but 
we cannot address these more complex realizations of assembly bias directly.

We built upon our demonstration of assembly bias in 
abundance matching estimating the degree to which assembly bias can represent a 
systematic error on the probabilities with which galaxies reside in halos of particular 
masses, the halo occupation distribution (HOD). In this first paper on the subject, we 
have chosen to limit the scope of our study by fixing cosmological parameters to the 
known, true, underlying cosmology of the simulation that we have used to construct 
our mock galaxy catalogues and studying systematic errors in HODs only. We will return 
to cosmological constraints in the presence of assembly bias and study additional 
observables in a follow-up paper.

Section \S~\ref{sec:results} details our results. In short, we find that neglecting 
assembly bias at the levels present in our mock catalogues leads to significant 
systematic errors in inferred halo occupation distributions (Figures~\ref{fig:hod19} to \ref{fig:alphas}) 
in nearly every case we have studied. For mock galaxy samples selected purely on a luminosity 
threshod, these systematic errors are of a modest absolute size. Systematic 
errors in $M_{\mathrm{min}}$ and $M_1$ are often $\lesssim 0.2\, \mathrm{dex}$, while 
the offsets in $\alpha$ are $\lesssim 0.2$. Nevertheless, these systematic errors are 
significant compared to the statistical errors on the inferred HOD parameters in most cases. 
Systematic errors in inferred HOD parameters are significantly more severe for 
color-selected subsamples of galaxies. These results indicate that traditional HOD 
analyses of galaxy clustering may be entirely blind to the systematic errors caused 
by assembly bias.

Motivated by this, in \S~\ref{subsec:vpf} we investigated the potential 
to detect assembly bias using the void probability function (VPF), 
an example of an auxiliary statistic that has been studied previously 
for precisely this purpose \citep{tinker06,tinker_etal08b}. Using our abundance matching 
and age matching mock galaxy catalogues, we constructed explicit examples in which strong levels of assembly 
bias leave no statistically significant imprint on the VPF, and/or would not be evident in standard HOD analyses 
of void statistics. This casts doubt that consistency between the observed and HOD-predicted VPF 
can be interpreted as ruling out assembly bias as a potential systematic \citep{tinker_etal08b}. 
The only case in which this test identifies a problem with the HOD inferred from clustering, 
is the blue, $M_r < -20$ sub-sample. This case represents a dramatic failure to infer the 
correct HOD (see Fig.~\ref{fig:hod20}) and would also easily be ruled out by a number of 
other observables, such as group conditional mass functions. Again, this suggests 
that the VPF is not an especially incisive tool for identifying the effects of 
reasonable levels of assembly bias in inferred HODs.

Although it is not commonly discussed in the context of assembly bias, 
we point out that the phenomenon of galactic conformity is squarely at odds with the notion 
that halo mass alone determines galaxy properties. Galactic conformity refers to 
the observed tendency for red central galaxies to host a redder satellite population 
than blue central galaxies residing in halos of the same mass \citep{weinmann06b}. 
This manifestly violates the ``halo mass only'' assumption of the standard HOD. As discussed 
in \S~\ref{subsec:colorab}, small-scale clustering may be influenced by this 
phenomenon in a statistically significant way, though a more focused investigation of 
this point would be required before more conclusive statements could be made. We leave 
this as a subject for future work.

Our primary results conclude in \S~\ref{subsec:quenching} with a case study of 
the potential threat assembly bias poses to standard HOD studies of 
galaxy evolution.  We presented $F_{\mathrm{q}}^{\mathrm{sat}}(\Mhost),$ the halo 
mass-dependence of the quenched fraction of satellite galaxies, as an example of a quantity that could 
be significantly mis-estimated from a standard HOD fit that has been compromised 
by assembly bias (Fig.~\ref{fig:quenching}). The drastic consequences that unknown levels of assembly bias 
may have on the relatively simple statistic $F_{\mathrm{q}}^{\mathrm{sat}}$ is particularly interesting in 
light of recent analyses of COSMOS data \citep{tinker_etal13}, in which standard HOD techniques are used 
to draw conclusions about complex characteristics of the galaxy distribution such as
 the characteristic quenching timescale of satellite galaxies, or the migration rate of centrals to the red sequence. 
 Of course, our analysis methods 
are not directly analogous to COSMOS analysis in \citet{tinker_etal13}: we have studied a different 
formulation of the HOD from theirs, and we have focused exclusively on galaxy clustering measurements, 
whereas they have included galaxy-galaxy lensing data (see below). Nonetheless, 
\citet{tinker_etal13} have demonstrated the potential of the HOD to provide rich information about 
the history of star formation in galaxies, and so the results in \S~\ref{subsec:quenching} provide 
strong motivation to constrain the true level of assembly bias in the data.

It has become increasingly common to fit simultaneously for statistics in addition 
to two-point galaxy clustering in halo model analyses. For example, many different 
approaches to galaxy-halo modeling have been brought to bear on galaxy-galaxy lensing measurements 
\citep[e.g.,][]{leauthaud11a,tinker_etal13,cacciato_etal13,yoo_seljak12,hearin_etal13}. 
The mass-to-number ratio of clusters may also provide additional constraining power on 
both halo model and cosmological parameters \citep{tinker_etal12,reddick_etal13}. One hopes 
that the independent information provided by additional statistics such as these would break 
the degeneracies evident in \S~\ref{sec:results}. However, the results shown in \S~\ref{subsec:vpf} 
illustrate that even the relatively strong levels of assembly bias present in our mock catalogues can go 
entirely undetected in alternative statistics, such as the VPF, that naively seem well suited 
to this purpose. While we have limited the scope of the present paper to projected 
two-point clustering only, it will be interesting to extend this analysis to 
itemize the ways in which additional statistics may mitigate systematic 
errors induced by assembly bias and we will pursue this avenue in a follow-up paper.

We stress that we have attempted to be conservative in our quantification of the potential 
systematic error induced by unknown levels of assembly bias. In particular, we have marginalized over 
the parameter $f_{\mathrm{b}}$, a nuisance parameter introduced in \citet{tinker_etal12} and designed to account, 
in part, for imperfect calibration of halo bias. This enables some of the large-scale clustering offset between 
mock galaxy samples with and without assembly bias to be absorbed into $f_{\mathrm{b}}$ and, indeed, this is 
reflected in the inferred values of $f_{\mathrm{b}}$ shown in Fig.~\ref{fig:hod19} through Fig.~\ref{fig:hod21}. 
However, we emphasize that this additional parametric freedom is particularly ineffective at mitigating against 
assembly bias in the color-selected samples because assembly bias increases the clustering strength of 
red galaxies while decreasing the large-scale clustering strength of blue galaxies and $f_{\mathrm{b}}$ 
cannot accommodate such countervailing demands. Moreover, the fact that assembly bias causes significant 
systematic errors in the luminosity threshold sample HODs suggests that assembly bias causes a 
scale-dependent shift in the projected correlation function that cannot be accommodated by a simple shift in 
large-scale clustering. In Appendix~\ref{sec:appendix}, we give examples of how our results change 
when $f_{\mathrm{b}}$ is not marginalized over. In Appendix~\ref{sec:appendix}, we also show that 
our results are robust to including additional parametric freedom in the radial distributions of 
satellites. Further, our jackknife error estimates on the clustering in our mock catalogues are significantly 
larger at all radii, and for all samples, than the corresponding statistical errors in, for example, the 
clustering measured in analogous samples by the SDSS \citet{zehavi11}. Thus if assembly bias is present in 
the real universe and has comparable strength to that which is present in our mock galaxy catalogues, systematic errors 
even more severe than what we present here would be present in the HODs inferred from galaxy clustering.

In the absence of definitive studies that constrain assembly bias to negligible levels, it seems prudent 
to consider inferences drawn about halo occupation statistics from large-scale clustering 
to be subject to a systematic error that is large compared to its statistical error.
In order to mitigate the possibility that assembly bias can induce a 
systematic error in the inferred statistics of the galaxy distribution, 
it will be necessary to model assembly bias in parameterized forms 
and in significantly greater detail than has been attempted before. 
Explicit, theoretical modeling of {\em halo} assembly bias has been attempted 
before \citep{wechsler06}. Achieving the necessary precision 
will require a significant effort involving, in part, precise 
calibration of halo abundance and clustering as a function of halo 
properties other than mass.

In addition to a precision calibration of halo assembly bias, 
a rigorous theoretical formulation of {\em galaxy} assembly bias will be necessary so that 
analytical parameters quantifying the character and strength of this effect can included in 
likelihood analyses. Indeed, it may be possible to recast the HOD in terms of only a single, distinct halo 
property (or combination of halo properties) $y$, $\mathrm{P}(N \vert y)$ as described in \S~\ref{sec:assembias}, 
such that the HOD is a more faithful representation of the relationship between galaxies and their host halos. 
In this case, assembly bias is a manifestation of the fact that $y$, rather than mass, is the halo property 
that is most directly related to the galaxy population within a halo. Itemizing halo abundance, 
clustering, and structure as a function of $y$ would enable HOD modeling in terms of this new halo 
property and mitigate the systematic errors induced by assembly bias. In the mock galaxy samples that 
we studied in this paper, recasting the HOD and all other ingredients of the halo model in terms of 
$V_{\mathrm{max}}$ (rather than halo mass) would have been sufficient to describe the galaxy-halo 
relationship in our mock luminosity threshold samples.

This challenge can be viewed as an opportunity. With the wealth of extant 
and forthcoming data on galaxy clustering, galaxy-galaxy lensing, and any 
number of other statistics, it may now be possible to cultivate and 
constrain a significantly richer empirical relationship between 
galaxies and their dark matter halos. This may lead to models that can 
associate galaxies with halos based on a number of halo properties, further 
bridging the gap between the vast amount of existing observational data 
and direct numerical simulations of galaxy 
formation in a cosmological context. It is our hope that this 
study provides motivation to pursue these goals.

Lastly, we note that our study is subject to several noteworthy caveats. 
First, while we know of no definitive study that rules out 
significant assembly bias, it is possible that assembly bias is 
far less prevalent in the true galaxy distribution than it is in 
models based upon abundance matching. We have constructed explicit 
examples in which assembly bias is large, induces large systematic errors, and 
is not easily diagnosed, but we know of no reason that assembly bias must be 
as large as abundance/age matching predict. Indeed, abundance matching 
is known to be an inadequate description of observed galaxy clustering 
statistics in their detail \citep[e.g.][]{hearin_etal12b}. Moreover, 
recent results \citep{behroozi_etal13c} indicate that the 
property we used in our abundance matching, $\vpeak,$ sometimes occurs during 
a non-equilibrium phase of halo evolution, and so it may be implausible for 
$\vpeak$ to correlate with stellar mass to high precision. In addition, 
subhalo incompleteness may also pose a problem for detailed predictions 
of galaxy clustering \citep{wetzel_white10,watson_etal12b,guo_white13}, 
even in state-of-the-art simulations such 
as Bolshoi and Millennium \citep[although see also][]{klypin_etal13}.
We were forced into using such an incomplete model precisely because no model 
exists that reproduces all of the known properties of the observed 
galaxy distribution. Related to these points is the fact that our covariance 
matrices have been estimated from the same mock galaxy catalogues that we have used 
in our fits. Again, this strategy was necessary because there are few high-resolution 
simulations available that can be used to construct mock galaxy catalogues over a 
wide range of luminosities using abundance matching. 
Lastly, the specific HOD parameterizations and priors used in previous 
HOD analyses vary greatly from one study to the next. In the results 
presented in the main body of this paper, as well as in Appendix~\ref{sec:appendix}, 
we have not placed any priors on our HOD parameters. We have experimented 
with a variety of priors and alternative parameterizations, 
and while the inferred HODs are altered significantly by such 
choices (emphasizing the fact that priors should be informative), 
our qualitative conclusions are robust to these choices.

%%%%%%%%%%%%%%%%%%%%%%%%%%%%%% SUMMARY %%%%%%%%%%%%%%%%%%%%%%%%%%%%%%

\section{SUMMARY}
\label{sec:summary}

%%%%%%%%%%%%%%%%%%%%%%%%%%%%%%%%%%%%%%%%%%%%%%%%%%%%%%%%%%%%%%%%%%%%%

We conclude this paper with the following summary of our primary findings.

\ben
\item Galaxy assembly bias of considerable strength is a generic prediction of the widely used abundance matching 
technique for assigning galaxies to halos. The same is true of the recently-introduced age matching algorithm for 
assigning colors to mock galaxies in dark matter simulations. Both of these methods make predictions for the 
observed galaxy distribution that are in good agreement with a rich variety of SDSS measurements. 
\item It is possible to obtain an acceptable fit to galaxy clustering data with a traditional HOD model, 
even when the strength of assembly bias in the galaxy sample is significant.
\item Assembly bias of the kind predicted by abundance/age matching causes there to be a significant 
systematic error on the halo-galaxy connection inferred from fits to galaxy clustering. As a similar level 
of assembly bias has not yet been ruled out, the halo-galaxy connection (whether HOD, CLF, or otherwise) 
inferred from clustering data should be subject to an additional systematic error that is large compared to 
statistical errors.
\item The void probability function (VPF) may be useful in constraining the color-dependence of 
galaxy assembly bias, but we have constructed explicit examples in which the VPF cannot 
detect assembly bias even when the systematic errors in HOD parameters derived from galaxy clustering fits are large. 
\item Uncertainty in the true level of galaxy assembly bias can have a dramatic effect on HOD modeling 
of the star formation histories of satellite galaxies and may even dominate the error budget in these 
applications.
\item For future studies of the galaxy-halo connection, including (re)analyses of existing datasets, 
we recommend a comprehensive effort to model and constrain the true level of galaxy assembly bias, 
both for color-selected galaxy samples and samples selected purely on luminosity. To aid this effort, 
we make publicly available all of the mock catalogues used in this study; these mock catalogues were specifically 
designed to isolate the effects on the galaxy distribution that are purely due to assembly bias, 
and can be found at http://logrus.uchicago.edu/$\sim$aphearin. 
\een

%%%%%%%%%%%%%%%%%%%%%%%%%%%%%% ACKNOWLEDGEMENTS %%%%%%%%%%%%%%%%%%%%%%%%%%%%%%

\section{ACKNOWLEDGEMENTS}
\label{sec:thanks}

We thank Andreas Berlind, Shaun Cole, Hiram Coombs, Carlos Frenk, Jeff Newman, 
Risa Wechsler, Idit Zehavi, Zheng Zheng, Ramin Skibba, and particularly 
Doug Watson for useful discussions throughout various stages of this work. 
We thank David Weinberg and Simon White for insightful email exchanges 
regarding an earlier draft of this manuscript. 
We thank Kristin Riebe for helping us navigate the MultiDark database. 
We are particularly grateful to Jeremy Tinker and Rachel Reddick for comparing the results of their 
halo model code to our own and for helping us with numerous technical questions related to implementations 
of the halo model and the halo occupation distribution. We thank John Fahey for {\em America}. 
The work of ARZ is supported by the U. S. National Science Foundation 
through grant AST 1108802 and by the University of Pittsburgh. Significant portions of this work were completed 
during visits to The Institute for the Physics and Mathematics of the Universe (IPMU) at the 
University of Tokyo and we are thankful to IPMU and particularly Alexie Leathaud, 
Surhud More, and Rie Ujita for their hospitality. The work of ARZ was also 
supported by the National Science Foundation under grant PHYS-1066293 and the hospitality 
of the Aspen Center for Physics. APH is supported by the U.S. Department of Energy under contract 
No. DE-AC02-07CH11359.

%%%%%%%%%%%%%%%%%%%%%%%%%%%%%%%%%%%%%%%%%%%%%%%%%%%%%%%%%%%%%%%%%%%%%

\bibliography{abhod}

%-----------------------------------------------------------------------------------------------------------------------------------------------
% Appendix
%-----------------------------------------------------------------------------------------------------------------------------------------------
\appendix
\section{Additional Nuisance Parameters and HOD Recovery}
\label{sec:appendix}

In the main body of the text, we explored the fidelity of the HOD recovered by fitting the 
projected two-point functions of a variety of mock galaxy samples. In any such fit, a variety 
of choices must be made regarding the parameters that are allowed to vary in order to describe 
the clustering. In \S~\ref{sec:results}, we presented results in which we marginalized over the 
halo bias, in order to account for imperfect calibration of host halo clustering, and held the spatial 
distributions of satellite galaxies fixed. In this Appendix, we provide examples of how our results 
change in detail when we alter these assumptions. However, we emphasize that our results do not 
change qualitatively. In particular, that the inferred HODs are significantly different in fits to samples 
with and without assembly bias is a robust conclusion.

\begin{figure}
\begin{center}
\includegraphics[width=8.5cm]{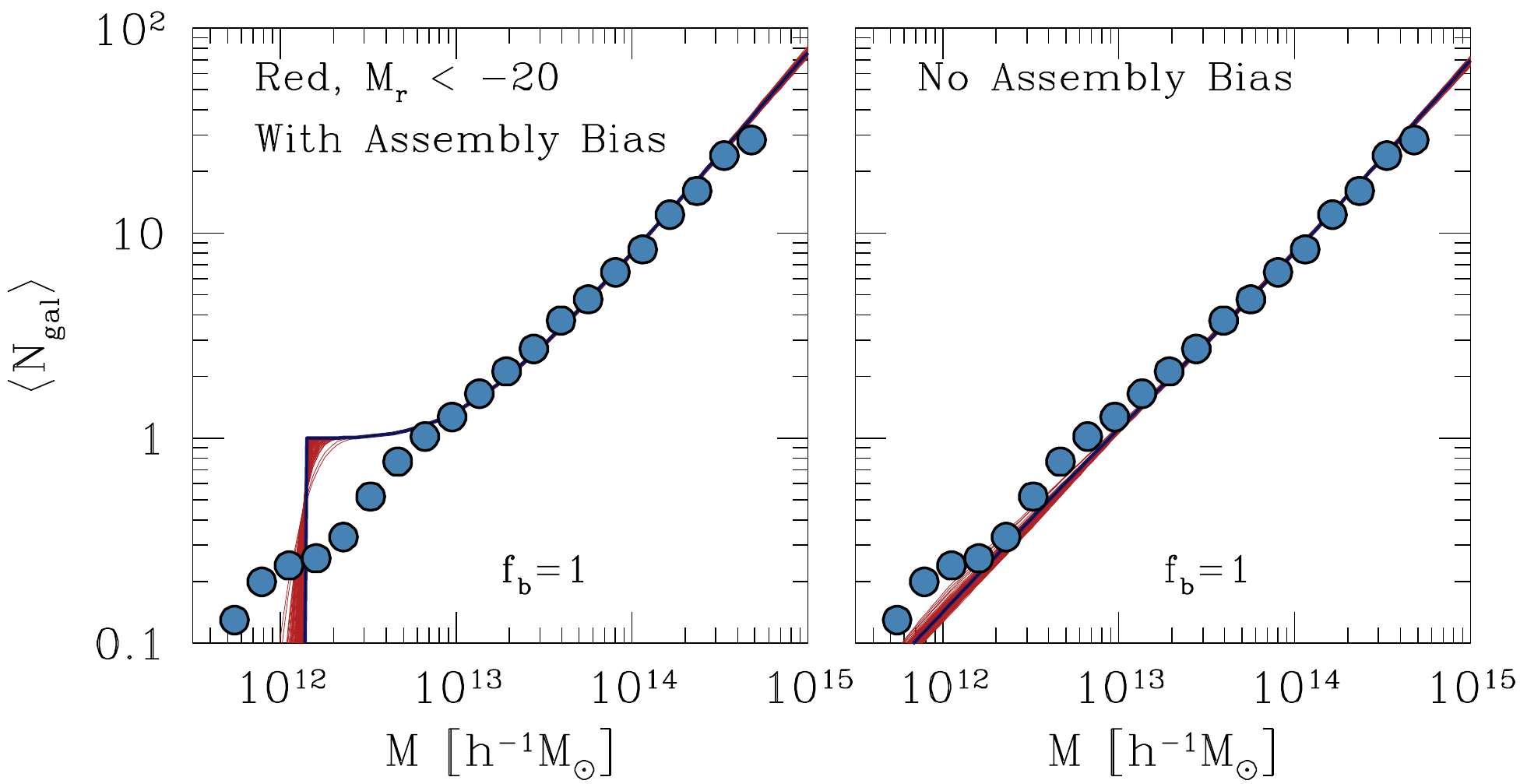}
\includegraphics[width=8.5cm]{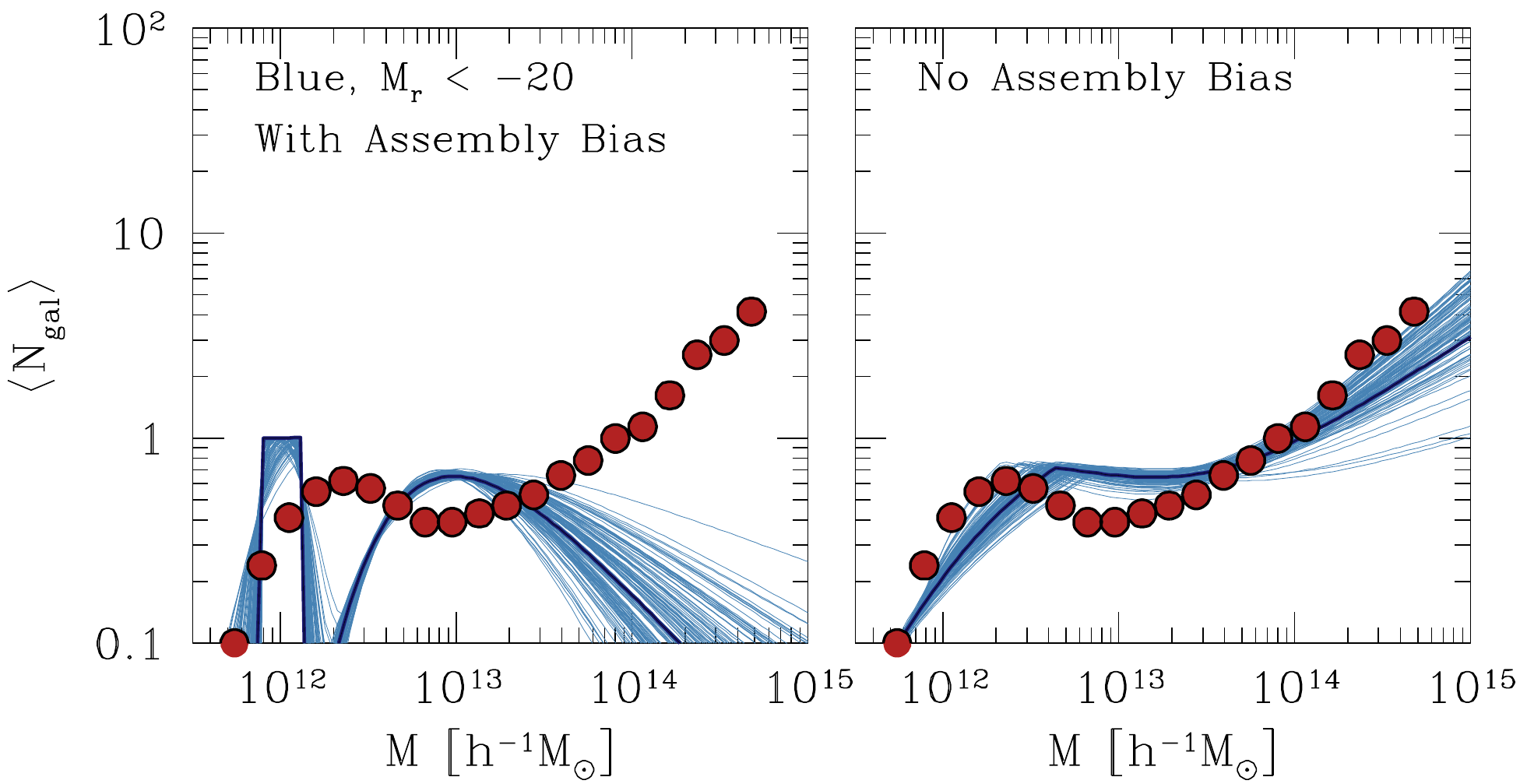}
\includegraphics[width=8.5cm]{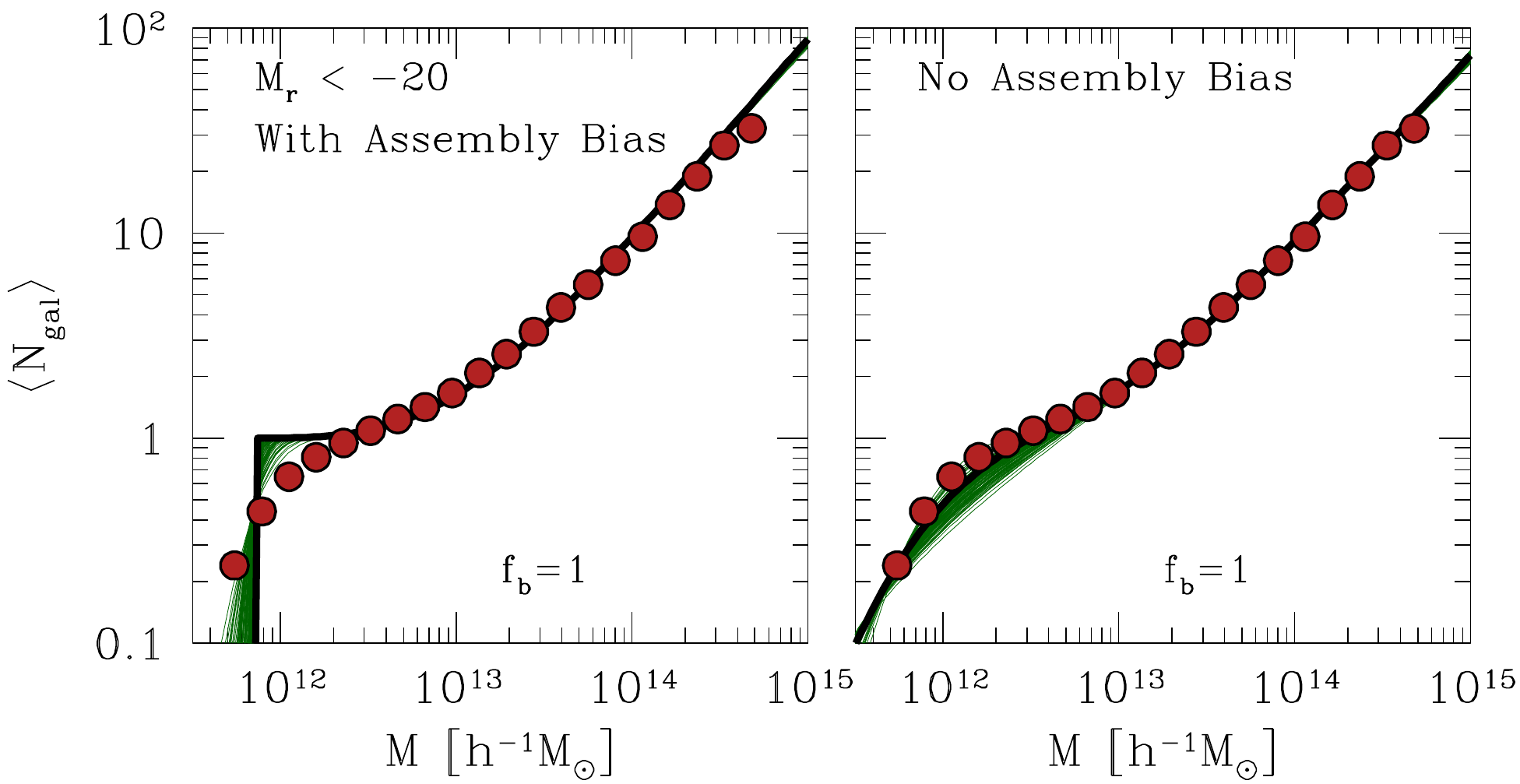}
\caption{
Comparison of the best-fit HODs for galaxies in the 
$M_r < -20$ sample with the true HOD in the simulation (points). 
This figure is the same as Fig.~\ref{fig:hod20} except that these 
fits were conducted with the halo bias parameter held fixed to $f_b=1$. 
}
\label{fig:hod20nofb}
\end{center}
\end{figure}

\begin{figure}
\includegraphics[width=7.5cm]{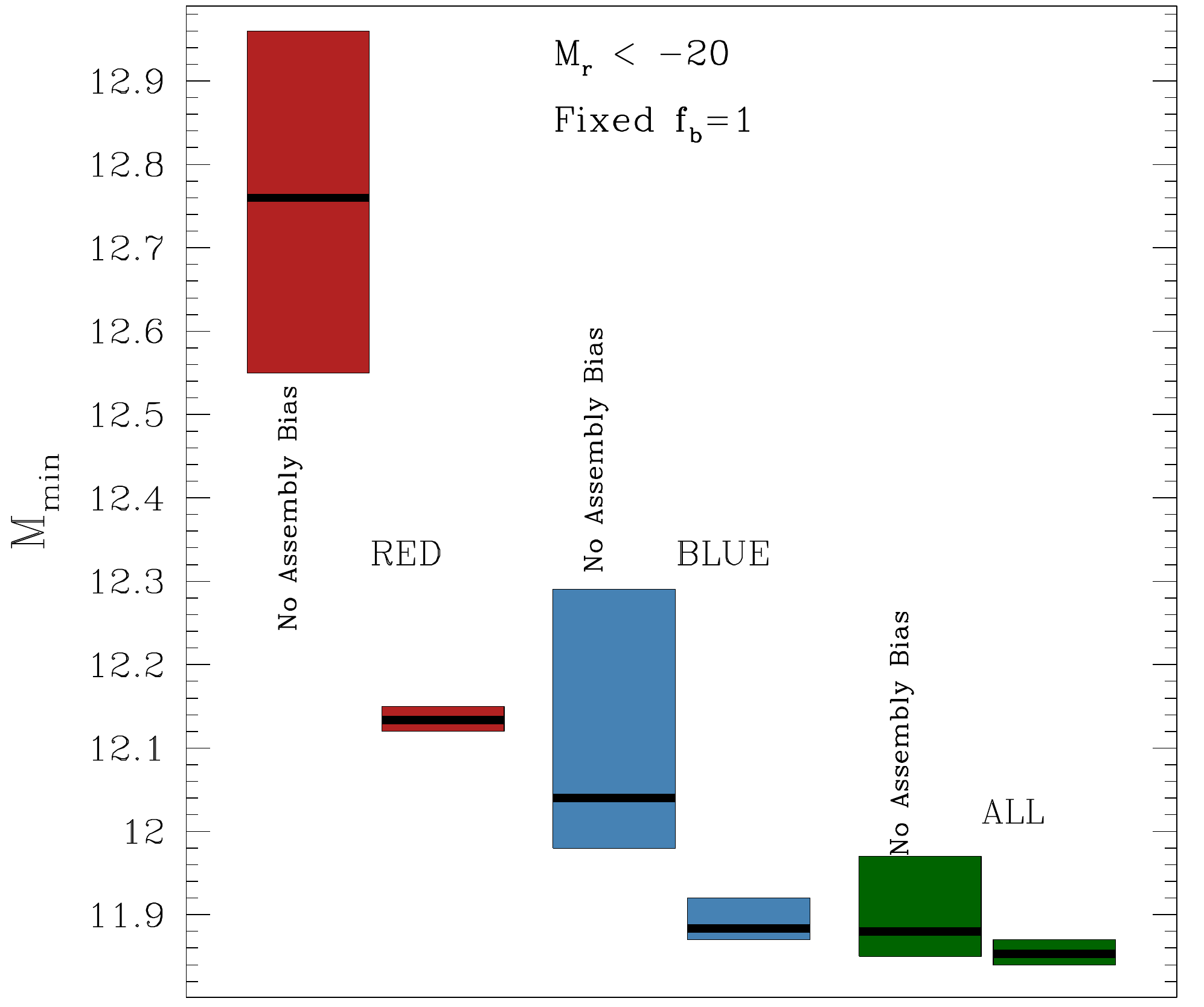}
\includegraphics[width=7.5cm]{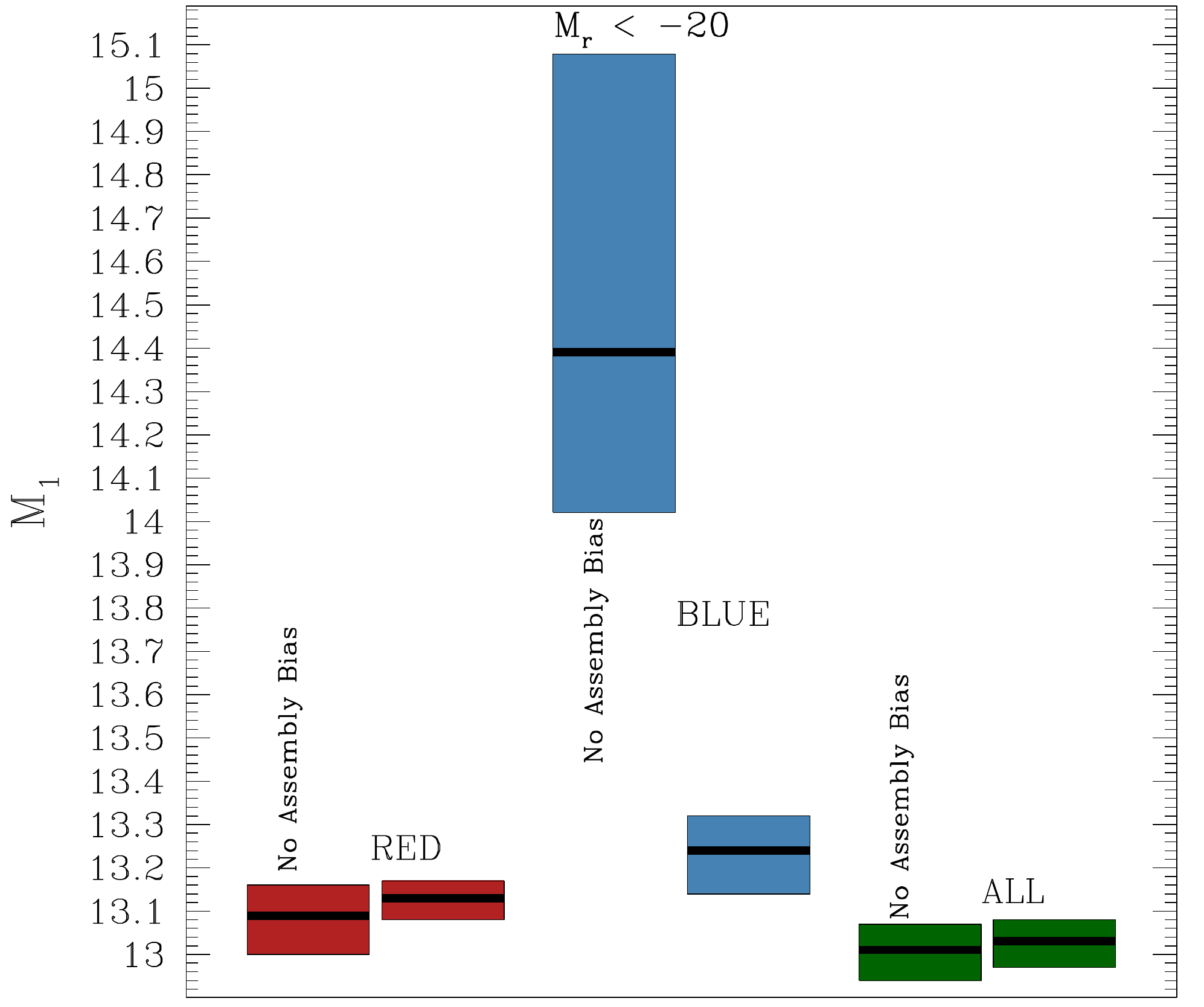}
\includegraphics[width=7.5cm]{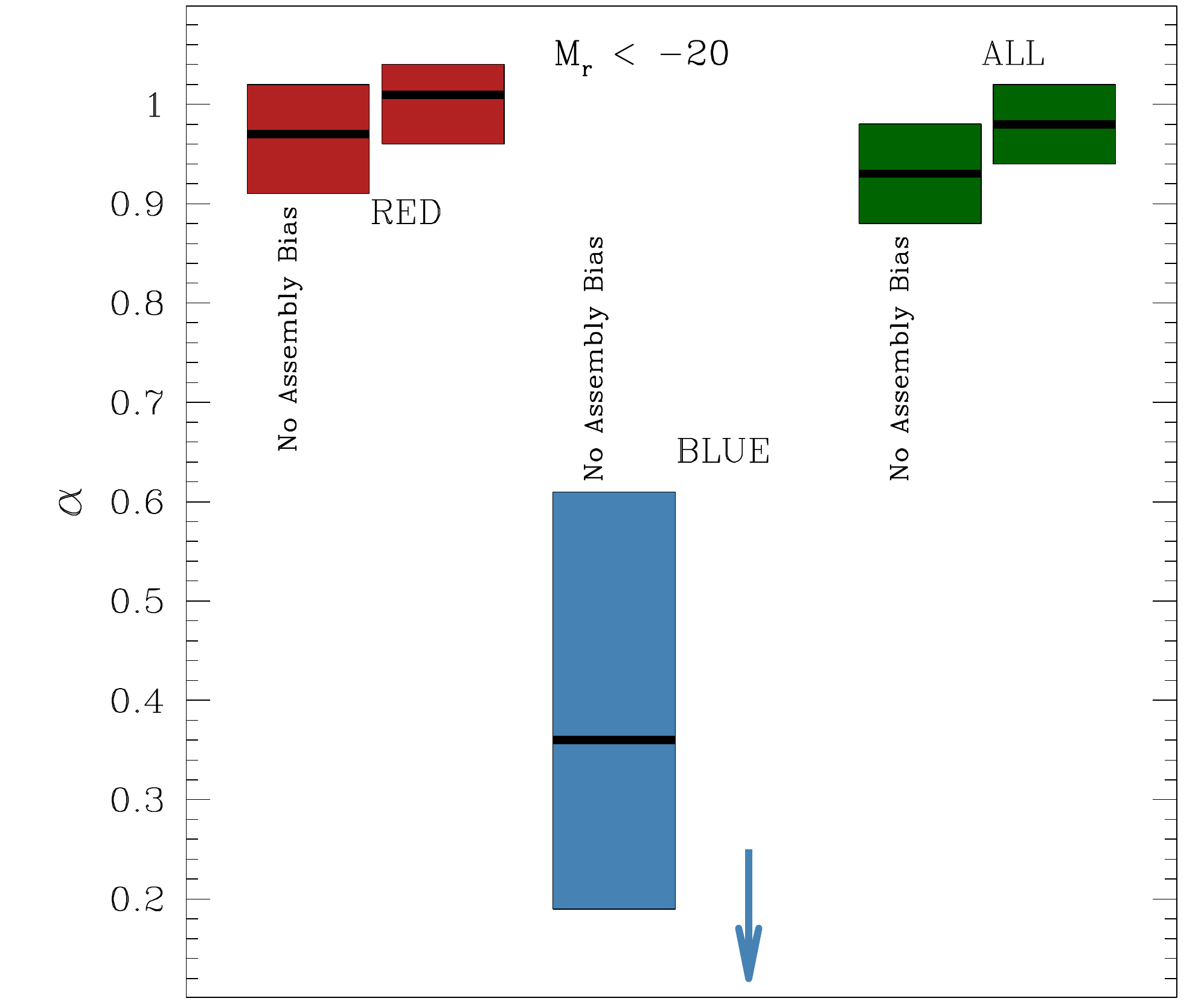}
\caption{
Constraints on HOD parameters inferred from fits with in which the bias nuisance parameter has 
been held fixed at $f_{\mathrm{b}}=1$, which corresponds to assuming perfect calibration of halo 
bias. The top panel shows the inferred constrains on $M_{\mathrm{min}}$, 
the middle panel shows inferred constraints on $M_1$, and the bottom panel 
shows inferred constraints on the power-law index of the satellite portion of the HOD, $\alpha$. 
}
\label{fig:con20l}
\end{figure}

As a first example, we show the effect of the marginalization over the halo bias nuisance 
parameter $f_{\mathrm{b}}$. In particular, 
Figure~\ref{fig:hod20nofb} shows the HODs inferred from fits to the projected two-point functions 
of the $M_r < -20$ samples with the halo bias parameter held fixed at $f_b=1$. This figure should 
be compared to Fig.~\ref{fig:hod20} in the main text in order to assess the influence of the bias 
parameter. Note several things about Fig.~\ref{fig:hod20nofb}. First, notice that the span of models 
that provide similarly acceptable fits to the galaxy clustering is narrower in this case, as should be 
expected because there is less parameter freedom. Second, notice that the HODs are recovered with 
similar fidelity in the fits to the samples with no assembly bias. Finally, compared to the 
fits to the galaxy samples without assembly bias, the fits to the samples with assembly bias exhibit 
the same systematic differences in the inferred HODs. This suggests that our qualitative conclusions 
are robust to relatively small uncertainties in the calibration of the halo bias and that the scale-dependence 
of the clustering is sufficiently different in the models with and without assembly bias as to drive 
significant differences in the inferred HODs. Figure~\ref{fig:con20l} depicts the marginalized HOD 
parameter constraints for the models with $M_r<-20$ and $f_{\mathrm{b}}=1$. These constraints 
exhibit the same fundamental trends as described in the main body of the this paper.

%%%%%%%%%%%%%%%%%%%%%%%%% FIGURE %%%%%%%%%%%%%%%%%%%%%%%%%%%%%

%%%%%%%%%%%%%%%%%%%%%%%%%%%%%%%%%%%%%%%%%%%%%%%%%%%%%%%%%%%%%%%%%%%%

%%%%%%%%%%%%%%%%%%%%%%%%%%%%%%%%%%%%%%%%%%%%%%%%%%%%%%%%%%%
%%%%%%%%%%%%%%%%%%%%%%%%% FIGURE %%%%%%%%%%%%%%%%%%%%%%%%%%%%%
%%%%%%%%%%%%%%%%%%%%%%%%%%%%%%%%%%%%%%%%%%%%%%%%%%%%%%%%%%%

We now move on to fits in which we have allowed the spatial distributions of the satellite galaxies to vary 
during the HOD fitting. This has been done in a number of recent publication in which small-scale galaxy 
clustering has been fit with similar models \citep[e.g.][]{tinker_etal12,vdBosch13,reddick_etal13}. There are  
at least two reasons for introducing additional parameter freedom to describe the spatial distributions of 
the satellite galaxies. Satellite galaxies may not necessarily follow the dark matter distribution and, 
indeed, neither satellite halos nor satellite galaxies trace the overall dark matter distributions of their host 
halos in simulations \citep[e.g.,][]{zentner03,zentner05,nagai_kravtsov05}. Moreover, satellite halos may 
be distributed in a triaxial configuration about their host halos and there is some hope that introducing a 
satellite distribution nuisance parameter can account for this effect without modeling triaxiality directly 
(though this remains to be demonstrated explicitly). We follow the recent literature and 
introduce an additional parameter defined to be the ratio 
of the NFW concentration assumed for the spatial distribution of satellite galaxies to the NFW concentration 
of the dark matter, $f_{\mathrm{conc}}=c_{\mathrm{sats}}/c_{\mathrm{dm}}$, where $c_{\mathrm{dm}}$ is 
the standard dark matter concentration from the Bolshoi simulation \citep{bolshoi_11} and $c_{\mathrm{sats}}$ 
is the concentration parameter used to describe the average radial distribution of satellite galaxies. In our 
standard fits described in \S~\ref{sec:results}, we held this parameter fixed to $f_{\mathrm{conc}}=0.6$. In 
this section, we allow $f_{\mathrm{conc}}$ to vary while holding $f_{\mathrm{b}}=1$ for simplicity.

%%%%%%%%%%%%%%%%%%%%%%%%% FIGURE %%%%%%%%%%%%%%%%%%%%%%%%%%%%%

\begin{figure}
\begin{center}
\includegraphics[width=8.5cm]{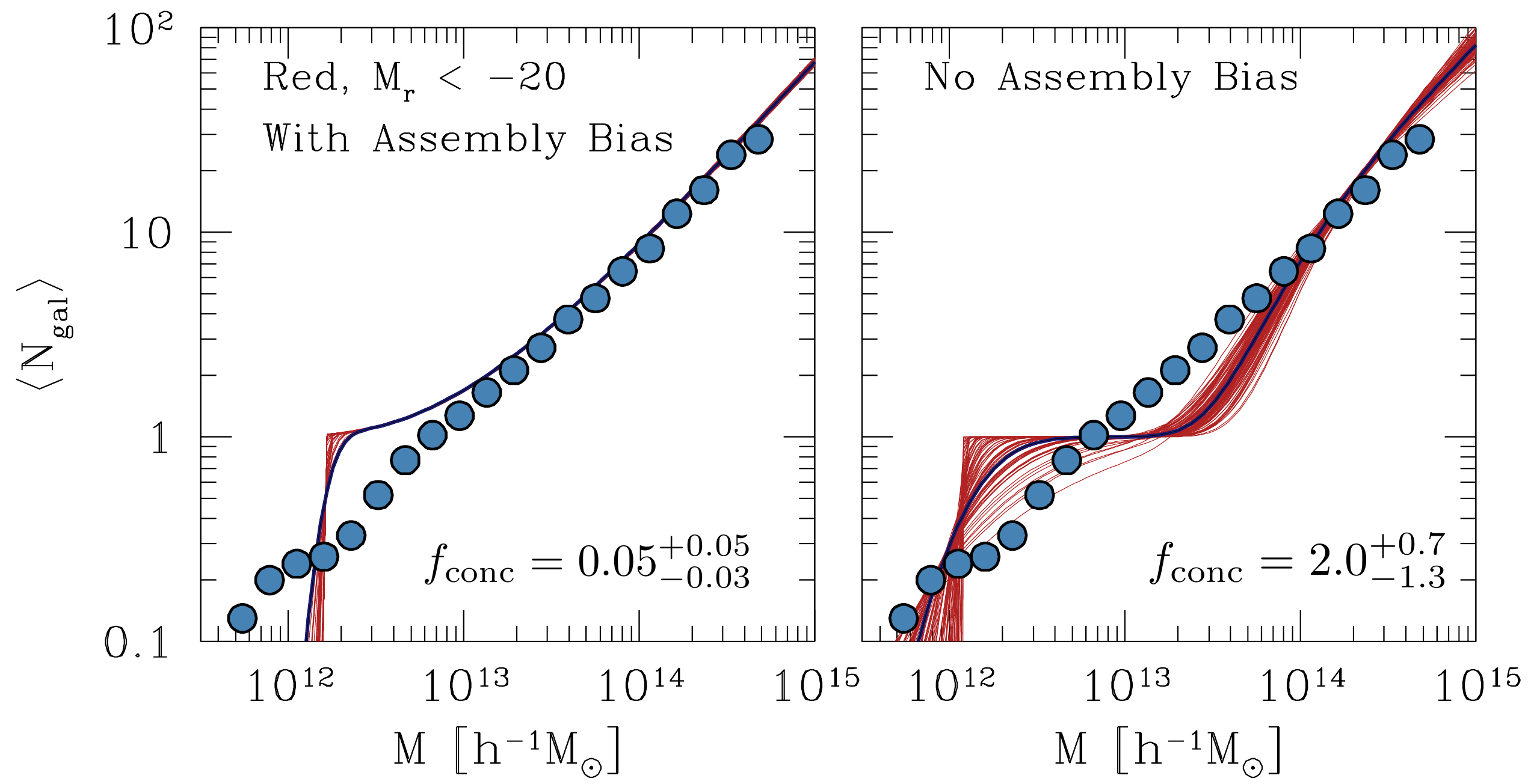}
\includegraphics[width=8.5cm]{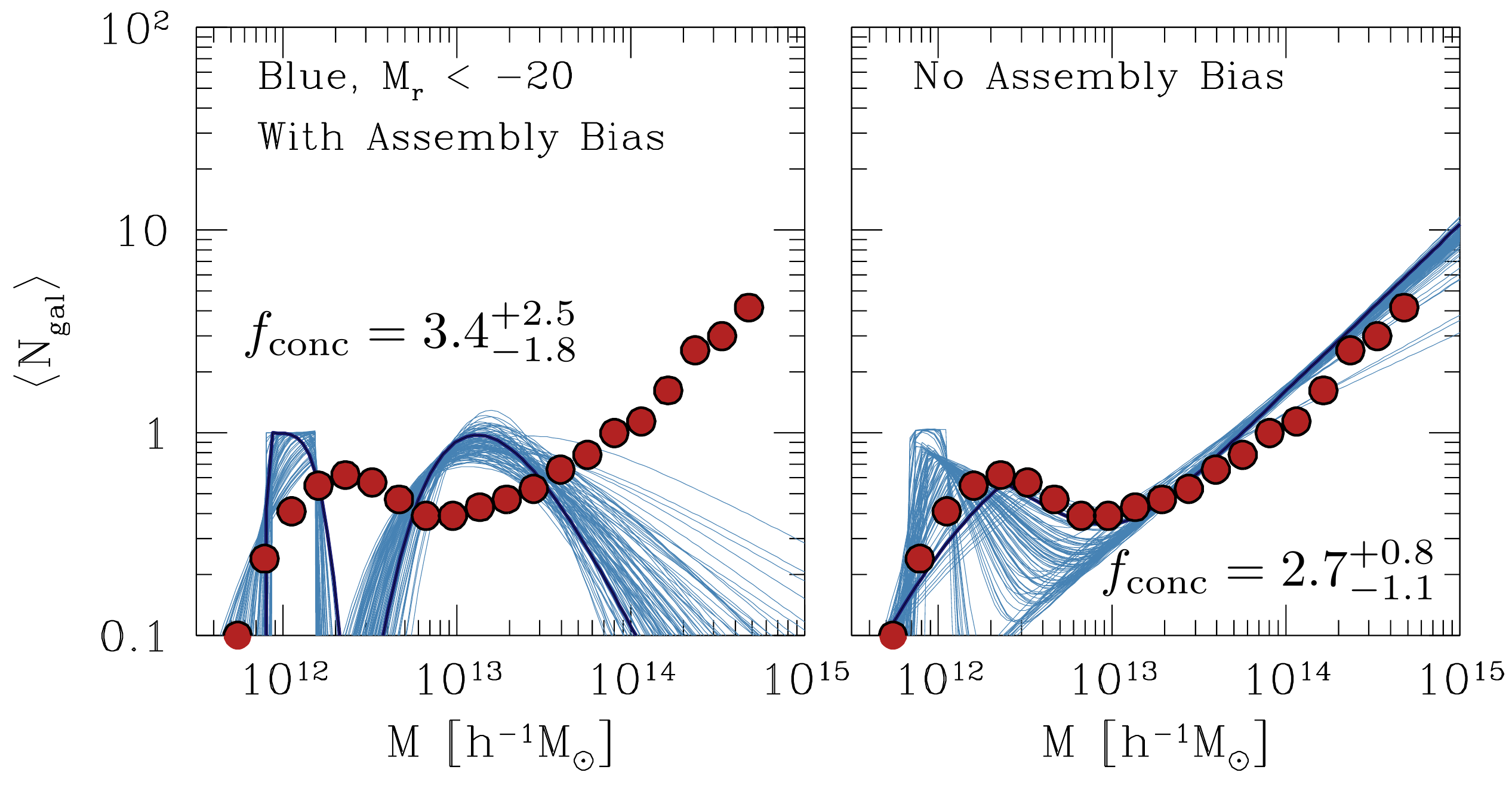}
\includegraphics[width=8.5cm]{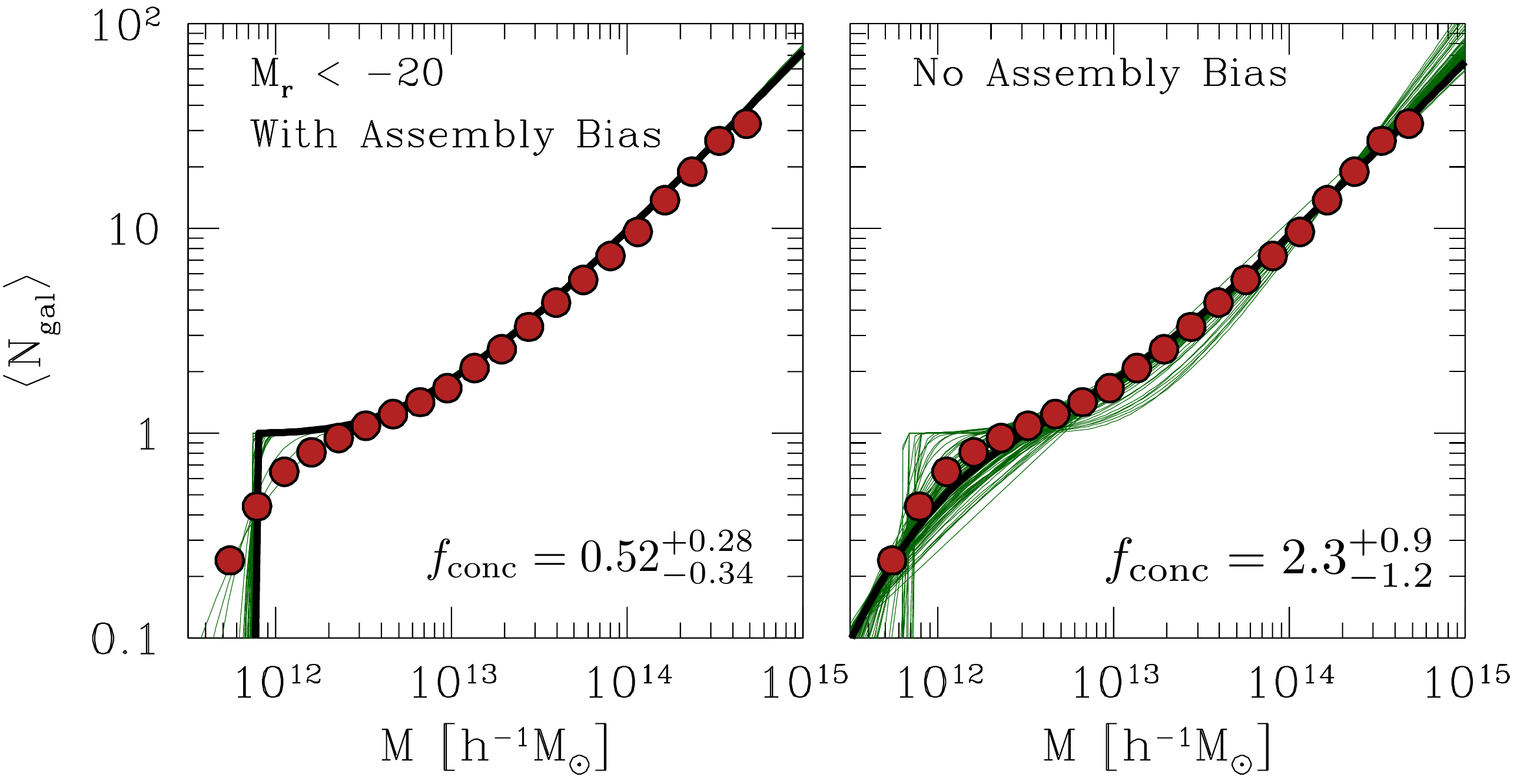}
\caption{
Comparison of the best-fit HODs for galaxies in the 
$M_r < -20$ sample with the true HOD in the simulation (points). 
This figure is the same as Fig.~\ref{fig:hod20} except that these 
fits were conducted while simultaneously allowing the concentrations 
of the satellite galaxy distributions to vary. The constraints on 
$f_{\mathrm{conc}}$ are shown in each panel and this parameter 
is generally poorly constrained.
}
\label{fig:hod20c}
\end{center}
\end{figure}

%%%%%%%%%%%%%%%%%%%%%%%%%%%%%%%%%%%%%%%%%%%%%%%%%%%%%%%%%%%%%%%%%%%%

Figure~\ref{fig:hod20c} shows the shift in the inferred HODs when the concentrations of the satellite 
galaxy distributions were allowed to vary simultaneously with the HOD parameters. In the case of the 
color fits, we allowed the blue and red galaxies to have distinct values of the concentration parameter 
$f_{\mathrm{conc}}$. The results shown in Fig.~\ref{fig:hod20c} make clear that including the additional 
parameter freedom from varying satellite galaxy concentrations does not eliminate the qualitative biases 
in the inferred HODs that we report in this paper. In fact, new biases are introduced due to the significant 
degeneracies that exist between the HOD parameters and the galaxy concentration parameters. The 
sense of the bias can be gleaned by comparing Fig.~\ref{fig:hod20c} to either Fig.~\ref{fig:hod20nofb} in this 
Appendix or Fig.~\ref{fig:hod20} in \S~\ref{sec:results}. Allowing concentrations to vary tends to drive an 
additional offset in inferred HODs such that satellites become abundant in relatively higher mass halos 
in the samples with no assembly bias and vice versa in samples with assembly bias. This is a relatively 
subtle effect in the luminosity threshold sample (bottom panel of Fig.~\ref{fig:hod20c}), 
but it is more evident in the red sub-sample (top panel of Fig.~\ref{fig:hod20c}). This demonstration 
suffices to show that varying the satellite galaxy spatial distributions 
within hosts does not change the qualitative conclusions of our paper 
that assembly bias can significantly bias inferred HODs from galaxy clustering. While it is possible to 
explore the degeneracies between satellite galaxy concentration and HOD parameters more thoroughly, 
such an exploration would be quite complex and we place it beyond the scope of the present work.

\begin{figure}
\includegraphics[width=7.5cm]{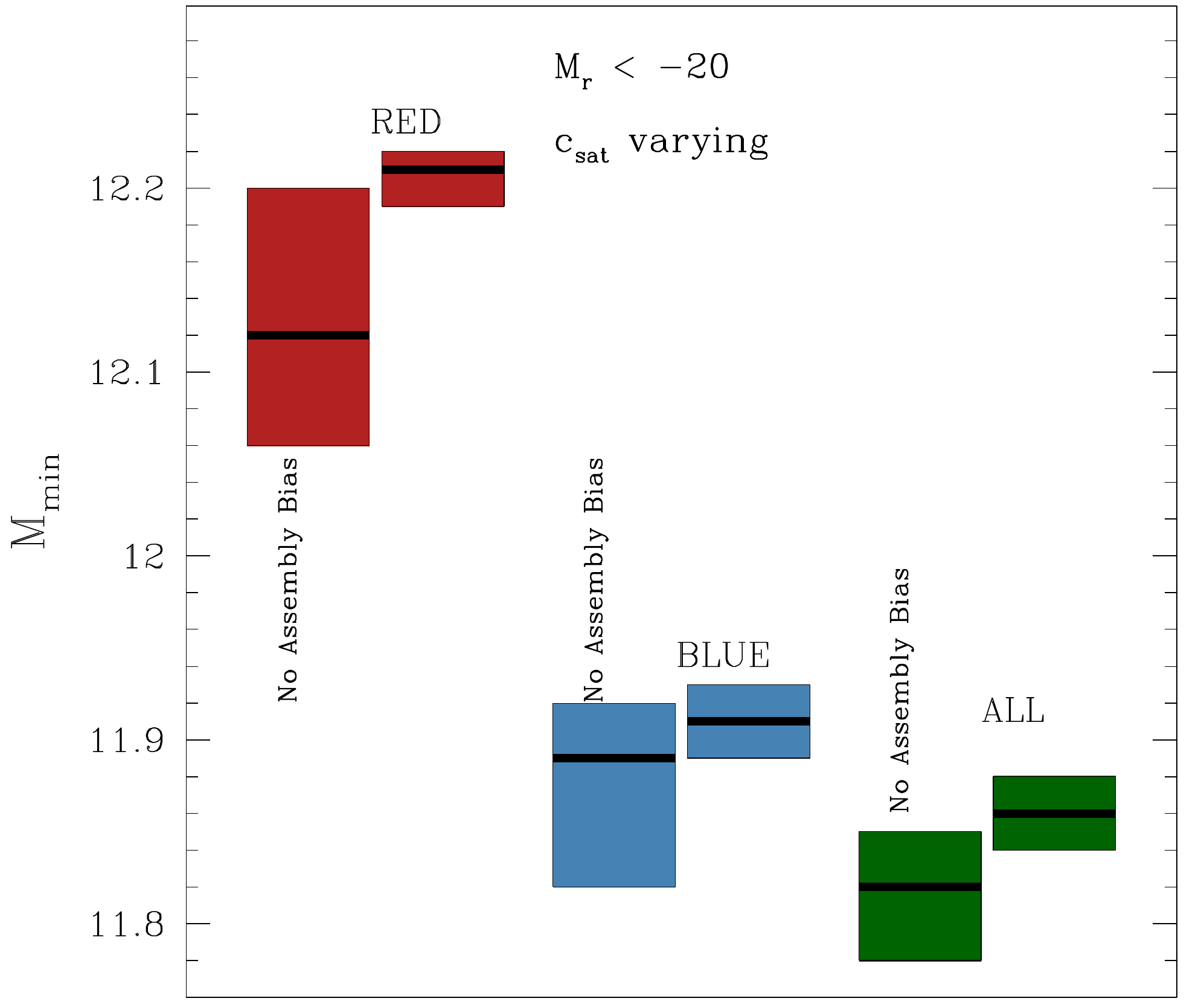}
\includegraphics[width=7.5cm]{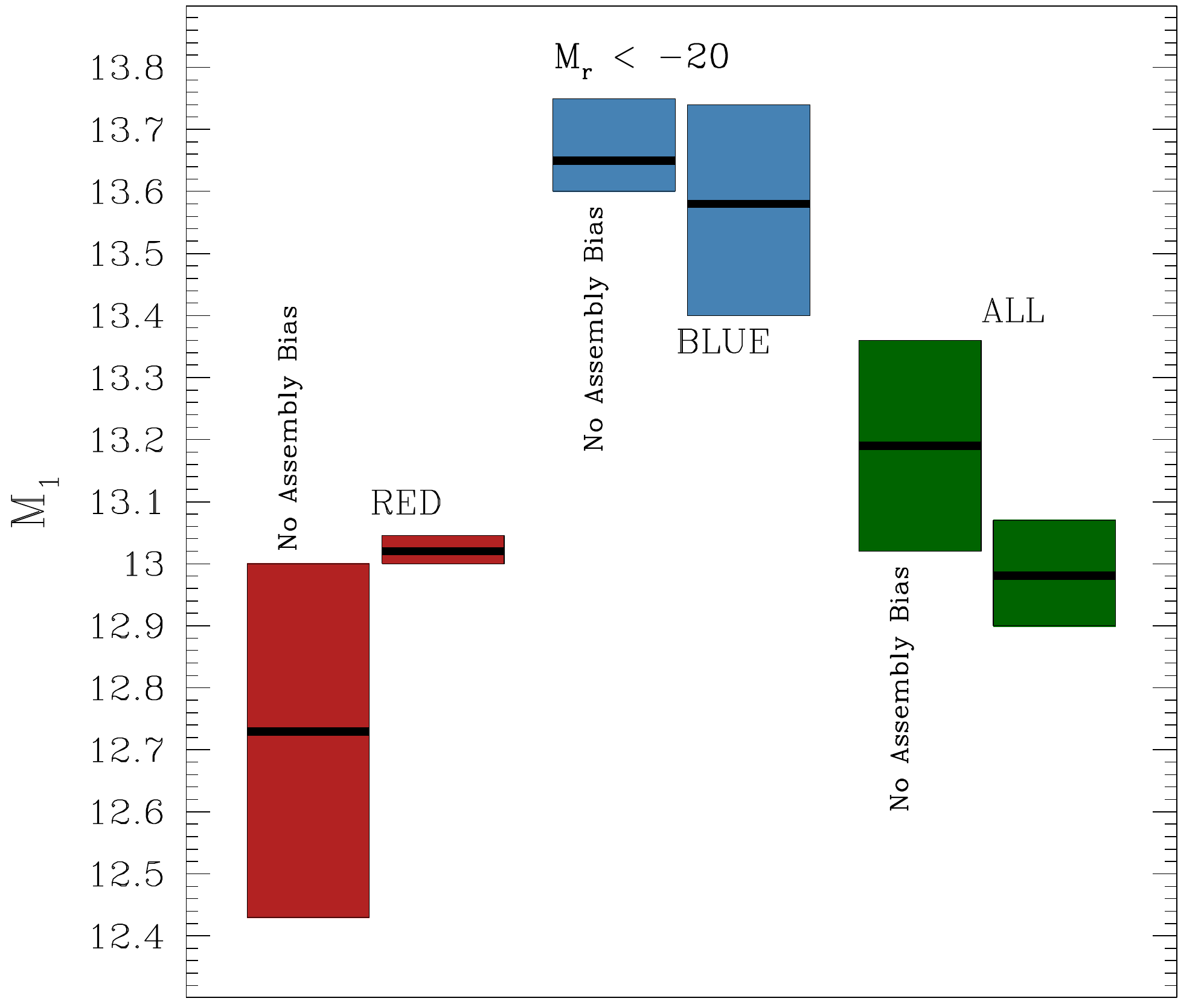}
\includegraphics[width=7.5cm]{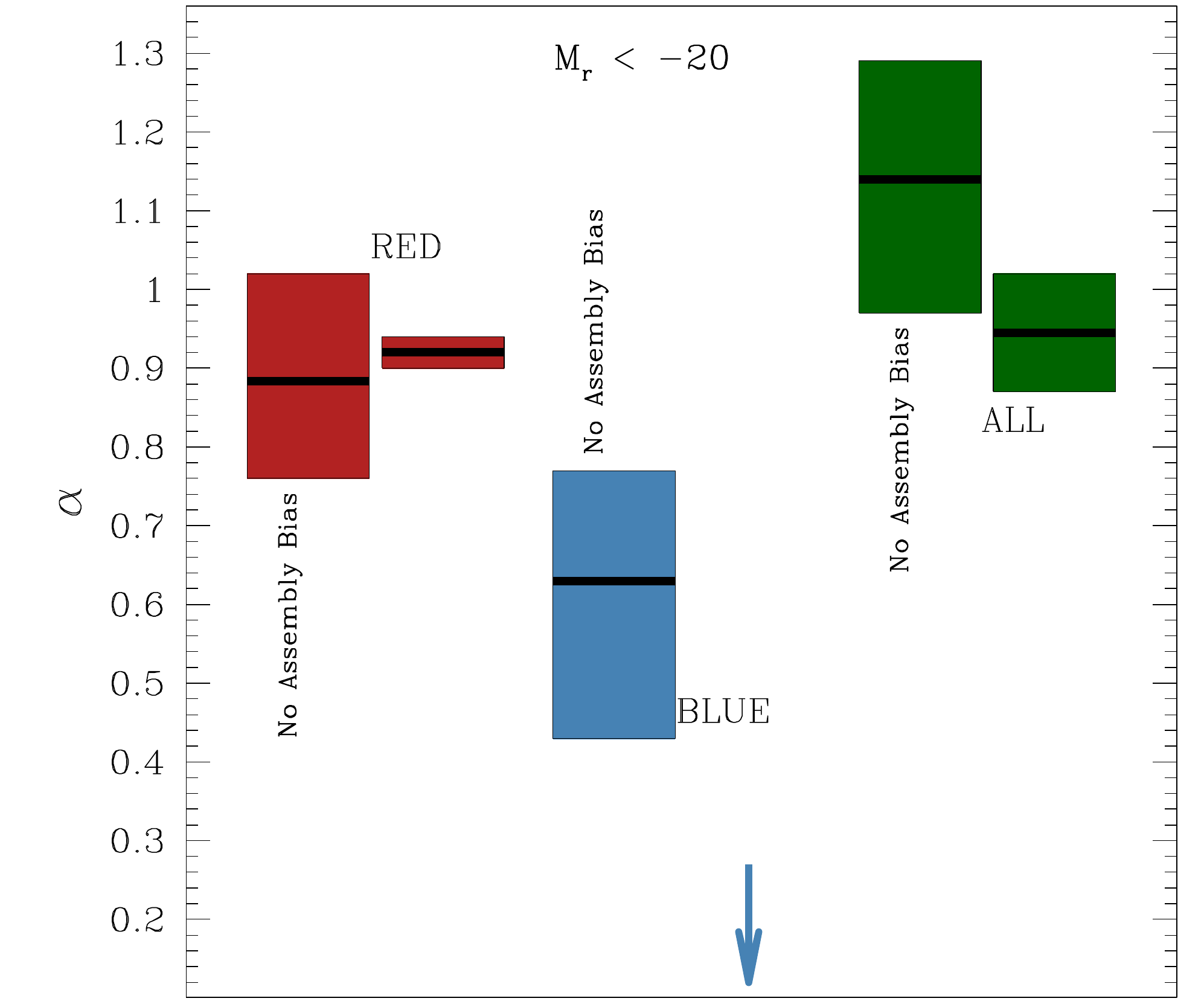}
\caption{
Constraints on HOD parameters inferred from fits with in which the concentration parameter describing 
the spatial distributions of satellite galaxies are allowed to fit. The top panel shows the inferred constrains 
on $M_{\mathrm{min}}$, the middle panel shows inferred constraints on $M_1$, and the bottom panel 
shows inferred constraints on the power-law index of the satellite portion of the HOD, $\alpha$. Fitting 
for concentrations in addition to the standard HOD parameters alters the fits notably (compare to 
Fig.~\ref{fig:mmins}-\ref{fig:alphas}), but does not alter the qualitative point that assembly bias introduces 
additional errors in HOD parameter inferences. 
}
\label{fig:con20c}
\end{figure}

We depict the marginalized constraints on the HOD parameters in Fig.~\ref{fig:con20c} for the $M_r<-20$ 
samples. While the constraints shift systematically from our fiducial case, 
notice that the constraints are offset 
from each other significantly due only to the effect of assembly bias. Indeed, the extra parameter freedom 
afforded by fitting concentrations partially overcompensates for the differences between the samples with 
and without assembly bias and causes the HOD constraints in the threshold samples not split on color to 
be {\em more} significantly offset from each other, rather than less. The conclusion is the same and not surprising. 
The satellite galaxy concentrations cannot serve as a nuisance parameter to guard against assembly bias 
effects and, indeed, may exacerbate systematic errors in inferred HODs induced by assembly bias.

%%%%%%%%%%%%%%%%%%%%%%%%%%%%%%%%%%%%%%%%%%%%%%%%%%%%%%%%%%%
%%%%%%%%%%%%%%%%%%%%%%%%% FIGURE %%%%%%%%%%%%%%%%%%%%%%%%%%%%%
%%%%%%%%%%%%%%%%%%%%%%%%%%%%%%%%%%%%%%%%%%%%%%%%%%%%%%%%%%%

\end{document}